\PassOptionsToPackage{dvipsnames}{xcolor}
\documentclass[preprint]{aastex62-no-date}
\pdfoutput=1
\usepackage{amsmath}
\usepackage{capt-of}   % don't use the package caption -- it causes some kind of conflict.  7/29/15
\usepackage{empheq}
\usepackage{enumitem}
\usepackage{hyperref}
\usepackage{hhline}
\usepackage{longtable}
\usepackage{numprint}
\usepackage{pifont}
\usepackage{xspace}
\usepackage{placeins}
\usepackage{graphicx}
\usepackage{mwe}
\newcommand{\rf}{\textcolor{black}}
\newcommand{\rff}{\textcolor{black}}
\newcommand{\ed}{\textcolor{black}}

\newcommand\pound{\scalebox{0.8}{\raisebox{0.4ex}{\#}}}

\newcommand{\ho}{\ensuremath{H_0}\xspace}

\newcommand{\coverage}{14,000\xspace}
\newcommand{\decalscover}{9000\xspace}
\newcommand{\mzlscover}{5000\xspace}

\newcommand{\lenstrain}{632\xspace}
\newcommand{\lenstot}{1312\xspace}
\newcommand{\lensDC}{1005\xspace}
\newcommand{\lensREX}{307\xspace}
\newcommand{\lenstotnew}{1210\xspace}

\newcommand{\decalsDCrecs}{\ensuremath{22,879}\xspace}
\newcommand{\mzlsDCrecs}{\ensuremath{8761}\xspace}

\newcommand{\cmuDCdecals}{726\xspace}
\newcommand{\cmuDCdecalsA}{115\xspace}
\newcommand{\cmuDCdecalsB}{110\xspace}
\newcommand{\cmuDCdecalsC}{501\xspace}

\newcommand{\cmuDCmzls}{154\xspace}
\newcommand{\cmuDCmzlsA}{29\xspace}
\newcommand{\cmuDCmzlsB}{22\xspace}
\newcommand{\cmuDCmzlsC}{103\xspace}

\newcommand{\totREXrecs}{\ensuremath{7039}\xspace}
\newcommand{\decalsREXrecs}{\ensuremath{5861}\xspace}
\newcommand{\mzlsREXrecs}{\ensuremath{1178}\xspace}

\newcommand{\cmuREXcands}{68\xspace}

\newcommand{\cmuREXdecalsA}{13\xspace}
\newcommand{\cmuREXdecalsB}{6\xspace}
\newcommand{\cmuREXdecalsC}{42\xspace}

\newcommand{\cmuREXmzlsA}{2\xspace}
\newcommand{\cmuREXmzlsB}{1\xspace}
\newcommand{\cmuREXmzlsC}{4\xspace}

\newcommand{\cmuREXA}{15\xspace}
\newcommand{\cmuREXB}{7\xspace}
\newcommand{\cmuREXC}{46\xspace}
\newcommand{\shieldnewtot}{364\xspace}
\newcommand{\shieldA}{57\xspace}
\newcommand{\shieldB}{60\xspace}
\newcommand{\shieldC}{247\xspace}

\newcommand{\lensA}{216\xspace}

\newcommand{\lensAnew}{193\xspace}

\newcommand{\zerrmaxA}{3.7\xspace}

\newcommand{\lensBnew}{175\xspace}

\newcommand{\lensCnew}{842\xspace}
\newcommand{\lensCplusnew}{360\xspace}

\newcommand{\lensaboveCplus}{728\xspace}

\newcommand{\lensknown}{102\xspace}
\newcommand{\lensknownC}{55\xspace}
\newcommand{\lensknownCplus}{27\xspace}
\newcommand{\tractor}{\textit{The Tractor}\xspace}
\newcommand{\twopr}{^{\prime \prime}}
\newcommand{\zl}{\ensuremath{z_{lens}\xspace}}

\accepted{December 22, 2020}
\submitjournal{ApJ}

\shorttitle{Strong Lenses in DESI Legacy Surveys}
\shortauthors{Huang, Storfer, Gu, Ravi, Pilon, Sheu, Venguswamy et al.}

\begin{document}
%\title{Discovering 1000 New Strong Gravitational Lens Candidates in the DESI Legacy Imaging Surveys}
\title{Discovering New Strong Gravitational Lenses in the DESI Legacy Imaging Surveys}

%%%% DECaLS Imaging Survey

\correspondingauthor{Xiaosheng Huang}
\email{xhuang22@usfca.edu}
%, august.muench@aas.org}

\author[0000-0001-8156-0330]{X.~Huang}
\affiliation{Department of Physics \& Astronomy, University of San Francisco, San Francisco, CA 94117-1080}

\author[0000-0002-0385-0014]{C.~Storfer}
\affiliation{Department of Physics \& Astronomy, University of San Francisco, San Francisco, CA 94117-1080}

\author{A.~Gu}
\affiliation{Department of Physics, University of California, Berkeley, Berkeley, CA 94720}
\affiliation{Department of Electrical Engineering \& Computer Sciences, University of California, Berkeley, Berkeley, CA 94720}

\author{V.~Ravi}
\affiliation{Department of Computer Science, University of San Francisco, San Francisco, CA 94117-1080}

\author{A.~Pilon}
\affiliation{Department of Physics \& Astronomy, University of San Francisco, San Francisco, CA 94117-1080}

\author{W.~Sheu}
\affiliation{Department of Physics, University of California, Berkeley, Berkeley, CA 94720}
\affiliation{Department of Electrical Engineering \& Computer Sciences, University of California, Berkeley, Berkeley, CA 94720}

\author{R.~Venguswamy}
\affiliation{Department of Computing, Data Science, and Society, University of California, Berkeley, Berkeley, CA 94720}

% \author{S.~Bailey}
% \affiliation{Physics Division, Lawrence Berkeley National Laboratory, 1 Cyclotron Road, Berkeley, CA, 94720}

% \author[0000-0002-5042-5088]{S.~Bailey}
% \affiliation{Physics Division, Lawrence Berkeley National Laboratory, 1 Cyclotron Road, Berkeley, CA, 94720}

\author{S.~Banka}
\affiliation{Department of Electrical Engineering \& Computer Sciences, University of California, Berkeley, Berkeley, CA 94720}

\author[0000-0002-4928-4003]{A.~Dey}
\affiliation{NSF's National Optical-Infrared Astronomy Research Laboratory, 950 N. Cherry Ave., Tucson, AZ 85719}

\author[0000-0003-1838-8528]{M.~Landriau}
\affiliation{Physics Division, Lawrence Berkeley National Laboratory, 1 Cyclotron Road, Berkeley, CA, 94720}

\author[0000-0002-1172-0754]{D.~Lang}
\affiliation{Dunlap Institute, University of Toronto, Toronto, ON M5S 3H4, Canada}
\affiliation{Department of Astronomy \& Astrophysics, University of Toronto, Toronto, ON M5S 3H4, Canada} 
\affiliation{Perimeter Institute for Theoretical Physics, Waterloo, ON N2L 2Y5, Canada}

\author[0000-0002-1125-7384]{A.~Meisner}
\affiliation{NSF's National Optical-Infrared Astronomy Research Laboratory, 950 N. Cherry Ave., Tucson, AZ 85719}

\author[0000-0002-2733-4559]{J.~Moustakas}
\affiliation{Department of Physics and Astronomy, Siena College, 515 Loudon Rd., Loudonville, NY 12211}

\author{\ed{A.D.~Myers}}
\affiliation{Department of Physics \& Astronomy, University of Wyoming, 1000 E. University, Dept 3905, Laramie, WY 8207}

\author{R.~Sajith}
\affiliation{Department of Physics, University of California, Berkeley, Berkeley, CA 94720}
\affiliation{Department of Electrical Engineering \& Computer Sciences, University of California, Berkeley, Berkeley, CA 94720}

\author[0000-0002-3569-7421]{E.F.~Schlafly}
\affiliation{Lawrence Livermore National Laboratory,
7000 East Ave., Livermore, CA 94550-9234}

\author[0000-0002-5042-5088]{D.J.~Schlegel}
\affiliation{Physics Division, Lawrence Berkeley National Laboratory, 1 Cyclotron Road, Berkeley, CA, 94720}

%\collaboration{(The DESI Collaboration)}

\begin{abstract}
\ed{We have conducted a search for new strong gravitational lensing systems in the Dark Energy Spectroscopic Instrument Legacy Imaging Surveys’  Data Release 8.  We use deep}
%%We search in the Dark Energy Spectroscopic Instrument Legacy Imaging Surveys for new strong lensing systems by using deep 
residual neural networks, building on previous work presented in \ed{\citet{huang2020a}}.
These surveys together cover approximately one third of the sky
%%%% $\sim$ 14,000~deg$^2$ 
visible from the northern hemisphere, reaching a $z$-band \ed{AB}
magnitude of $\sim 22.5$. 
We compile a training sample that consists of  
known lensing systems as well as  
non-lenses 
in the Legacy Surveys and the Dark Energy Survey.
After applying our trained neural networks to the survey data, we visually inspect and rank images 
with probabilities 
above a threshold.
Here we present \lenstotnew new strong lens candidates.

\end{abstract}

\section{Introduction}
\label{sec:intro}
%Strong gravitational lensing is a powerful tool in cosmology and astrophysics.
%\citep{walsh1979a, lynds1986a, soucail1987a, soucail1988a, paczynski1987a} 
%are powerful tools in astrophysics and cosmology.
%the most precise method for measuring the mass structure and substructure of galaxies at cosmological distances. 
%Strong gravitational lensing systems are a powerful tool for cosmology.  
%Strong gravitational lensing is a stunning manifestation of the warping of space-time and a powerful tool for cosmology.  
Strong gravitational lensing systems 
%%%%--- stunning manifestations of the warping of space-time --- 
are a powerful tool for astrophysics and cosmology.
They have been used to study how dark matter is distributed in galaxies and \rf{galaxy} clusters 
\citep[e.g.,][]{kochanek1991a, koopmans2002a, 
bolton2006a, koopmans2006a, vegetti2009a, 
tessore2016a, monna2017a, jauzac2018a, 
shajib2019a, meneghetti2020a}, 
and are uniquely suited to probe 
%\rf{low-mass dark matter halos 
dark matter substructure 
%in the lensing galaxies and along the line of sight} and to test the predictions of CDM
beyond the local universe 
\citep[e.g.,][]{vegetti2014a, vegetti2018a, ritondale2019a, diazrivero2020a}.
%%%% Add this here:  galaxy mass profile modeling (gas, star, DM) and evolution (may not mention the added benefit to lensed QSO modeling)
\rf{Furthermore, by modeling galaxy-scale strong lenses 
%typically with comparable amounts of dark and luminous matter within the radius of the brightest arc 
%or the Einstein ring \citep[e.g.,][]{auger2010a},
 %combining the lensing constraints with the stellar kinematics, and 
 %modeling each lens 
as the sum of luminous and dark components,  
 % use the combined dataset to 
the mass-to-light ratio and inner density profile of the dark matter halo can be simultaneously constrained  \citep[e.g.,][]{auger2010a, sonnenfeld2019b, shajib2020b}.
%the stellar and dark components of lensing galaxies can be differentiated \citep[e.g.,][]{shajib2020b}  
%, are well-suited for the study of both dark and luminous matter
%collett2018a},
%This can be done in practice by modeling each lens as the sum of a stellar and a dark matter component and 
%The population distribution of these parameters can be inferred statistically from a larger number lenses (Sonnenfeld et al 2019). 
%The average stellar mass to light ratio of the population of lenses and the mean central dark matter density profile slope , 
%and the amount of adiabatic contraction in the dark matter halos \citep{shajib2020a}. 
%Cite Birrer 2020a properly.
% With systematic error under control \cite[e.g.][]{birrer2020a, shajib2020a}, for lensed time variable sources (quasars or SNe) to truly make progress on resolving the current tension, statistical error must be well under 1\% (Treu \& Marshall 2016).
% Or say that to control systematic uncertainties, one needs more lenses that cover a wide range of redshift.
 % the typical Einstein radius for the lensed qso's used in TD measurements is 1" in Wong+ 2019.
%The determination of the inner dark matter density profile can provide unique constraints both on dark matter physics, for example by testing the self-interaction scenario, and on baryonic physics: the dark matter distribution responds to processes such as adiabatic contraction from gas infall (Gnedin et al 2011) and feedback from the central supermassive black hole (Martizzi et al 2013) and galaxy mergers (Tortora et al 2018, 2019).
%Measurements of mass distributions
Such measurements for a large number of lensing systems 
%(which can be combined with dynamical mass measurements) 
over a wide range of redshifts makes it possible to  study the structural evolution 
%of the internal structure 
of massive elliptical galaxies, 
and possibly in the future, lenses of any Hubble type
%the mass-to-light ratio and dark matter concentration and profile slope
%galaxy mass profiles
\citep[e.g.,][]{sonnenfeld2015a, nightingale2019a}.
%, shajib2020a}.
For nearby strong lensing galaxies, 
extra-galactic tests of General Relativity can be performed 
by combining lens modeling with 
%rotation curves %or other 
spatially resolved 
%dynamic and
stellar kinematic observations \citep{collett2018a}. 
%significantly constrain measurement uncertainties and enable the dissection of the 
%Spiniello et al. 2011; Barnab`e et al. 2012) 
}

\rf{For redshifts beyond the range that lensing galaxies can typically probe, 
together with high resolution imaging (\rff{from the \textit{Hubble Space Telescope}, adaptive optics, or the \textit{James Webb Space Telescope} in the near future}), 
strong lensing as an cosmic telescope magnifies spectral \citep[e.g.,][]{cornachione2018a} 
and spatial features \citep[e.g.,][]{marshall2007a, patricio2019a, vanzella2020a} of the lensed distant galaxies,
providing the only way to study the morphology and internal structures of galaxies at sub-kpc scales at high redshifts that can extend to $z > 2$.}
%sometimes at physical scales only possible through lensing.} 
%for comparisons with
%sometimes to a physical scale comparable to 
%studies of local galaxies.}

\rf{
Recent measurements of the Hubble constant \ho span a range of $\sim$10\% \citep[e.g.,][]{abbott2017a, abbott2018b, riess2019a, wong2019a, freedman2019a, freedman2020a, planck2020a, khetan2020a, philcox2020a, choi2020a}, 
and significant tension between predictions for \ho based on early-universe observables 
%% don't remove: Planck -- sound horizon; Philcox -- matter-rad equality scale
and direct late-universe measurements remain \citep[e.g.,][]{verde2019a}.
%%%% with due to unidentified systematics or new physics.
%with the value inferred from CMB by assuming \lcdm at the low end and some of the most precise local measurements at the other \citep{riess2019a}.
%  direct measurement from Cepheid-calicbrated SNe~Ia and the value inferred from \lcdm from CMB (riess, Planck).  
%The time delays of multiply-lensed QSOs provide one of the most competitive direct measurement of \ho that have the potential to address this 
%discrepancy \citep[e.g.,][]{wong2019a}}.
Multiply-lensed supernovae (SNe) are 
%potentially superior 
ideal for measuring time delays and \ho because of their well-characterized light curves,
%on a time scale of months, 
and in the case of Type~Ia, with the added benefit of standardizable luminosity \citep{refsdal1964a, treu2010a, oguri2010a}.}
In recent years, strongly lensed supernovae, 
both core-collapse \citep{kelly2015a, rodney2016a} and Type Ia \citep{quimby2014a, goobar2017a}, have been discovered.
Time-delay \ho measurements from multiply imaged supernovae  
% together with 
% existing approaches of direct measurement of \ho
% measurements from distance ladders \citep[e.g.,][]{riess2019a, freedman2019a, freedman2020a}
% and lensed quasars with the assumption of flat $\Lambda$CDM \citep[e.g.,][]{suyu2010a, suyu2013a, treu2016a, bonvin2017a, wong2019a} 
 can therefore be an important independent approach to address the discrepancy between \ho measured locally and the value inferred from the CMB \citep[e.g.,][]{goldstein2017a, goldstein2018a, goldstein2018b, wojtak2019a, pierel2019a, suyu2020a}.
 
 \rf{Furthermore time delay \ho measurements are a powerful complement to other independent measurements of the dark energy equation of state
 %, including BAO, CMB, and SNe~Ia
\citep[e.g.,][]{linder2011a, treu2016a}.
%\rf{%it would be nice if I can cite tristan, but not necessary
%Such a tension, if not resolved by systematic effects, indicates new physics beyond flat $\Lambda$CDM \citep[e.g.,][]{poulin2019a, agrawal2019a}.
%assuming the Friedmann–Lema\^{i}tre–Robertson–Walker (FLRW) metric, 
%\rf{Among the many significant application of strong gravitational lensing:
%1. curvature; 2. galaxy mass profile modeling (gas, star, DM) and evolution (may not mention the added benefit to lensed QSO modeling); 3 using it as a telescope.
%Remember, I don't have say everything, or cite everything.}
Beyond the flat $\Lambda$CDM cosmological model, 
by combining strong lensing time delay of multiply-imaged time-varying sources 
and SNe~Ia distance measurements, 
one can 
determine \ho in a model-independent way and  
measure the spatial curvature \citep[e.g.,][]{li2018a, taubenberger2019a, collett2019a}, 
and test the 
Friedmann–Lema\^{i}tre–Robertson–Walker (FLRW) metric \citep{rasanen2015a}.
}

%%and say what one can do with those -- constrain slope at different radii: does Shajib say that?.}  
%Combining lensing with rotation curves or other dynamic and kinematic data can further differentiate the mass distributions of stellar, 
%gas, and dark matter components of galaxies 
%%%%\rf{====$>$ Does Taubenberger  determine curvature?? (I think it is model-independent -- and does that mean it also determines the curvature, or does it?  Or should I lump it with the other lensed qso papers?) $<$====} 
%Furthermore, strong lensing systems can provide constraints on the dark energy equation of state \citep[e.g.,][]{oguri2010a, treu2010a, collett2014a, linder2011a, treu2016a, suyu2018a} 
%%cao2015a -- there's a cao 2014 on lensing by BH of GW
%\citep{linder2011a;}.
%%Strong lensing can be a powerful tool in astrophysics: 
%detailed study of the background source \citep[e.g.,][]{patricio2019a} and supernova progenitor physics
%\citep[e.g.][]{kasen2010a, Maeda2018a}.

\rf{For many of these analyses, 
the available sample sizes of confirmed strong lenses is a major limiting factor.}
%\rf{Many of the studies above would benefit from a large strong lens sample.}
%The application of strong lensing to cosmology has been limited by the available sample size of the lenses.
In the last few years, several groups have used \ed{convolutional} neural networks to search for strong lensing systems in photometric surveys including, in \ed{increasing} sky coverage, 
%(and descending depth), 
CFHTLS \citep{jacobs2017a}, KiDS \citep{petrillo2017a, petrillo2019a, li2020a}, 
DES \citep{jacobs2019a, jacobs2019b}, and
Pan-STARRS \citep{canameras2020a}.

Data release 8 (DR8) of the DESI Legacy Surveys\footnote{\url{http://www.legacysurvey.org/}}
\citep{dey2019a},
for which at least $z$ band is observed with a 4-m telescope,
covers \ed{$\sim$\coverage~deg$^2$},
three times the size of the DES footprint.
In \citet[][H20]{huang2020a}, 
we identified hundreds of new strong lenses in the Legacy Surveys Data Release~7 (DR7) 
by using a residual neural network. 
%Apart from PanSTarrs this is the largest optical imaging survey in the northern hemisphere.  
%it is how many magnitudes deeper than PanStarrs.
In this paper, 
building on H20, 
%we make two significant improvements that 
we have significantly improved the efficiency of \ed{the neural network} 
and report the discovery of new strong lensing systems \rf{over a wide ranged of redshifts} in DESI Legacy Surveys DR8.
%\ed{\footnote{\url{http://www.legacysurvey.org/dr8/description/}}}.
 
%%Finally cluster lensing can be used to directly test LCDM (e.g., Kneid \& Natarajan 2012, and Natarajan et al. 2017) [later in the paper (and in the proposal), make a separate section on cluster lenses.  Also another section to highlight some interesting cases --- such as clear Einstein crosses.]

This paper is organized as follows.  
A brief description of the Legacy Surveys is given in \S~\ref{sec:observations}.  
In \S~\ref{sec:model-train}, 
we describe our methodology,
%and training sample, 
including the improvements we have made on H20.
In \S~\ref{sec:results}, we show the inference results and present our best strong lensing system candidates.  
We discuss our results in \S~\ref{sec:discussion}, and conclude in \S~\ref{sec:conclusion}.

%\vfill

\FloatBarrier
\section{Observations}
\label{sec:observations}
The Legacy Imaging Surveys consist of three projects: the Dark Energy Camera Legacy Survey (DECaLS), observed by the Dark Energy Camera \citep[DECam;][]{flaugher2015a} on the 4-m Blanco telescope at the Cerro Tololo Inter-American Observatory; 
the Beijing-Arizona Sky Survey (BASS), by the 90Prime camera \citep{williams2004a} on the Bok 2.3-m telescope owned and operated by the University of Arizona located on Kitt Peak;
and the Mayall $z$-band Legacy Survey (MzLS), by the Mosaic3 camera \citep{dey2016a} on the 4-meter Mayall telescope at Kitt Peak National Observatory. 
Together they cover
$\sim$\coverage deg$^2$ of the extragalactic sky visible from the northern hemisphere with at least three passes in each of the three bands,  $grz$.
The $5\,\sigma$ $z$-band median limiting \ed{AB} magnitude is 22.5~mag for galaxies with an exponential disk profile with $\mathrm{r}_{\mathrm{half}} = 0.45\twopr$.
%%$\mathrm{r}_{i, \, \mathrm{half}} = 0.45\twopr$. 

The combined survey footprint is split into two contiguous areas by the Galactic plane. 
%%%%The imaging is conducted using a unique strategy of dynamic observing that results in a survey of nearly uniform depth. 
DECaLS covers the $\sim$\decalscover~deg$^2$ $\delta \lesssim +32^\circ$ sub-region of the Legacy Surveys, while BASS/MzLS the $\sim$~\mzlscover~deg$^2$ northern sub-region. 
Figure~\ref{fig:dr8-footprint} shows the different regions in the the Legacy Surveys footprint and the depth of the $z$-band observation.

\begin{minipage}{\linewidth}
\makebox[\linewidth]{
  \includegraphics[keepaspectratio=true,scale=0.4]{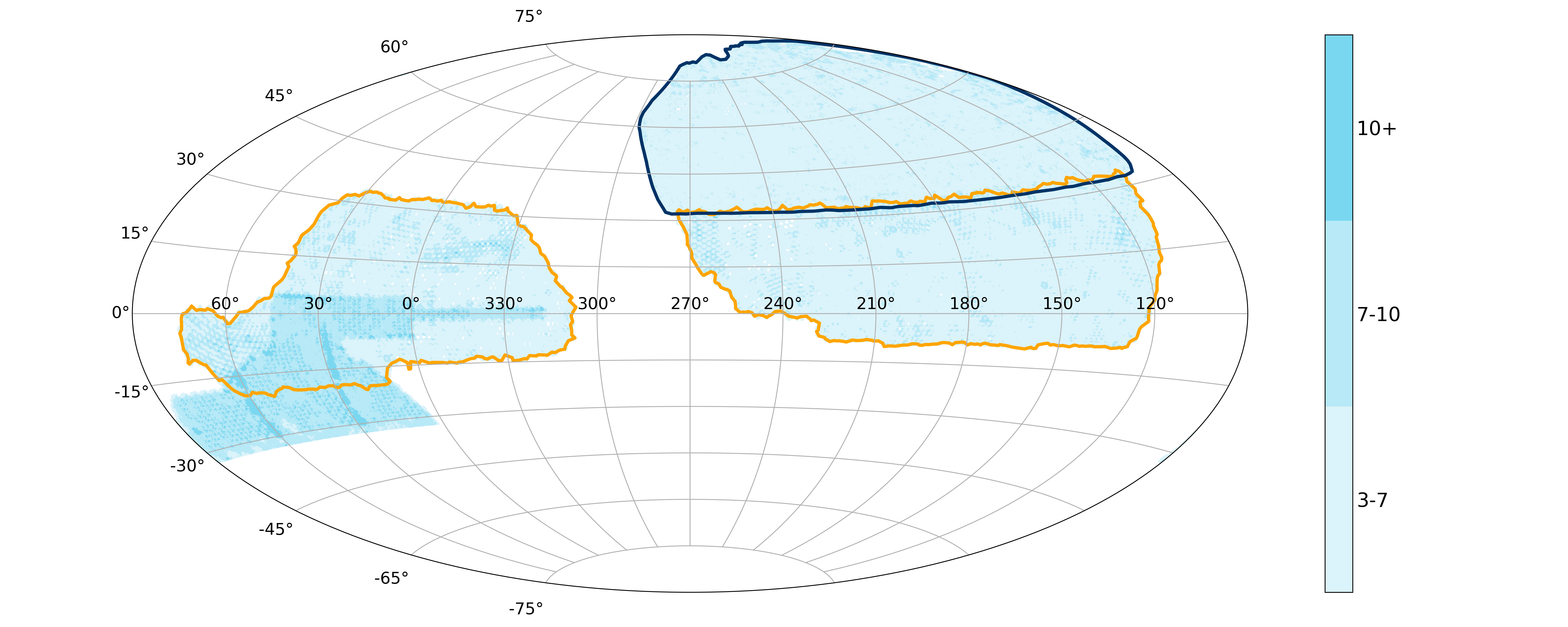}}
\captionof{figure}{
The DESI Legacy \ed{Imaging} Surveys footprint in an equal area Aitoff projection in equatorial coordinates.
%%%% , approximately between \ed{$-18^\circ < \delta < +84^\circ $  ($|b| > 18^\circ $ in Galactic coordinates)}.
\ed{The blue and gold borders approximately outline the north (coinciding with MzLS/BASS) and south (residing within DECaLS) regions of the spectroscopic survey, respectively.}
%%The sky coverage 
Slightly above $\delta = 32^\circ$, 
there is a small amount of overlap between the imaging surveys.
Patches with different shades of blue indicate the depth in $z$ band: 
light for between three and seven passes;
medium, between seven and ten, 
and dark, greater than ten. 
\ed{Note that DECaLS includes the DES footprint, but has incomplete coverage below $\delta \approx-32^\circ$ in Data Release 8.}
%% \ed{Even though DECaLS includes all of DES, at DR8, the re-processing of the image data with \tractor source extraction and typing below $-32^\circ$ was incomplete.}
%Note that observations in $grz$ bands of the northern part ($\delta > -18^\circ$) of the DES (black dotted outline) is included in DECaLS.
%The region that extends further south than DECaLS (the red outline) is also part of DES.  
% Even though it is not part of the Legacy Surveys, 
%The objects in this region have been reliably cataloged and typed by \tractor down to $\delta \approx-32.375^\circ$.  
%thanks to the efforts by the Legacy Surveys team.
%The gray region south of the black dotted outline was not part of the the Legacy Surveys at the time of DR7 \citep[see Figure 1 of][]{dey2019a}.
%% Do I want to say this -- The exposure is shorter and the number of passes is greater with comparable depth (citation).
}
\label{fig:dr8-footprint}
\end{minipage}

For DECaLS (\ed{gold} outline in Figure~\ref{fig:dr8-footprint}), 
the delivered image quality has FWHM of approximately 1.29, 1.18, $1.11\twopr$ for $g$, $r$, and $z$ bands, respectively.  
For the $\delta \gtrsim +32^\circ$ (\ed{blue} outline in Figure~\ref{fig:dr8-footprint}) footprint of the Legacy Surveys, 
MzLS has imaged in $z$-band that complemented the BASS $g$- and $r$-band observations in the same sub-region, 
with median delivered image quality of approximately 
% of the delivered image quality 
$1.61\twopr$, $1.47\twopr$, and 1.01$\twopr$ for $g$, $r$, and $z$ bands, respectively.

The Legacy Surveys used \tractor package \citep{lang2016a} 
as a forward-modeling approach to perform source extraction on pixel-level data. 
\tractor takes as input the individual images from multiple exposures in multiple bands, with different seeing in each. 
%%%%12 Publicly available at https://github.com/dstndstn/tractor
After source detection,
% As described on the Legacy Surveys website
the point source (``PSF") and spatially extended (``REX", round exponential galaxy) models are computed for every source and the better of these two is used when deciding whether to keep the source.
The spatially extended sources (REX) are further classified if $\chi^2$ is improved by 9 
%%(i.e., approximately a $3\,\sigma$ improvement) 
by treating it as a deVaucouleurs (DEV), an exponential (EXP) profile, or a composite of DEV + EXP, 
or COMP\footnote{\url{http://legacysurvey.org/dr8/description/}}.  
The same light profile (EXP, DEV, or COMP) is consistently fit to all images in order to determine the best-fit source shape parameters and photometry.

The categories of DEV and COMP 
%definitively 
indicate the classification of elliptical galaxies.  
Given that the vast majority of lensing events are caused by massive early type galaxies, 
we decided to first target objects with DEV and COMP classifications, and then REX, which tend to be smaller and/or fainter galaxies.

\section{The Training Sample and Residual Neural Networks}
\label{sec:model-train}
Deep convolutional neural networks (CNNs) 
and their variations have been shown to be highly effective in image recognition.  
In recent years, this technique has been successfully applied to recognize instances of strong lenses in simulations \citep[e.g.,][and references therein]{metcalf2018a}.
As mentioned in \S~\ref{sec:intro}, several groups have searched for and found strong lenses in existing imaging surveys.  
In all these efforts, at least the positive examples (lenses) in the training samples were constructed from simulated lenses, typically on the order of $\mathcal{O} (10^5)$.
This is because the number of known lenses, on the order of several hundred, 
is thought to be too small to effectively train CNN models.
Building on H20, we continue to 
% take a different approach by using 
use only \textit{observed} data for lenses and non-lenses in our training sample.
In this section (and \S~\ref{sec:results}), 
we show we can train deep neural networks with a much smaller sample and far fewer positive examples 
and achieve comparable if not superior results.
%that are on par with, if not better than, 
%the results of other groups.
In \S~\ref{sec:train}, we describe our training sample.
We show the training results using the residual neural network from \citet{lanusse2018a} in \S~\ref{sec:L18-model}.
Finally in \S~\ref{sec:shielded-model}, 
we present an improved neural network model.

%%%%-------------------Training Sample------------------------
\subsection{Training Sample}\label{sec:train}

% A catalog of known lenses in the Legacy Surveys is also necessary in order to identify new lens candidates.  Both DECaLS and DES used DECam %[do I need a referecne?   (see D19).   DES has $grizY$ observations with greater depths (see Figure~\ref{fig:dr8-depth-map}) in the three bands common with Legacy Surveys.  
The Master Lens Database\footnote{\url{http://admin.masterlens.org/index.php}} \ed{\citep{moustakas2012a}}, which contains hundreds of lensing events 
\ed{discovered prior to 2016},
provided the initial list for the lens training sample.
\setcitestyle{notesep={; }} \hspace{-5mm} 
We have since added several hundred more lenses \ed{and lens candidates} from more recent publications \citep[][and H20]{carrasco2017a, diehl2017a, pourrahmani2018a, sonnenfeld2018a, wong2018a, jacobs2017a, jacobs2019a, jacobs2019b}.
%The known lensing events to-date is around 1000. 
\setcitestyle{notesep={, }} \hspace{-5mm}
Initially our primary goal was to find lenses in DECaLS, 
part of which was observed by \ed{DES}.
%% and DES (using only $grz$ band data).
Therefore in total we have identified 632 previously known lenses or lens candidates in DECaLS and DES, to be used in our training sample. 
%Through human inspection, we deem \rf{613} as discernible lenses in the Legacy Surveys (199) and DES (414) footprints. %(the full list is in Appendix A).  
%Of these, 240 are in the LS.  %The rest are in the DES footprint.  DR1 for DES is available on the legacysurvey.org skyviewer.  
For the lenses in the DES footprint, we only include $grz$ bands.
%, to be consistent with the LS.
We also assemble $\sim21,000$ non-lens image cutouts from DECaLS and DES, 
all with at least three passes in each of the $grz$ bands, a $z$-band mag $<20.0$, and typed as DEV or COMP, 
randomly distributed in the footprint.
%%. This is 60\% larger than the number of the non-lenses used in the training sample in H20.
% in the training set.
%Of these, 5000 are galaxies categorized as DEV or COMP in D19 (see \S~\ref{sec:observations}), which are elliptical galaxies, and another 5000 of all types of galaxies. 
%For both cases, we apply a $z$-band magnitude cut of 22.5~mag. 
%(i.e., everything except for the PSF type in Tractor). 
Given that on average we expect one strong lens in $\mathcal{O} (10^4)$ galaxies \rf{\citep[e.g.,][]{collett2015a, jacobs2019b}}
%\citep[e.g.,][]{oguri2010a},
% to be a strong lens,  
incidental inclusion of a lens or two in these randomly selected galaxies is not a significant concern.  
% Though to be thorough one of the co-authors did perform a quick visual inspection and identified and removed only a few suspicious cases. 

In the training sample of H20, 
we found that the images for the lenses tend to be much deeper than the non-lenses.
This led to the neural net during the inference stage  preferentially assigning high probabilities to images with deeper  observations 
whether they are lenses or not.
To correct for this bias, given that many (359) of our lenses in the training sample are from \ed{DES} 
south of $\delta = -18^\circ$ with deeper observations
(see Figure~\ref{fig:lenses-nonlenses-training}), 
we have included in the non-lens sample five thousand random cutout images from the same region.
%%the DES region extending down to $\delta = -37.375^\circ$, with deep observations.  
%% These are especially needed because 
% statistically representative the training sample is.
%The reason for including non-elliptical galaxies is to provide more non-lens configurations for the neural net.  %%Based on the lens sample from Legacy Surveys, this includes 97\% of the lenses.
%  Given we plan to apply our model to real observations, 
%  and given our lower tolerance for false positives (FP's), 
%  we want to be as thorough as possible, 

As with H20, included in the non-lens sample are cutouts selected by eye so as to cover as many non-lens configurations as possible, 
 especially cases that can potentially be confused by the neural net.  
These include spiral galaxies of different sizes and spiral arm configurations, elliptical galaxies, galaxy groups, 
%cutout 
images having objects with different colors (typically a blue galaxy next to a red one), 
cosmic rays appearing in different bands (some of which have curved trajectories), unusual arrangements of galaxies or stars, 
and finally certain data reduction artefacts.

\begin{minipage}{\linewidth}
\makebox[\linewidth]{
	%%%% \begin{minipage}[t]{0.45\textwidth}
  	\includegraphics[keepaspectratio=true,scale=0.35]{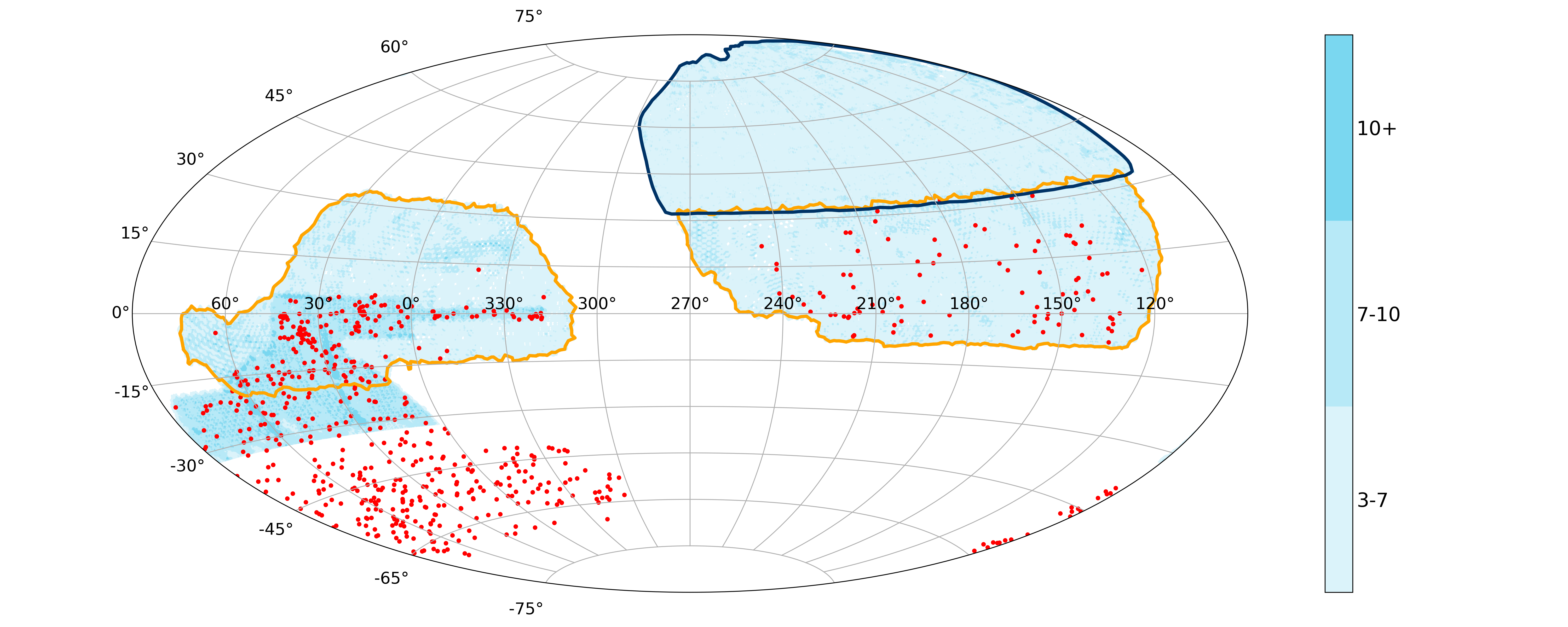}  
}
	%%%% \end{minipage}
	%%%% \begin{minipage}[t]{0.45\textwidth}
    %%%%\includegraphics[keepaspectratio=true,scale=0.2]{{figures/training_samp_nlens}.png}
	%%%%% \end{minipage}
\captionof{figure}{\ed{Previously known lenses or lens candidates 
	%%%%, left panel) and non-lenses (black dots, right panel) 
	%%%% %%%% I decided to not include the figure --- the part below -32.375 needs more explaining.  I can add this later if referee asks for it.]
in our training sample shown as red dots}, 
against the background of the depth map of Legacy Surveys DR8 
(see the caption for Figure~\ref{fig:dr8-footprint}).
The lenses south of the 
%%DECaLS 
\ed{DESI spectroscopic}
footprint (gold outline) are from DES.
%%all with similar depth (10+ passes in $z$ band).
}
\label{fig:lenses-nonlenses-training}
\end{minipage}

\begin{minipage}{\linewidth}
\makebox[\linewidth]{
\includegraphics[width=.40\linewidth]
{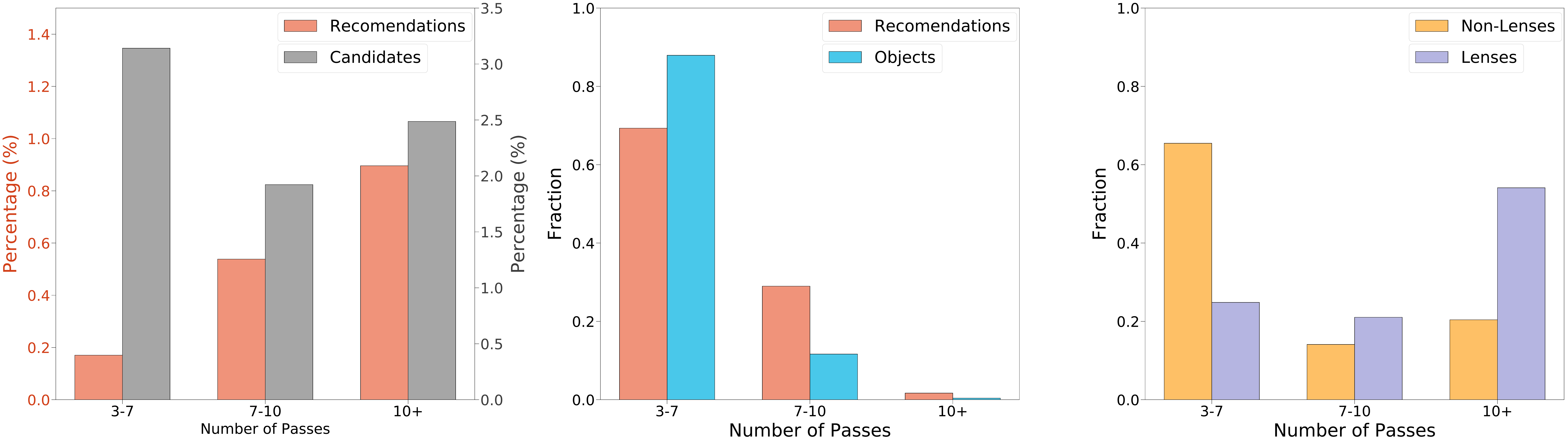}
}	
\captionof{figure}
{The yellow and violet columns show the fractions of lenses and non-lenses in the training sample, respectively\ed{, for the three bins of z-band depth}.  
%% Overall, the images containing lensing systems tend to be deeper.
}
\label{fig:lens-nlens-training-hist}
\end{minipage}

The distribution of the lenses and non-lenses in our training sample is shown in Figure~\ref{fig:lens-nlens-training-hist}.
While fractionally there are still more non-lenses in the shallowest bin and more lenses in the deepest bin, 
overall the disparity between the relative proportions of lenses and non-lenses in each depth bin is much improved 
compared with the training sample in H20.
%%%% (see \ed{their} Figure~10).

\newpage
\subsection{Residual Neural Networks}\label{sec:L18-model}

We use the Residual Neural Network (ResNet) model of \citet[][L18]{lanusse2018a},
%\footnote{\url{https://github.com/McWilliamsCenter/CMUDeepLens}}  
after re-implementing it in TensorFlow\footnote{\url{https://www.tensorflow.org/}}.
We have left their architecture and hyperparameters unchanged (for details, see Section 3.3 of L18), 
except that we double the batch size to 256.
The lens and non-lens images in the training sample are cutouts with a dimension of 101~$\times$~101 pixels, following the specification in the Lens Challenge \citep{metcalf2018a}.  

We split the training sample into training and validation \emph{sets}, with a \ed{ratio} of 7:3.
We then train the ResNet on Google Colab\footnote{\url{https://colab.research.google.com/}} using a GPU (NVIDIA Tesla v100). 
The 120 epochs of training took 4 hours.
% and results are shown in Figure~\ref{fig:loss-epoch} and Figure~\ref{fig:roc}.

The ResNet attempts to minimize the cross entropy loss function:

\begin{equation}\label{eqn:loss}
    \displaystyle-\sum_{i=1}^{N} y_i \log \hat{y}_i+(1-y_i) \log (1-\hat{y}_i)
\end{equation}

\noindent
where $y_i$ is label for the $i$th image (1 for lens and 0 for non-lens), and $\hat{y}_i \in [0,1]$ is the model predicted probability.

While the loss function is monitored during the training process to determine the point of termination, the overall performance of the trained model is typically assessed by the Receiver Operating Characteristic (ROC) curve.  The ROC curve shows the True Positive Rate (TPR) vs. the False Positive Rate (FPR) for the validation set, where P(ositive) indicates a lens and N(egative), a non-lens.
With the definitions \ed{True Positive (TP)} = correctly identified as a lens, 
False Positive (\ed{FP}) = incorrectly identified as a lens, 
True Negative (\ed{TN}) = correctly rejected, and False Negative (\ed{FN}) = incorrectly rejected,
\begin{equation*}
    \rm{TPR} = \frac{\rm{TP}}{\rm{P}} = \frac{\rm{TP}}{\rm{TP} + \rm{FN}}
\end{equation*}

and
\begin{equation*}
    \rm{FPR} = \frac{\rm{FP}}{\rm{N}} = \frac{\rm{FP}}{\rm{FP} + \rm{TN}}
\end{equation*}

\noindent
The curve is generated by gradually increasing the threshold probability for a positive identification from 0 to 1.  
Random classifications will result in a diagonal line in this space with an area under the ROC curve (or AUC) equal 0.5. 
For a perfect classifier, AUC = 1.

%%--- defined as pushing the whole training set through the neural net.

In Figure~\ref{fig:loss-roc}, left panel, 
we show how the cross entropy loss functions vary as training progresses.  
For the validation set, we show the value at every epoch.  
For the training set, the cross entropy was reported for every step, which we have boxcar smoothed with a window size of 57.
This is because the training set has a total of 14,725~images,
with a batch size of 256 images, 
it takes approximately 57 steps to complete one full training epoch.
Figure~\ref{fig:loss-roc} shows that the AUC for the validation set has plateaued well within the 120 epochs of training.  
We achieve an AUC of 0.992 for the validation set (Figure~\ref{fig:loss-roc}, right panel).  
This is a significant improvement over an already high AUC of 0.977 from H20.

\noindent%
 \begin{minipage}{\linewidth}% to keep image and caption on one page
 \makebox[\linewidth]{%        to center the image
   \includegraphics[keepaspectratio=true,scale=0.4]{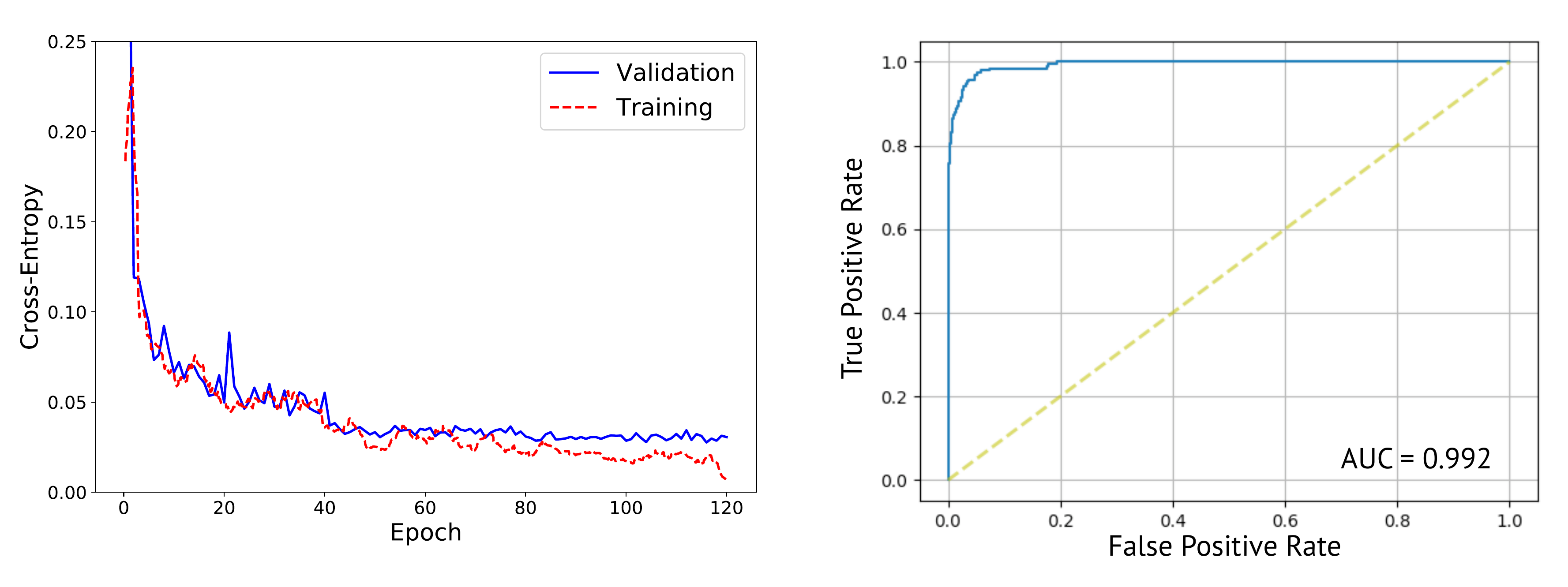}}
 \captionof{figure}{ 
Left: The cross entropy loss functions for the training and validation sets over 120 epochs.
Right: The receiver operative characteristic (ROC) curve for the validation set with the area under the curve (AUC) = 0.992
\rf{for the last model after 120 epochs of training}.}\label{fig:loss-roc}%      only if needed  
 \end{minipage}

%%%% Regarding the imbalance between the two classes ($\sim$~20:1), L18 pointed out that adjusting the relative weights between the two classes is equivalent to adjusting the threshold.  

\subsection{Improvement on the L18 Model}\label{sec:shielded-model}

%%In H20, we stated that we would like to \ed{explore modifications to}  
%%see if we can modify the neural network model to improve performance.
We have experimented with a variety of ways to improve on the model in L18, 
including transfer learning and domain adaptation \citep[e.g.,][]{tzeng2017a}, among other techniques. 
We will provide the full results of the comparisons from these different approaches in a future publication.

So far we have run inference and visually inspected the results for one of the variants.  
In this modification on the original L18 model was the addition of  ``shielding" layers, 
inspired by \ed{the InceptionNet architecture} of \citet{szegedy2014a}. 
These ``shields" are $1 \times 1$ convolutional layers inserted between every three blocks of the L18 architecture (see their Figure~4), 
so named because they have the effect of reducing dimensionality and mitigating the exponential increase in computational complexity present in the original architecture. 
With appropriate adjustments to the number of channels in the shielding layers, 
we reduce the number of trainable parameters by a factor of 50 (from 3 million to 60 thousand), 
thereby shortening the training time by 17\%. 
Moreover, the validation AUC has increased from 0.992 (using the original L18 model; \ed{\S~\ref{sec:L18-model}}) to 0.997. 
Thus the reduction in model complexity does not appear to have an adverse impact on performance, 
and in fact has improved it.
This is likely because the problem at hand (to tell lenses apart from non-lenses)
although complex, 
does not require a large number of dimensions in the underlying latent space. 
The addition of ``shielding" layers compresses dimensionality by more than an order of magnitude, 
forcing the network to learn only the most salient features. 
For example, in the final block of the architecture in L18 (see their Figure 4) we experimented with reducing the output from 512 channels to 256, 128, 64, 32, and 16 channels.  
We find that ``shields" that keep the output to 32 channels perform the best.

In \S~\ref{sec:results}, 
we will show lens candidates from both the original model in L18 and the ``shielded" model (the one with 32 output channels),
to achieve greater completeness for the lens search in DR8 
and to demonstrate that a different neural network model can identify new lens candidates. 

% \newpage
\section{Results}
\label{sec:results}
In this section we present the lens candidates. 
In \S~\ref{sec:L18-cands}, we present all the candidates found by using the ResNet model of L18, specifically: 
\S~\ref{sec:L18-DC-DECaLS}, candidates that are DEV or COMP in DECaLS;  
\S~\ref{sec:L18-DC-MzLS}, candidates that are DEV or COMP found in BASS/MzLS, 
and \S~\ref{sec:L18-REX-LS}, candidates that are typed as REX in DECaLS and MzLS.
In \S~\ref{sec:shieled-cands}, we show candidates that are found with the ``shielded" model (see \S~\ref{sec:shielded-model}).
To determine the probability threshold for human inspection,
we consult the precision-recall curve (PRC), 
where  precision = TP/(TP+FP) and recall = TP/(TP + FN), which is the same as TPR (\S~\ref{sec:L18-model}).
The PRC for the validation set, with probability threshold values marked, is shown in Figure~\ref{fig:precision-recall}.

\begin{minipage}{\linewidth}
\makebox[\linewidth]{
  \includegraphics[keepaspectratio=true,scale=0.4]{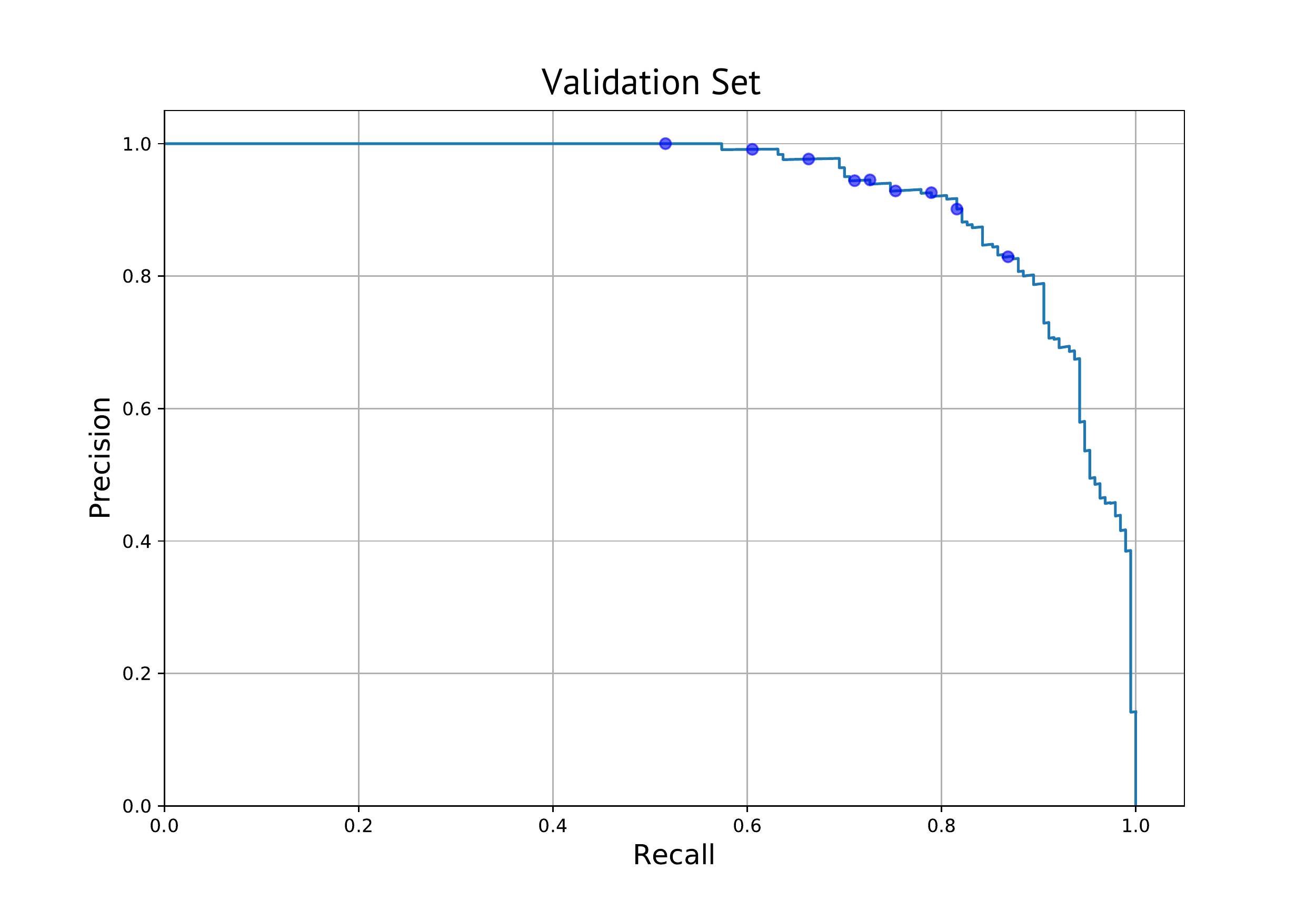}}
  \captionof{figure}{
 The precision-recall curve for the validation set.  
 The blue points from left to right correspond to probability threshold values from 0.9 to 0.1 with an interval of 0.1.
  }
  \label{fig:precision-recall}
\end{minipage}

We recognize that different terms have been used for the same quantities.
To avoid confusion, in this paper:
\begin{equation*}
 \mathrm{recall = TPR = completeness} 
\end{equation*}

\noindent
and 
\begin{equation*}
 \mathrm{precision = purity} 
 %\approx \ed{\mathrm{effective\,purity}}   
\end{equation*}

\noindent
This redundancy in terminology in part stems from fairly standard usage 
(e.g., recall or TPR depending on the context) 
and in part from the difference in terminology between machine learning and astrophysics (recall or completeness, precision or purity).

While ideally we would like to identify all the lenses that are discoverable in the data set, 
there is a ceiling to the number of images that can be inspected in a reasonable amount of time.
We choose the threshold of 0.1 because 
it seems to be a reasonable compromise between purity (precision) and completeness (recall). 
%%%% we determined this empirically by examining the result from a small scale deployment (clearly diminishing return below this threshold) and because 
\ed{Keep in mind that the PRC provides completeness and purity for the validation set.}
\ed{For deployment on the whole data set, 
it is not possible to determine the completeness 
%%or purity
without inspecting the entire data set, which is infeasible.}
%%Completeness is very difficult to assess for observed data.
We will address the question of completeness 
%%for observed data
in the context of comparing the results of different neural network models in a future publication (see \S~\ref{sec:shielded-model}).
\ed{
%%Clearly there is a discrepancy between the expected approximately 83\% purity (see Figure~\ref{fig:precision-recall}) 
%%at the chosen probability threshold of 0.1 and the actual purity of 3.2\% at the time of deployment.
Since our training sample has a lens to non-lens ratio ($\sim$ 1 in 33) that is
%%%% 300 times 
\ed{much} higher than expected for the data set as a whole ($\sim$ 1 in $10^4$),
%The true purity at the time of deployment depends on the expected frequency of lensing events.
we estimate the expected purity for deployment at our chosen probability threshold of 0.1 in the following way.
Given the 7:3 training and validation split, 
there are approximately 
$N_l=190$ lenses and $N_{nl}= 6300$ non-lenses
%are the number of lenses and non-lenses, respectively, 
in the validation set. 
The number of non-lenses misclassified as lenses is then $\sim 33 (= N_{l} \times r \times \frac{1 - p}{p} $),
%%%%\end{equation}
%%%%}
%%%%
%%%%\noindent
where $r(=0.87)$ and $p (= 0.83)$ are the recall (or, completeness) and precision (or, purity), respectively.
%%\ed{For deployment, 
%%%% on DEV and COMP, we started with $\sim 10$ million cutouts.
%% as stated earlier, we expect one in $\mathcal{O} (10^4)$ to be a strong lens.}
%and found $\sim 1000$ lens candidates, 
%or approximately 1 in 10,000.  
%This is is consistent with the estimate from \citep{oguri2010a}.
%We can therefore extrapolate based on the number of lenses in the validation set, 190, 
%and estimate the number of non-lens contamination at the time of deployment.
The fraction of non-lenses that are misclassified as lenses is $33/N_{nl} \approx 0.00052$.
%%%%, or the fractional contamination is}
%%%%
%%%%\ed{
%%%%\begin{equation}
%%%%f_{contam} = N_{l} \times r \times \left(\frac{1 - p}{p} \right)/N_{nl}
%%%%\end{equation}
%%%%}
%%%%
%%%%\noindent
%%%%\ed{where $N_l(=190)$ and $N_{nl}(= 6300)$ are the number of lenses and non-lenses, respectively, in the valdiation set, and $r(=0.87)$ and $p (= 0.83)$ are the recall (or, completeness) and precision (or, purity), respectively,
%%%%at the probability threshold of 0.1.
%%%%Then $f_{contam} = 0.0053$.
With the expectation of one strong lens in \ed{$\mathcal{O} (10^4)$} galaxies, 
this translates to a purity of 1 in 52, 
%%\ed{(= f_{contam} \times 10^4)}$, 
or 1.9\%.}
We will refer to all cutout images with probabilities above this threshold as the ResNet ``recommendations".
%%Practically speaking, for observed data, 
\ed{Below, through human inspection, we will compare the percentage of lens candidates relative to the ``recommendations''
%%that takes place in a reasonable amount of time provides the best available \ed{check on 
with this estimated purity for deployment.
%%%% We \ed{therefore} introduce \ed{``effective purity", defined as the ratio of the lens candidates determined by human inspection and the number of ResNet recommendations}.
(Note that in H20, 
%(\S~4.3), 
we used the term ``human inspection efficiency'' for this quantity).}
%% In reality, however, the purity will not be above 80\%. 
%%(see \S~\ref{sec:purity} below).  Nevertheless the PRC provides guidance for the choice of the probability threshold.

%%%% The best way to back this up is by plotting number of candidates against proba, and also to correlate human grades w/ proba.
%%%% The implication seems to be that there are hundreds more lenses to be discovered in lower probability bins. But the solution may be to build a better neural network model and/or better training sample, rather examining a rapidly increasing number of neural net recommendation at lower probabilities.
%%%% Don't say the following until after I re-read Lanusse
%%%% That our probability threshold is far below 0.5 is to be expected: 
%%%% our training sample is highly asymmetric, 
%%%% with the fraction of positive example far below 50\% (see \S~\ref{sec:model-train}).

Throughout this section, all objects we run inference on
have $\geq 3$ passes for all three bands and $z$-band mag $<20.0$.
For the ensuing human inspection, we follow this process.
Co-authors S.B., A.G., A.P., V.R., C.S., W.S., and R.V. make the ``first pass" selections,
according to these criteria, erring on the generous side:
small blue galaxy/galaxies (red galaxies are rare but certainly acceptable) next to the red galaxy/galaxies at the center that 
    \begin{itemize}%[label=$\circ$]
        \item are typically 1 - 5$\twopr$ away   
        \item have low surface brightness
        \item curve toward the red galaxy/galaxies
        \item have counter/multiple images with similar colors (especially in Einstein-cross like configuration)
        \item are elongated (including semi- or nearly full circles)

    \end{itemize}

\noindent
Typically, most candidates do not have all these characteristics. 
In general, the greater the number of characteristics listed above an image has, the higher they are ranked by humans.
For the ``second pass",
co-authors X.H. and A.D. examine all ``first pass" selections and assign an integer score between 1 and 4. 
These two scores are averaged.
We assign a letter grade according to the average,
\rff{using the following criteria}.
\rff{For the third criterion below, and for the rest of the paper, 
we define the \textit{angular scale} of the a candidate system as the angular separation between the lens and the most prominent putative arc. 
Note that this can be somewhat different from (and in the case of a single large tangential arc, 
tends to be larger than)
the extent of the critical curve
%and for a single large arc, 
%it tends to be larger than the extent of the critical curve 
\citep[e.g.,][]{narayan1996a, kneib2011a}.}
%%%% Meneghetti GL lecture notes, p. 42, fig 3.8: sometimes the arc would straddle and pretty much sits right on the critical curve.  That's why I say "tends to".

\begin{itemize}
    \item $ \geq3.5$: Grade~A.  We have a high level of confidence of these candidates.  
    Many of them have one or more prominent arcs, usually blue.
    The rest have one or more clear arclets, sometimes arranged in \ed{multiple}-image configurations with similar colors (again, typically blue).  However, there are clear cases with red arcs.
    
    \item $= 3.0$: Grade~B.   
    They have similar characteristics as the Grade A's.  
    For the cutout images where there appear to be giant arcs they tend to be fainter than those for the Grade A's.  
    Likewise, the putative arclets tend to be smaller and/or fainter, or isolated (without counter images).
    
    \item $ = 2.5$ or 2.0: Grade~C.  
    They generally have features that are even fainter and/or smaller than what is typical for Grade~B candidates, but that are nevertheless suggestive of lensed arclets.  
    \ed{Counter images are often not present or indiscernible.} 
    In a number of cases,
    %, if these are indeed lensing systems, 
    the \rff{angular scales of the candidate systems}
    %source image(s)} 
    are comparable to or only slightly larger than the seeing.
    \ed{Therefore, for some of these candidates, to attain a higher level of certainty, higher spatial resolution or deeper data would be required.}
\end{itemize}

\noindent
%%%% In the revision stage?  If the the average is below 2.0 but receives at least one score above 2, they are included in the Appendix.
For Grade~B and C candidates, we have included a small percentage of cases where it is difficult to judge whether it is a lensing event vs. a coincidental placement of galaxies, a spiral galaxy, a ring galaxy,
\ed{or tidal features associated with galaxy interactions}. 

%%%% ----------------------------------------------------------------------
%%%%                            L18 Candidates
%%%% ----------------------------------------------------------------------

\subsection{Lens Candidates from the L18 ResNet}\label{sec:L18-cands}

Below we present all the strong lens candidates found by using the ResNet model in L18.

%%%% -------------------------- L18 DECaLS DC ----------------------------
\subsubsection{Candidates from DEV and COMP in DECaLS}\label{sec:L18-DC-DECaLS}

Searching for strong lenses among the DEV and COMP objects in the DECaLS region originally was our primary goal. 
Our training sample is selected from the same region
(see Figure~\ref{fig:lenses-nonlenses-training}).
%% Since source extraction and typing by \tractor are reliable beyond the southern edge of DECaLS (Figure~\ref{fig:dr8-footprint}) down to $\delta = -32.375^\circ$, even though this part of the DES footprint is not officially part of DECaLS, our search extends to the same declination.
%%$\delta = -32.375^\circ$, as well.
We deploy our model on $\sim$~10~million cutouts centered on galaxies typed as DEV or COMP.
%%%% 10,073,687
With the probability threshold set at 0.1, 
in total we have examined \decalsDCrecs ResNet recommendations.

We have found \cmuDCdecalsA Grade~A, \cmuDCdecalsB Grade~B, and \cmuDCdecalsC Grade~C candidates.
The locations of these candidates in the sky are shown in Figure~\ref{fig:L18-DC-candidates}.
In total, we have identified \cmuDCdecals candidates, 
\ed{achieving a purity of} approximately 1 in 31 ResNet recommendations.
%This is the highest human inspection efficiency we are aware of.

\begin{minipage}{\linewidth}
\makebox[\linewidth]{
  \includegraphics[keepaspectratio=true,scale=0.35]{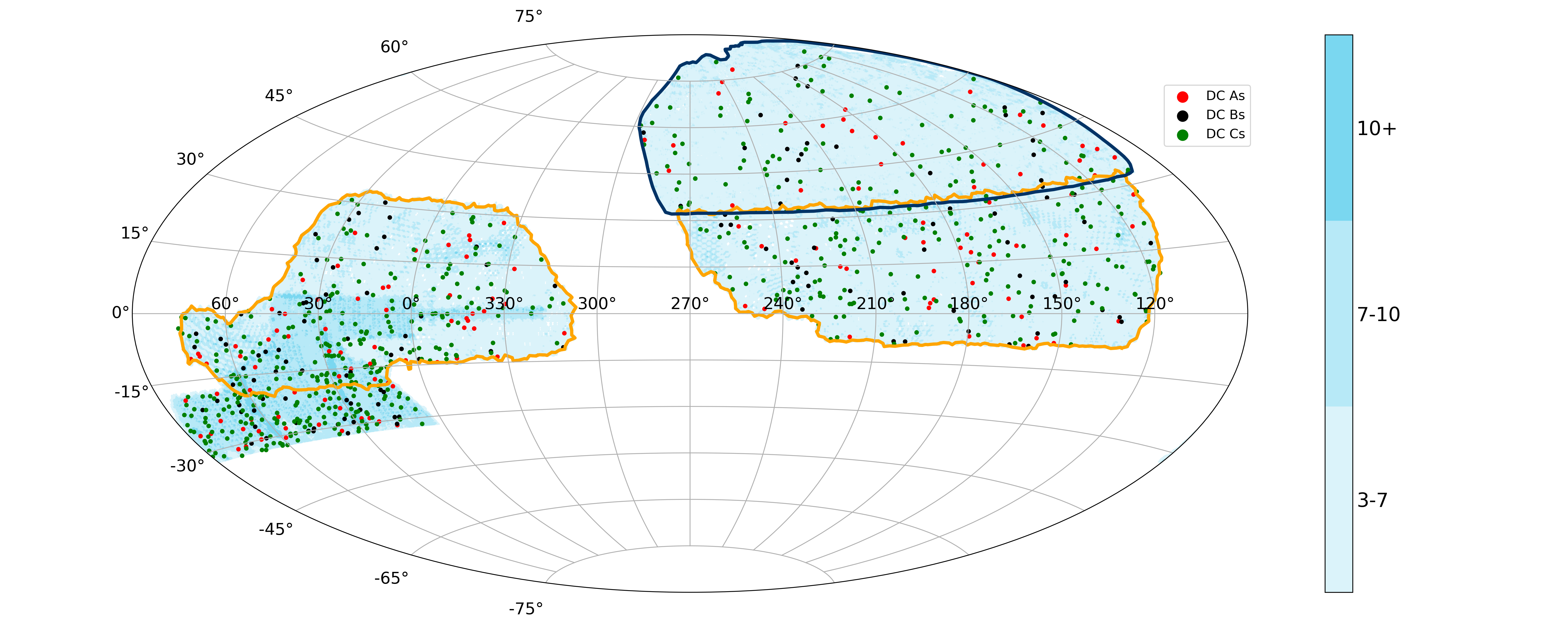}}
   \captionof{figure}{
The new candidate lensing systems typed as DEV and COMP by \tractor in the DECaLS and BASS/MzLS regions (see Figure~\ref{fig:dr8-footprint} caption) are shown as red (Grade A), black (Grade~B), and yellow (Grade~C) circles.
%%The footprint of DECaLS is highlighted by red outlines and the $z$-band depth is represented by different shades of blue (for more details, see the caption for Figure~\ref{fig:dr8-footprint}).
%% Note that our search extends further than the southern edge of DECaLS, down to $\delta = -32.375^\circ$.
}
  \label{fig:L18-DC-candidates}
\end{minipage}

%%%%This magnitude cut was chosen because it 
%includes \ed{92\%} of the known lenses in the Legacy Surveys and results in a manageable number of images for human inspection.   

%%%% -----------------------  Inspection Efficiency -----------------------
%%%% \subsubsection{\ed{Effective Purity}}\label{sec:purity}
%%%% Inluding during revision: High efficiency (without significantly sacrificing completeness) is important because for LSST, Euclid, WFIRST if order ($10^4$ - $10^5$) lenses are expected, the efficiency needs further improvement.  We have started the preliminary work to more fully investigate the completeness and how to improve it, which we will report in a future publication along with comparing different deep learning techniques (which was mentioned earlier in \S~\ref{sec:shielded-model}.

We \ed{now} briefly discuss the \ed{purity} of the ResNet results thus far, 
since this is the primary data set in which we originally planned to search for lenses.

\begin{minipage}{\linewidth}
\makebox[\linewidth]{
\includegraphics[width=.5\linewidth]{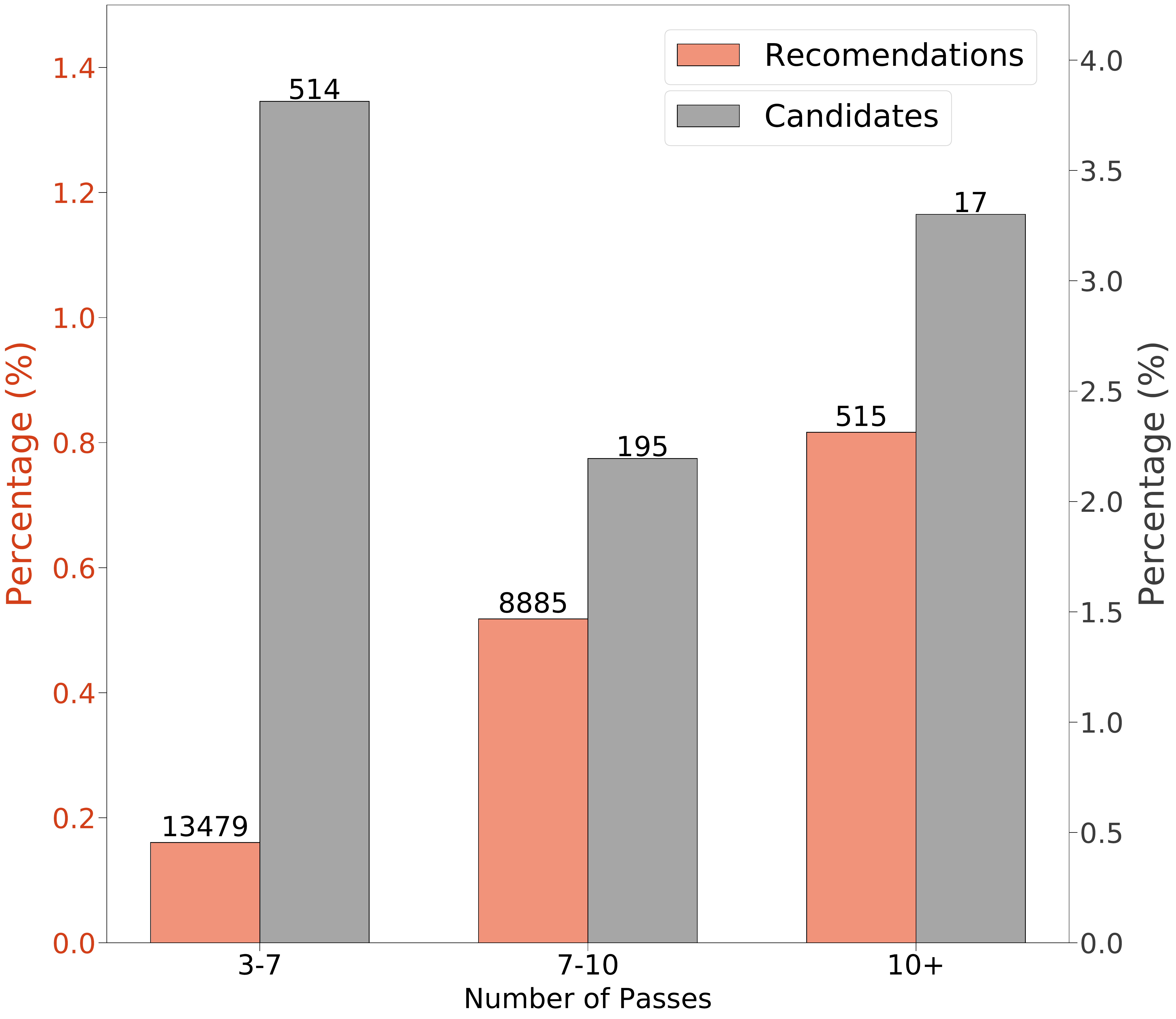}}
\captionof{figure}{
%Left: 
The orange columns (left $y$-axis) show the percentages of objects given a greater than 0.1 probability by our ResNet model (or ``recommendations") for the three bins of $z$-band depth.
The gray columns (right $y$-axis) show the percentages of ResNet recommendations that are selected as lens candidates through human inspection, \ed{or, the purity}.
%Right: The orange columns show the fraction of ResNet recommendations for each bin of $z$-band depth. The blue columns show the fraction of objects in each bin.
\ed{The number of recommendations or candidates for each bin is shown atop the corresponding column.} 
}
\label{fig:recs-cands}
\end{minipage}

In H20, we noted that due to the composition of our training sample (comparatively smaller number of non-lens images with deep observations), the neural net showed a preference for image\ed{s} with deep observations,
whether they contain lensing systems or not. 
%(H20, Figure~10).
For the inference results in this paper, 
Figure~\ref{fig:recs-cands} shows how \ed{a) the percentage of the ResNet recommendations relative to the objects and b) the percentage of candidates (determined by human inspection) relative to recommendations depend} on the observational depth
%%We use the number of passes for $z$-band as a proxy for depth 
(see Figure~\ref{fig:dr8-footprint}).
For the three depth bins, the percentages of lens candidates relative to the neural net recommendations are similar, approximately between 2.2 - 3.8\%.
%%even though the percentage for the shallowest bin is slightly higher than the other two.
This indicates that the neural net \ed{now} makes recommendations largely free of bias with regard to depth.
%This is further supported by the right panel of Figure~\ref{fig:lens-nlens-training-hist},
%while there is --- do I really need to discuss this? [At this stage, if there is any doubt, then the answer is no: cut this figure.] 
%% Thus while there may still be a bias toward deeper images, it is likely small.
This is consistent with our expectation based on the composition of the training sample used in this paper. 
%%(see Figure~\ref{fig:lenses-nonlenses-training}). 
The orange columns show that \ed{0.82\%} of the objects in the 10+ pass bin receive probability $>0.1$ (``recommendations"), 
\ed{five times the value of} 0.16\% in the 3 - 7 passes bin. 
%%do.
%% From examining the percentages of candidates relative to the number of recommendations in each depth bin (Figure~\ref{fig:recs-cands}, gray columns),
This trend in the ResNet recommendations 
%%is likely not due to the neural net's preference for deeper images,but rather the fact 
indicates that, not surprisingly,
there are more lenses to be discovered for deeper images.
%%\ed{Multiplying the percentages represented by the orange and gray columns in Figure~\ref{fig:recs-cands} shows that} 
In fact, approximately one in 16,337, 8795, 3710 galaxies is a lens, 
from the shallowest to the deepest bin, assuming 100\% completeness.
These values are consistent with the expectation of one strong lens in $\mathcal{O} (10^4)$ galaxies.

\ed{Overall, our ResNet model achieves a purity of 3.2\%, 
broadly consistent with our estimation of 1.9\% (see the introduction to \S~\ref{sec:results}).
%% This is OK, because 1 in 10,000 is only an order of magnitude estimate.   The level of completeness would affect the actually number as well.
%%%% Next time, we will use a validation set that has the ratio that matches the data set on which we will run inference. Even if that's against how train and validation sets are usually set up (a 7:3 split, e.g.).
}
Compared with H20, \ed{this much improved purity}
%(which was called ``human inspection efficiency'' in H20}) 
likely stems from three factors: 1) a larger (by about $\sim 60\%$) training sample; 
2) the lenses in the training sample are all well observed in DECaLS with clearly discernible lensing features; and 3) the non-lenses in the training sample includes a large number of images from DES that have observations with comparable depth as the lenses from DES in our training sample, 
which significantly reduced, if not eliminated, 
the ResNet's bias toward images with greater depth.
%%\ed{check Suyu's paper and say this is among the best.}

%\ed{However we cannot simply increase the number of non-lenses in the training sample to match the 1:10,000 lens to non-lens ratio.
%As this would mean that during training, 
%many batches would simply not have positive examples.}
%
%\ed{Other groups have similar discrepancy between the purity between validation and deployment \citep[e.g.,][]{jacobs2019b, canameras2020a}.
%They use simulated images, so it's not clear how to address this issue.
%In our case, a partial solution is add more lenses in training.
%With all the new lenses being discovered, we can increase the number of lenses in our training sample by a factor of two or more
%We are pursuing another solution through an ongoing investigation about the failure modes 
%and this may help us to address how to further increase the purity.
%We now know to achieve a purity that is one order of magnitude higher at deployment time, 
%with our current training sample we need to achieve a purity of xxx\% at the chosen probability threshold to achieve a purity of one in a few at the time of deployment.
%}
%
%\ed{At the present, at a minimum, we can assess the performance (purity) by using a validation set that has the realistic proportion of lenses and non-lenses.}

%%%% -------------------------- L18 MzLS DC ----------------------------
\subsubsection{Candidates from Deployment on DEV and COMP in BASS/MzLS}\label{sec:L18-DC-MzLS}

%%In H20, we stated that we would search for lensing systems in 
\ed{For the northern MzLS/BASS region, the $gr$ band observations have worse seeing.}
%%for the DEV and COMP categories.
%%%% by possibly making modification of our neural network model and/or training sample, to account for the different seeing in $gr$ bands.
Given the success of the deployment in DECaLS, \ed{however,} 
we decide to proceed with applying our trained ResNet model, without modification or re-training, \ed{to this} region. 
%%%% After a small scale experiment that yielded promising results, we proceeded with full deployment.

We run inference on 5.4~million cutouts centered on DEV and COMP objects,
\ed{with $z$-band magnitude $<20.0$}.
%%%% 5,014,851  
%%The ``first pass'' inspection was performed by co-authors C.S. and V.R.
The inspection of \mzlsDCrecs ResNet recommendations finds \cmuDCmzlsA A's, \cmuDCmzlsB B's, and \cmuDCmzlsC C's. 
The locations of the candidates in the sky are shown, together with the candidates found in DECaLS, in Figure~\ref{fig:L18-DC-candidates}.
In total, we have identified \cmuDCmzls candidates, 
approximately 1 in 57 ResNet recommendations. 
As expected, the \ed{purity of the ResNet recommendations}  
%%%% ResNet efficiency 
is worse than for DECaLS, but is still competitive.
Keep in mind that we used the same ResNet trained for DECaLS without any modification.
Furthermore, as we mentioned in \S~\ref{sec:observations}, the $gr$ band seeings are $1.61\twopr$ and $1.47\twopr$, respectively. 
%%%% by far the worst compared with any survey for which nn has been used to find lenses..
To our knowledge, 
%no other group has attempted lens search with neural networks has been attempted with seeing $\gtrsim 1.5\twopr$.
%We are the first and we have successfully found strong lenses (Figure~\ref{fig:candidates}), still with competitive efficiency \citep[e.g.,][]{jacobs2019b}.
this is the first time a lens search has been attempted and successfully carried out, with competitive \ed{neural network recommendation purity}, for a survey with seeing $\gtrsim 1.5\twopr$.
This is a remarkable result.

%%%% -------------------------- L18 REX ----------------------------
\subsubsection{Lens Candidates from Deployment on REX in Legacy Surveys}\label{sec:L18-REX-LS}

%% In H20, we stated that we would like to find lenses for the REX type.
%%The largest number of galaxies, by far, in the Legacy Surveys catalog are classified as the morphological type REX, i.e., the best-fit source model has a round exponential profile. 
The REX category contains an order of magnitude more objects than the DEV and COMP types combined, since most faint, extended galaxies are 
modeled by the REX profile (see \S~\ref{sec:observations}).
% The largest category in Legacy Surveys is REX (with an order of magnitude more objects than DEV and COMP combined) 
This category likely includes many elliptical galaxies,
though the percentage is unknown.  

Given the success with DEV and COMP in both DECaLS and BASS/MzLS, 
without modification of the model or additional training, 
we deploy our trained ResNet on 6.7~million cutouts centered on REX (5 million in DECaLS and 1.7 million in BASS/MzLS)\ed{, with $z$-band~mag $< 20.0$}.
When we performed this inference run, the source extraction and typing by \tractor became available for certain patches below $\delta = -32^\circ$.
These objects are included in the deployment.
%%%% DECaLS: 5064653; MzLS: 1679144; Total: 6743797
%%%% The ``first pass'' inspection was performed by co-authors S.B., A.G., A.P., C.S., W.S., and R.V.

In total, we have inspected \totREXrecs
(\decalsREXrecs in DECaLS and \mzlsREXrecs in BASS/MzLS) ResNet recommendations and identified 
%% 219 
\ed{168} candidates.
Of these, 156 are in DECaLS and 12 in BASS/MzLS, resulting in 
\ed{recommendations with purities} of 1 in 38 and 1 in 98, respectively.
%Since the vast majority of the candidates have been identified in DECaLS,
The average \ed{purity} is $\sim$ 1 in 42.
%%%% As expected, all the putative lenses in these images turn out to be elliptical galaxies.
We have removed candidates that have already been found in DEV and COMP  
(these lensing systems are ``discovered" again because the cutout images containing the same systems are centered on different objects this time).

In the end, we identify 
\cmuREXA A's (\cmuREXdecalsA in DECaLS and \cmuREXmzlsA in BASS/MzLS), 
\cmuREXB B's (\cmuREXdecalsB in DECaLS and \cmuREXmzlsB in BASS/MzLS), 
and \cmuREXC C's (\cmuREXdecalsC in DECaLS and \cmuREXmzlsC in BASS/MzLS), 
 for a total of \cmuREXcands new candidates.
The locations of the candidates in the sky are shown
in Figure~\ref{fig:L18-REX-candidates}.

\begin{minipage}{\linewidth}
\makebox[\linewidth]{
\includegraphics[keepaspectratio=true,scale=0.3]{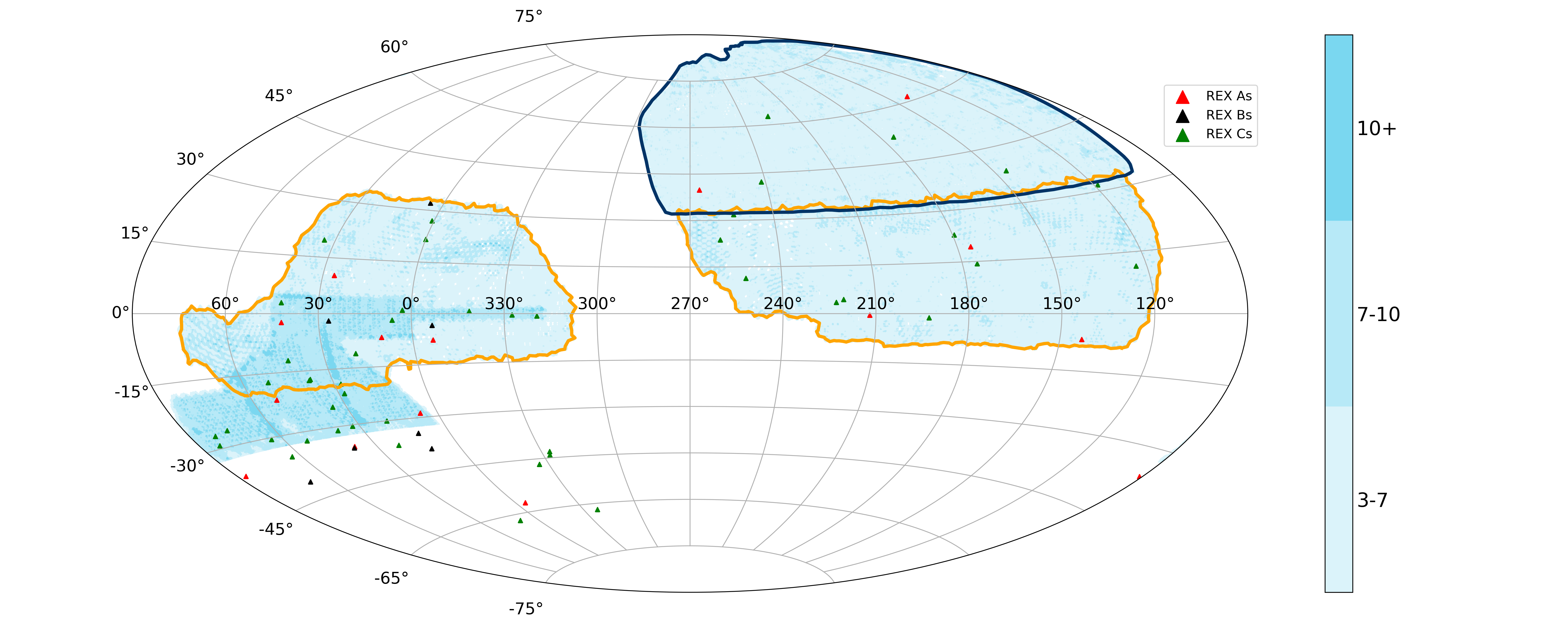}}
\captionof{figure}{
The new candidate lensing systems typed as REX by \tractor in the Legacy Surveys are shown as red (Grade A), black (Grade~B), and green (Grade~C) \ed{triangles}.
%The DECaLS and BASS/MzLS regions are highlighted by red and green outlines, respectively, and the $z$-band depth is represented by different shades of blue (see Figure~\ref{fig:dr8-footprint}).
}
\label{fig:L18-REX-candidates}
\end{minipage}

\ed{All the lens candidates found by the L18 model are summarized in Table~\ref{tab:L18-cands}.}

%\begin{minipage}{\linewidth}
%\makebox[\linewidth]{
\begin{deluxetable*}{lccccccc}[h]
\tablewidth{0pt}
\tabletypesize{\scriptsize}
\tablecaption{L18 Model\label{tab:L18-cands}}
\tablehead{\multicolumn{1}{c}{Grade} & \multicolumn{1}{c}{A}& & \multicolumn{1}{c}{B} & &
\multicolumn{2}{c}{C} &\colhead{Total by Type (DECaLS,MzLS)}\\
\cline{2-2}
\cline{4-4}
\cline{6-7}
\colhead{Human Score} & \colhead{$\geq3.5$} & & \colhead{$3.0$} & &  \colhead{$2.5$} & \colhead{$2.0$} & \colhead{}}
\startdata
  DC (DECaLS,MzLS)  &144 (115,29)& &132 (110,22) &&280 (242,38)& 324 (259,65)& 880 (726,154)\\REX (DECaLS,MzLS) &15 (13,2) & &7 (6,1) &  &22 (20,2)  &24 (22,2)  &68 (61,7)\\
 \hline
 Total by Grade (DECaLS,MzLS) &  159 (128,31)& &139 (116,23)& &302 (262,40) &348 (281,67) &948 (787,161) \\ 
\enddata
%\tablecomments{}
\end{deluxetable*}
%}
%\end{minipage}

%%%% ----------------------------------------------------------------
%%%%                              Shielded 
%%%% ----------------------------------------------------------------

\subsection{Candidates Found with the ``Shielded" Model in Legacy Surveys}\label{sec:shieled-cands}

As mentioned in \S~\ref{sec:shielded-model}, we have experimented with modifications on the L18 ResNet model to optimize performance, 
but so far using the same training sample (although we will experiment with the makeup of the training sample as well).
Here we present the lens candidates found by one of these attempts.

We deploy the ``shielded" model on the entire Legacy Surveys footprint on objects that satisfy the same criteria as for the L18 ResNet model.
We achieve a similar level of \ed{purity}, 
and have found \shieldnewtot \emph{new} lens candidates, 
including \shieldA A's, \shieldB B's, and \shieldC C's. 
This demonstrates that a different neural network is capable of finding new lenses in the same footprint.
\ed{These lens candidates are summarized in Table~\ref{tab:shielded-cands} with their locations on the sky shown in Figure~\ref{fig:shielded-candidates}}.

\begin{deluxetable*}{lccccccc}[h]
\tablewidth{0pt}
\tabletypesize{\scriptsize}
\tablecaption{Shielded Model\label{tab:shielded-cands}}
\tablehead{\multicolumn{1}{c}{Grade} & \multicolumn{1}{c}{A}& & \multicolumn{1}{c}{B} & &
\multicolumn{2}{c}{C} &\colhead{Total by Type (DECaLS,MzLS)}\\
\cline{2-2}
\cline{4-4}
\cline{6-7}
\colhead{Human Score} & \colhead{$\geq3.5$} & & \colhead{$3.0$} & &  \colhead{$2.5$} & \colhead{$2.0$} & \colhead{}}
\startdata
 DC (DECaLS,MzLS)  &19 (16,3)& &20 (19,1) &&36 (34,2)& 50 (45,5)& 125 (114,11)\\
 REX (DECaLS,MzLS) &38 (34,4) & &40 (37,3) &  &49 (47,2)  &112 (109,3)  &239 (227,12)\\
 \hline
 Total by Grade (DECaLS,MzLS) &  57 (50,7)& &60 (56,4)& &85 (81,4) &162 (154,8) &364 (341,23) \\ 
\enddata
%\tablecomments{}
\end{deluxetable*}

\begin{minipage}{\linewidth}
\makebox[\linewidth]{
\includegraphics[keepaspectratio=true,scale=0.3]{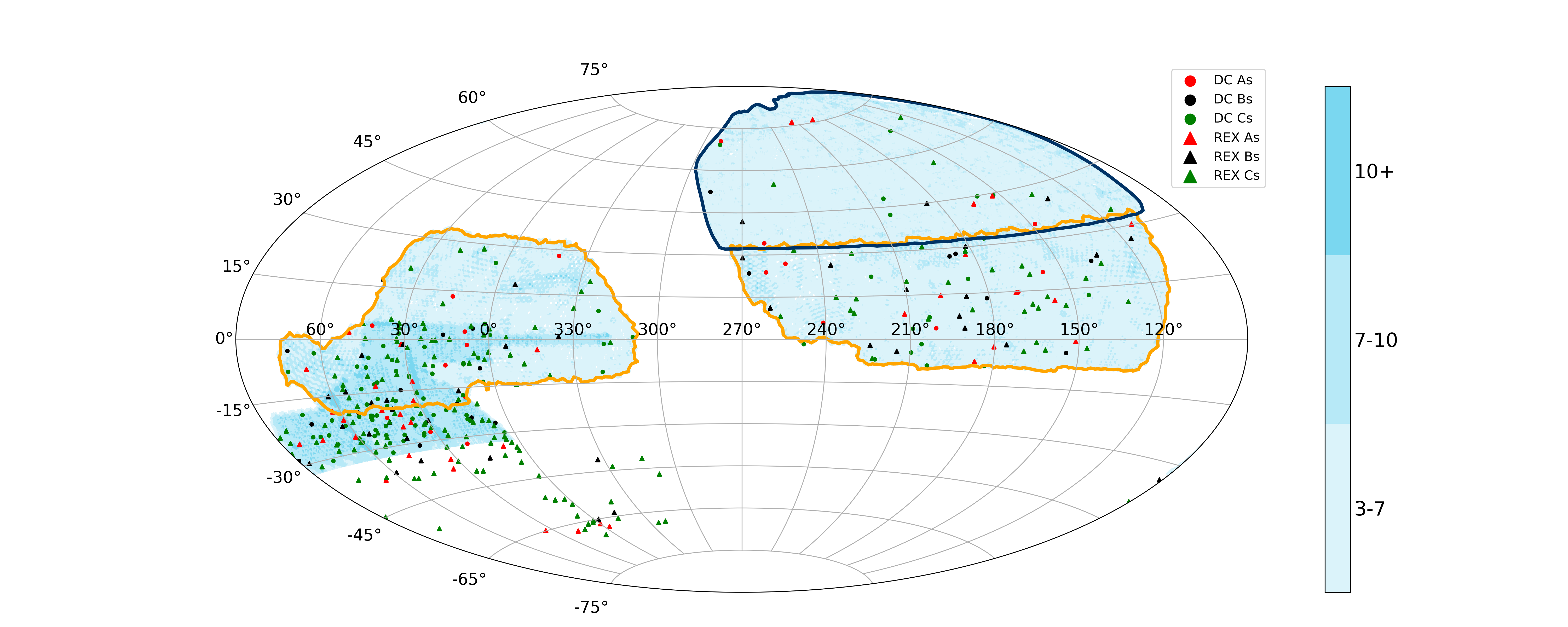}}
\captionof{figure}{
The new candidate lensing systems found by the ``shielded" model are shown as red (Grade A), black (Grade~B), and yellow (Grade~C) circles \ed{(DEV or Comp) and triangles (REX)}.
%% The DECaLS and BASS/MzLS \ed{regions} are highlighted by red and green outlines, respectively, and the $z$-band depth is represented by different shades of blue (see Figure~\ref{fig:dr8-footprint}).
}
\label{fig:shielded-candidates}
\end{minipage}

%%%% ----------------------------------------------------------------
%%%%                    Examples of Candidates
%%%% ----------------------------------------------------------------

\subsection{Summary of \S~\ref{sec:results}}\label{sec:lens-examples}

Altogether, we have found \lenstot strong lens candidates \ed{(Table~\ref{tab:all-cands})}.
\ed{Of these, \lensknown} have been found by other groups, none of which were included in our training sample.  
%The citations for these systems are given in Tables~\ref{tab:grade-a} to \ref{tab:grade-c}.
%with 18 from \citet{jacobs2019b} and one from \citet{inada2003a}.
This leaves \lenstotnew new lens candidates.
Of these, there are \lensAnew A's,
%(Figures~\ref{fig:grade-a-1} - \ref{fig:grade-a-3}; Tables~\ref{tab:grade-a})
\lensBnew B's,
%(Figures~\ref{fig:grade-b-1} - \ref{fig:grade-b-3}}; Table~\ref{tab:grade-b}), 
and \lensCnew C's.
%(Figures~\ref{fig:grade-c-1} - \ref{fig:grade-c-9}; Table~\ref{tab:grade-c}).
For each candidate system,
%presented in the figures, 
we report the average numerical scores from A.D. and X.H. and the absolute difference, the region where it is found, its type from \tractor, 
and the neural network model used.
\ed{The strong lens candidates discovered in this work are summarized in Table~\ref{tab:all-cands}}.
We have checked our candidate list against the spectroscopic database from SDSS DR16\footnote{\url{https://www.sdss.org/dr16/}}
%%%%~I and II \citep{york2000a}, SDSS~III/BOSS \citep{eisenstein2011a}, and SDSS~IV/eBOSS \citep{blanton2017a} 
and found that for approximately half of them 
the putative lensing galaxy has a spectroscopic redshift.
\ed{For the rest, we have found photometric redshifts from \citet{zhou2020a}}.
%%For the lenses xxx (or 20\%) of our candidate systems have $z_d$ between 0.75 and 1.0, much higher than the redshift range of confirmed galaxy lenses (typically 0.2 - 0.7).} 
%The available \ed{spectroscopic or photometric} redshifts are included in \ed{Tables~\ref{tab:grade-a} to \ref{tab:grade-c}.}

We believe we have held a high standard in grading our candidates.
Many of our Grade~C systems are in fact likely lensing candidates.
Among our candidates, of the \ed{\lensknown systems} that have been identified by other groups 
(but were not in our training sample), 
\ed{\lensknownC} are in Grade~C, \ed{\lensknownCplus} of which have a score of 2.5
\ed{(see Table~\ref{tab:all-cands})}.
%and the other two, 2.0.
This speaks to the quality of our Grade~C candidates.
% , perhaps especially of those with a score of 2.5.
We would 
%% especially 
like to note that 42\% (\lensCplusnew) of our Grade~C candidates have a human inspection score of 2.5.
As shown in the examples in Figure~\ref{fig:example-cands} below, 
many of these systems are high likelihood candidates.
In total, there are \lensaboveCplus new candidates with a score $\geq 2.5$.

%%%% (I can cite other papers: by Bolton).
\ed{Many of our} lens candidates 
%%are fainter and 
have \ed{spectroscopic or photometric redshifts}
%%, and mostly have optical and infrared colors consistent
$z \gtrsim 0.8$, 
%% \citep[e.g.,][]{jacobs2019a}, 
greater than the typical redshifts of 0.3 to 0.8 for the current known lensing sample \citep[e.g.,][]{brownstein2012a, wong2018a}.  
In fact, the highest spectroscopic redshift from SDSS~DR16 is 0.8924
(DESI-241.7346+42.1102) \rf{and the highest photometric redshift \citep{zhou2020a} is $1.021 \pm 0.061$ (DESI-072.0873-19.4173).}
%\ed{and the highest photometric redshift is 1.232 (DESI-116.3092+33.6326).}
%%%% during revision, add Zhou et al. Photometric redshifts (see Zhou et al. 2020) for the lenses indicate that 60 (or 20\%) of our candidate systems have zd between 0.75 and 1.0 (Figure 4), much higher than the redshift range of confirmed galaxy lenses (typically 0.2 - 0.7). Therefore the targets in this proposal will significantly increase the number of high redshift galaxy lenses. 
%%%% Higher lens redshifts substantially enhance the power for constraining the mass function of low-mass CDM halos, due to the greater optical depth for perturbations by low-mass halos associated with a longer path length along the LOS \citep{despali2018a, vegetti2018a, r/itondale2019a, diazrivero2020a}.
In addition, the 
%\rf{lens-image separations} 
\rff{angular scales}
of our systems are typically between 1.5- 5$\twopr$ (see Figure~\ref{fig:example-cands}), 
significantly larger than the typical value for previously known galaxy lensing systems ($\lesssim 1.5\twopr$). 
This translates to longer time delays and a smaller relative \ed{uncertainty} per system 
\ed{for quasars and supernova events in the background galaxy}, and therefore higher precision in the measurement of \ho \citep[e.g.,][]{suyu2020a}.
% Therefore, we expect a large number of our candidates to be
% to produce a well-studied sample of lenses 
% in the higher-redshift interval of $0.8 < z < 1.2$.

We end this section by highlighting in Figure~\ref{fig:example-cands} four examples each for four types of strong lens candidates that we have discovered.
Among the 16 candidates shown, 
four have a average human inspection score of 2.5,
and therefore are given a C grade, 
but they are nevertheless very likely lensing candidates.
%% \rf{XH: unfortunately, we can't claim multiple sets of mutliple images -- not yet anyway; HST may reveal more.  Note that DESI-010.9854-35.9683) is a spectacular system: the first system ever discovered with four sets of multiple images.  Systems with two \citep{gavazzi2008a, shajib2020a} are extremely rare.}
%%%% To be included in the future!!!!
%%%%, for example (pick the red lens with near Einstein ring and and the Einstein cross and the reddest one with a $g - r = 2.9$ (point this out!))
%%%% \textbf{DESI-xxx}, a system having a lens with $g - r = 3.3$, likely indicating a high redshift \citep[e.g.,][]{jacobs2019a}; \textbf{DESI-xxxx}, a nearly perfect Einstein Cross; 
%%%% and \textbf{DESI-xxxx}, with a red arc, which
%%%% would have a redshift of $\sim 4$ if it is a Lyman-break galaxy.

%%%%In addition, the lensed sources will tend to have higher redshifts than in known lensing systems as well. 

\begin{deluxetable*}{lccccccc}[h]
\tablewidth{0pt}
\tabletypesize{\scriptsize}
\tablecaption{Lens Candidates\label{tab:all-cands}}
\tablehead{\multicolumn{1}{c}{Grade} & \multicolumn{1}{c}{A}& & \multicolumn{1}{c}{B} & &
\multicolumn{2}{c}{C} &\colhead{Total}\\
\cline{2-2}
\cline{4-4}
\cline{6-7}
\colhead{Human Score} & \colhead{$\geq3.5$} & & \colhead{$3.0$} & &  \colhead{$2.5$} & \colhead{$2.0$} & \colhead{}}
\startdata
 L18+Shielded Models &216 & &199 &  &387  &510  &1312\\
 Known Lenses or Candidates  &23& &24 &&27& 28& 102\\
 New Lens Candidates in this work &  193& &175& &360 &482 &1210 \\ 
\enddata
%\tablecomments{}
\end{deluxetable*}

\noindent%
 \begin{minipage}{\linewidth}% to keep image and caption on one page
 \makebox[\linewidth]{%        to center the image
   \includegraphics[keepaspectratio=true,scale=0.36]{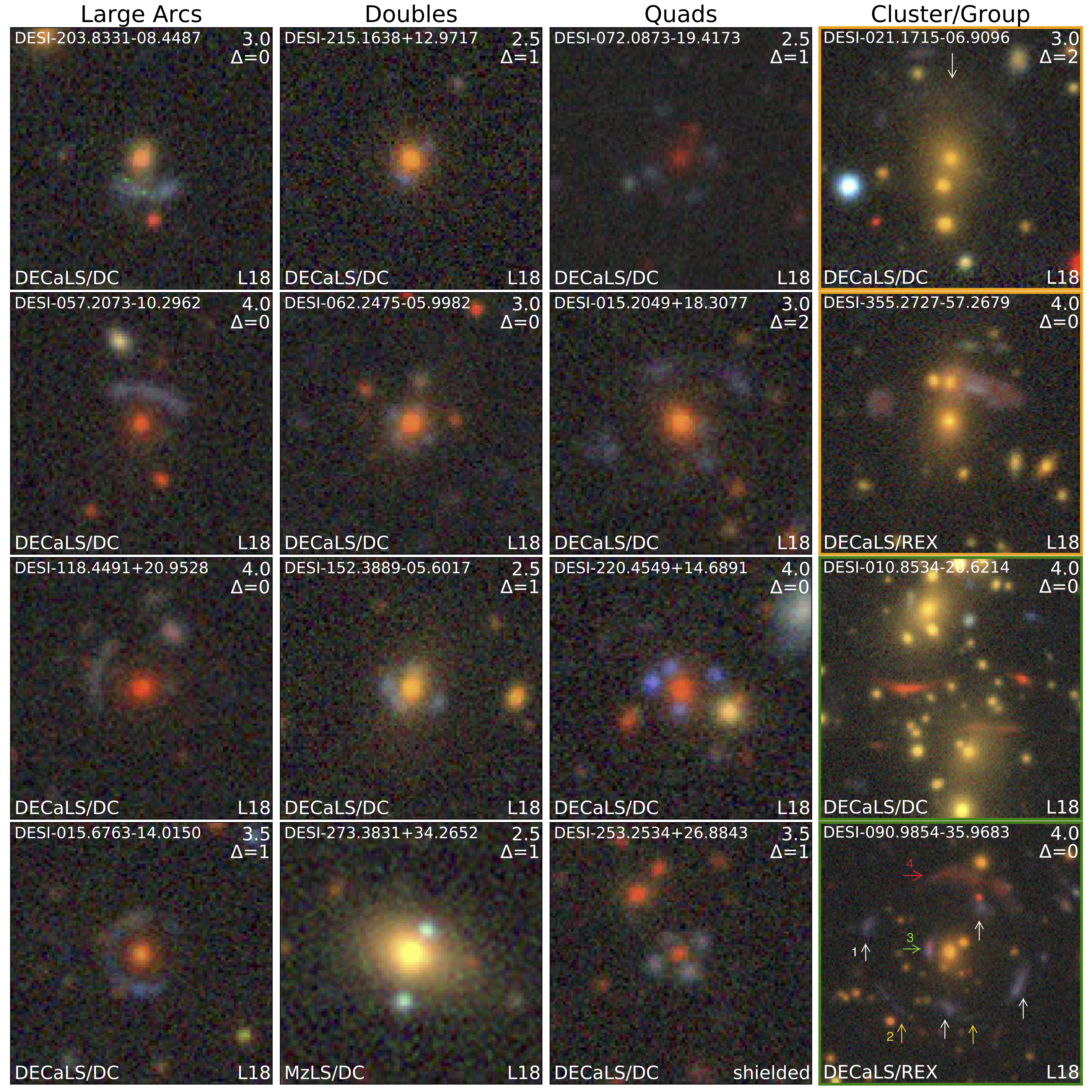}}
 \captionof{figure}{ 
Sixteen of the \lenstotnew new lensing candidates discovered in this paper. 
The naming convention is RA and Dec in decimal format.
Top right corner of each image indicates the average human inspection score with $\Delta$ being the absolute difference;
bottom left corner, the region and \tractor type (REX or DC = DEV or COMP);
and bottom right, the neural network model.
North is up, and east to the left.
The images without rims have a width of 101 pixels $\approx 26.5 \twopr$;
with orange rims, 151 pixel $\approx 39.6\twopr$; and 
green rims, 201 pixel $\approx 52.7\twopr$.
First Column: large arcs. 
The third system (DESI-118.4491+20.9528) clearly has a counter-image and the fourth one (DESI-015.6763-14.0150) is a near Einstein ring. 
Second Column: doubly lensed systems. 
The second (DESI-062.2475-05.9982) and third (DESI-152.3889-05.6017) systems hint at a possible Einstein cross (or a quad) 
and the fourth one (DESI-273.3831+34.2652) is a likely doubly lensed \ed{quasar} system. 
Third column: quadruply lensed systems.
\rf{The first system, DESI-072.0873-19.4173, has the highest photometric redshift, $1.021 \pm 0.061$, among our candidates.}
These 12 systems have a single galaxy as the main lens. 
Fourth Column: cluster/group lensing systems.
\ed{The first one (DESI-021.1715-06.9096) has a faint, giant blue arc 
(white arrow).
The second (DESI-355.2727-57.2679) and third (DESI-010.8534-20.6214) systems show one and two sets of red arcs, respectively.}
The fourth one \ed{(DESI-090.9854-35.9683)} is a spectacular system:
at least four lensed sources at different redshifts are apparent, including a quad (1, white arrows), 
a ``broken" long arc (2, yellow arrows), 
one red arc near the core of the group (3, green arrow), and a giant red arc at approximately 14$\twopr$ away from the lens center (4, red arrow). 
%% The galaxy at the lens center has a photometric redshift of 0.53 $\pm$ 0.12, and the brightest image in the quad system (the rightmost white arrow) has a photometric redshift of 0.90$\pm$0.37 (Zhou R., private communication).
Note that the four candidates receiving a score of 2.5, and therefore a grade of C, are nevertheless very likely lensing candidates.
}\label{fig:example-cands}
\end{minipage}

\rf{The first eighty (sorted by RA) of the \lensA Grade~A lenses are shown in Figure~\ref{fig:grade-a-1} and Table~\ref{tab:grade-a-1}.
All \lenstot candidates reported in this paper can be found on the website 
for this project\footnote{\url{https://sites.google.com/usfca.edu/
neuralens}}
in this \href{https://sites.google.com/usfca.edu/neuralens/publications/lens-candidates-huang-2020b}{online table}. 
%contains the following, 
%by column number:
%\begin{enumerate}[label={[}\arabic*{]}]
%    \item Cutout image
%    \item Name of the system
%    \item \tractor type (DC or REX)
%    \item Human inspection average score and the absolute difference ($\Delta$)
%    \item ResNet Probability
%    \item Photometric Redshift \citep{zhou2020a}
%    \item Spectroscopic Redshift (SDSS~DR16)
%    \item ResNet Model (L18 or ``shielded")
%    \item Reference (for the 102 known lenses or candidates)
%    \item $g$~band magnitude
%    \item $r$~band magnitude
%    \item $z$~band magnitude
%    \item Region of the footprint (DECaLS or MzLS)
%\end{enumerate}
}

%\newpage
\section{Discussion}\label{sec:discussion}
%%---------Discussion------------
In our training sample there are \lenstrain lenses.  
This is generally considered too small a number for training a neural network.  
Even our non-lens sample is much smaller than what is typically used \citep[e.g.,][]{jacobs2019a, jacobs2019b}. 
Nevertheless, we have succeeded in finding \lenstotnew new lens candidates in the three band Legacy Surveys
with nonuniform depth (see Figure~\ref{fig:dr8-footprint}).
The training sample was designed for searching among the DEV and COMP types in one of the two regions of the footprint, DECaLS,
and our neural network model performed well for this category.
\ed{The purity of our neural network recommendations}
%% The human inspection efficiency 
is at least on par with the best in the literature.
Compared with H20, using a larger training sample that includes a larger proportion of non-lenses with deep observations, 
%%% (see Figure~\ref{fig:lens-nlens-training-hist}), 
we have improved the \ed{performance of our neural network model (as measured by recommendation purity)} by a factor of 5 for DEV and COMP in DECaLS (from 1 in 150 to 1 in 31),
where the majority of our lenses are found.
%%%% This is not a surprise given the better seeing and the training sample being tailored for this region.

\rf{Just as significantly, 
our results show that our trained neural network models can be applied to data sets beyond the scope of the training sample.}
%namely to a different region with worse seeings (MzLS)
%and a different type by \tractor with smaller apparent sizes (REX).}
For DEV and COMP in BASS/MzLS, which has inferior $gr$ band seeing, 
and for REX in the entire Legacy Surveys footprint, 
\rf{which typically have smaller apparent sizes,}
we applied the exact same trained model, 
and the purities %efficiency 
are only slightly lower: 1 in 57 and 1 in 42, respectively.
%%for the two cases mentioned above.

 \rf{To the best of our knowledge, to-date most of the confirmed lensing systems are from spectroscopic searches in the SDSS 
\citep[e.g.,][]{bolton2008a, treu2011a, brownstein2012a, shu2017a}.
These systems typically have 
%lens-image separations 
\rff{an angular scale of} $\sim 1\twopr$.
Our candidates, 
discovered from ground-based imaging surveys,
have 
%lens-image separations 
\rff{angular scales} $\gtrsim 1\twopr$.
Thus they not only significantly expand the number of lensing systems but are complementary to most of the confirmed systems.
We will leave a detailed comparison to a future publication,
between the confirmed lensing systems and results from our next search.}

\subsection{Magnitude Distributions}\label{sec:mag-distr}

\rf{For a comparison in apparent magnitudes,
Figure~\ref{fig:mags-cands-train} shows the distributions of the $grz$ magnitudes for the candidate lensing galaxies and lenses in the training sample.
Recall that for both, we have imposed a cut of $z$-band mag~$< 20.0$.
Fractionally, there are more brighter lenses in the training sample.
This is in part because the brighter lenses are more likely to have been discovered before and are included in the training sample.
Beyond this trivial difference,
it is important to note the following.
%there are two important points to note.
%the distributions of our candidates are fainter than the lenses in the training sample.
%The peaks of $r$ and $g$ bands for our candidates are 0.5~mag and $\sim 1$~magnitude, respectively, fainter than for the training sample lenses.
%Therefore it is important to note from this comparison that 
1) The distributions of our candidates are fainter than the lenses in the training sample, that is,
we can discover lenses that have a fainter distribution than the training lenses.
2) There are more lenses to be found fainter than $z$-band mag = 20.0 (Figure~\ref{fig:mags-cands-train}, third panel).
In our next search we will use a fainter magnitude limit. %for cutouts included deployment.
We will also add to our training sample lenses that are fainter than this limit, 
though this number is likely to be small.
However, we do not expect this to be a major limitation to our ability to find fainter lenses, for two reasons:
a) we have shown that we can find lens candidates with a fainter distribution than training lenses,
%; B) we have shown that our trained ResNet models can be extended beyond the scope of the training lenses in other ways (see previous paragraph); 
and b)
%Our expectation is that by using the existing trained ResNet models, without retraining by including fainter training lenses,
%we would be able to find lens candidates with $z$-mag $> 20$.
we will include the lens candidates found in this paper,
%with score $\geq 2.5$, 
which will greatly boost the number of lenses near the faint end of the magnitude distributions and likely will help significantly with discovering even fainter lenses.
We expect to find many more lenses fainter than $z$-band mag $< 20.0$.
With the expectation that the neural net recommendation purity is likely to be worse for these fainter systems, 
we are developing new neural network models for our next search, as mentioned before.
We may also carry out experiments using a fainter magnitude cut for the search than the training lenses,
to see how the neural net purity and completeness depend on the search magnitude cut.
%near the faint end of the $z$-mag distribution,
%This is why using observed data for training works.
}
%Thus we expect that fractionally there are more fainter lenses among our candidates, by comparison.}

\begin{minipage}{0.9\linewidth}
\makebox[\linewidth]{
\hspace{0.4 in}
\includegraphics[width=.42\linewidth]{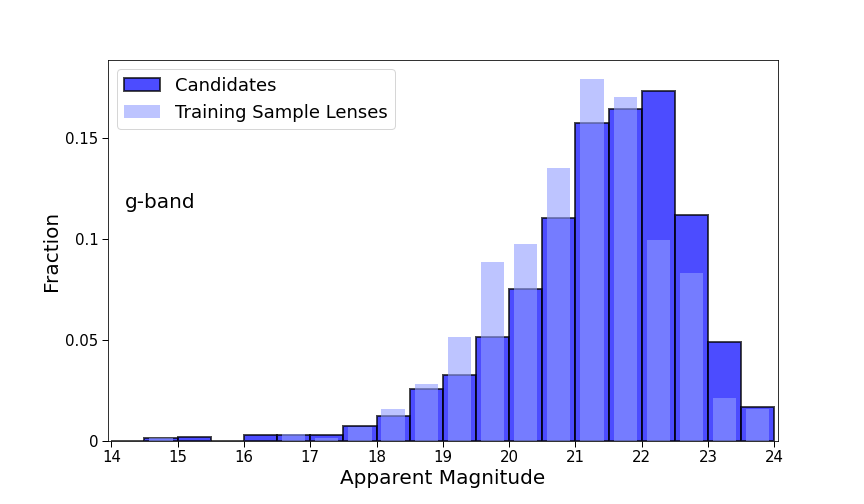}\hspace{-0.4 in}
\includegraphics[width=.42\linewidth]{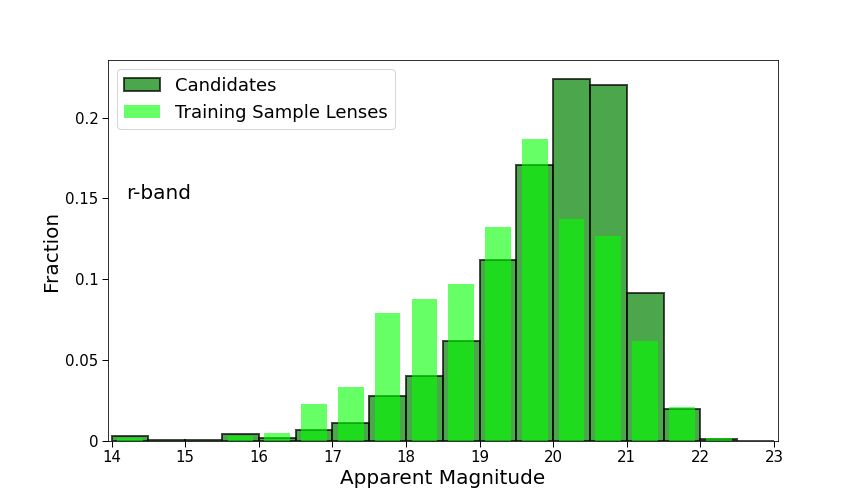}\hspace{-0.4 in}
\includegraphics[width=.42\linewidth]{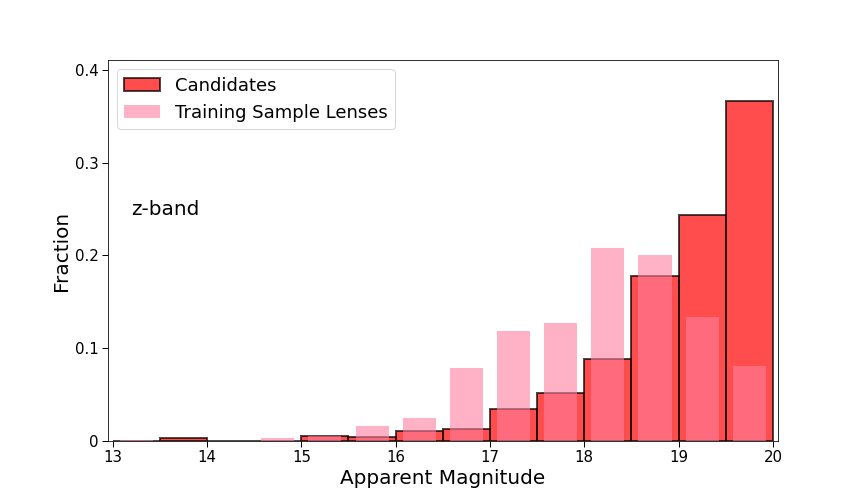}}	
	\captionof{figure}{
\rf{Magnitude distributions for $grz$ bands of our candidate lensing galaxies and lenses in our training sample.
}}
  \label{fig:mags-cands-train}
\end{minipage}

\vspace{0.2 in}

\rf{A more detailed comparison by galaxy type is presented in Figure~\ref{fig:mags-DC-REX}, 
for DEV+COMP (DC) and REX.
The REX type has a narrower distribution for all three bands.
Recall that in the classification scheme of \tractor, 
%the first distinction is galaxy vs. stars.
REX is the designation for those objects classified as galaxies that cannot be confidently promoted to DEV, COMP, or EXP.
They are generally fainter and smaller in apparent size.
This explains the smaller fractions for REX on the bright side of the distributions.
Due to their smaller apparent sizes, if they are too faint, 
they may not be classified by \tractor as spatially extended (see \S~\ref{sec:observations}) at all.
Hence the dearth on the faint side of the $gr$ distributions.
%As for the faint side, 
%The smaller fractions of REX at the faint end for $gr$ band is possibly due to the fact that some of them are below the \tractor detection limit for galaxies.
The exception is the $z$ band: 
there is a significant contribution of REX near the magnitude limit of 20.0.}

\begin{minipage}{0.9\linewidth}
\makebox[\linewidth]{
\hspace{0.4 in}
\includegraphics[width=.42\linewidth]{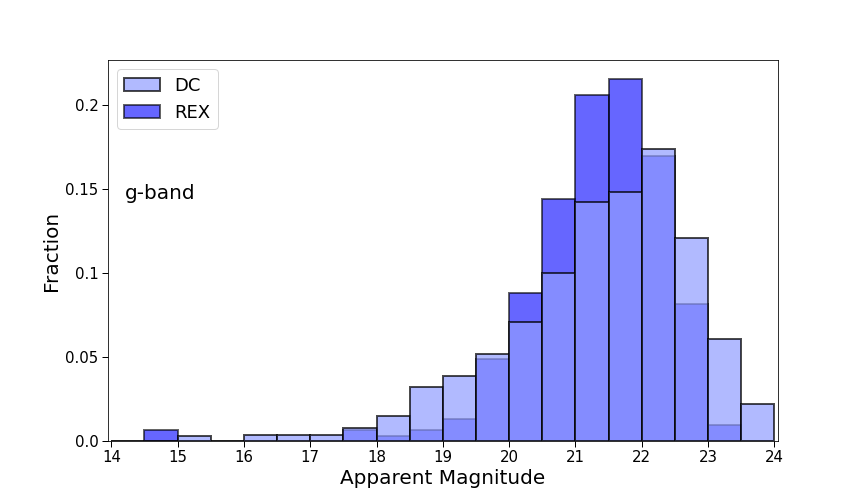}\hspace{-0.4 in}
\includegraphics[width=.42\linewidth]{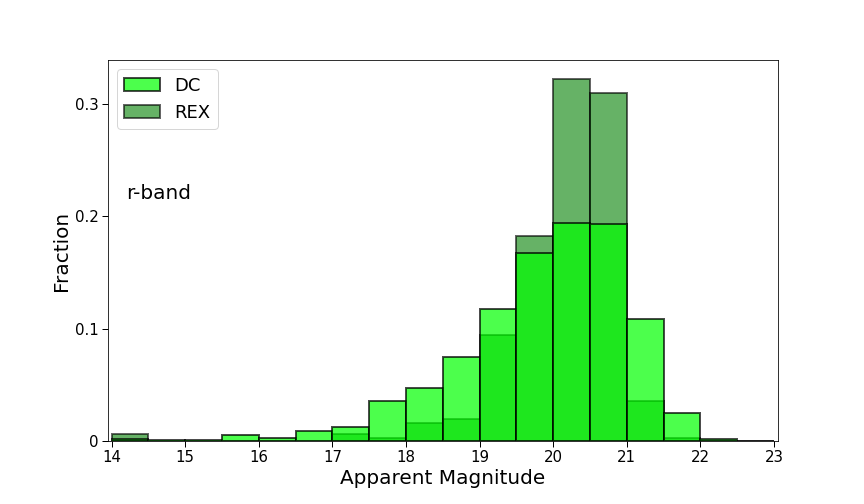}\hspace{-0.4 in}	
\includegraphics[width=.42\linewidth]{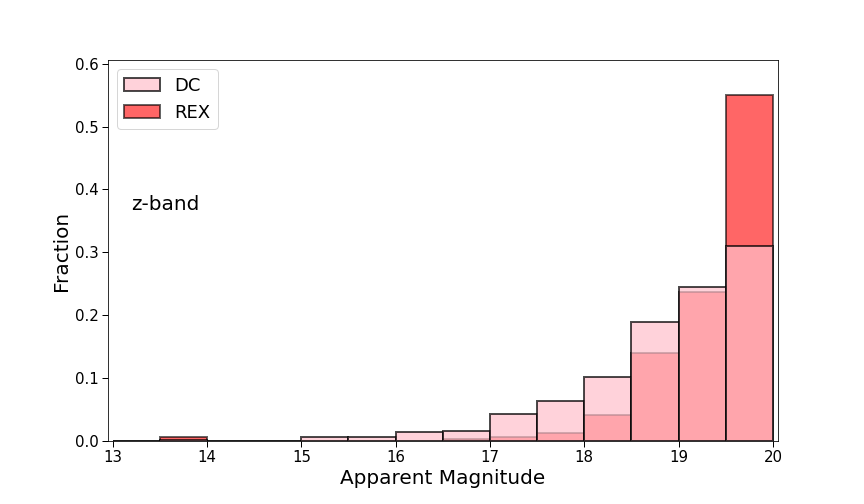}}	
	\captionof{figure}{
	%Illustration of how the ResNet preferentially gives high probability to objects with greater depth, as indicated/measured by the number of passes in $z$-band.
\rf{Apparent magnitude distributions for DEV and COMP (DC) and REX for $grz$ bands. 
}}
  \label{fig:mags-DC-REX}
\end{minipage}

\subsection{Redshift Distributions}\label{sec:redshift-distr}

\rf{We now compare the redshift distributions of our candidates and the lenses in the training sample (Figure~\ref{fig:redshifts-cands-train}).
%The relative dearth of candidates in the low redshift ($leq 0.2$)bins is due to the fact most of them were discovered before.
For spectroscopic redshifts, 
our candidates approximately have the same distribution as the training lenses.
One notable difference is the higher fraction of training lenses for $\zl \lesssim 0.25$,
which is expected: a large fraction of these relatively nearby lenses have been observed spectroscopically in SDSS.
A more detailed discussion about the behavior of the mid-range ($0.25 \lesssim \zl \lesssim 0.6$) is given below in the context of comparing the distributions of DC vs. REX candidates.}
%For bins at 0.75 or higher,
%However, there are only a small fraction of galaxies at $z > 0.75$ in SDSS.
%It is not clear whether firm conclusions can be drawn based on these small numbers at this time.}
%This is due to the large number of DES lenses and candidates in our training sample, the vast majority of which do not have spectroscopic redshifts.
%As noted earlier, DES has deeper observations than the Legacy Surveys.
%For spectroscopic redshift $\gtrsim 0.6$, 
%we have found many lens candidates, even though there are relatively few confirmed lensing systems in that redshift range.
%That there are higher fractions of candidates in the high redshift bins are consistent with the magnitudes of our candidates having a fainter distribution than the training lenses. 
%It is a possible indication that we can  discover lenses that have a higher redshift distribution than the lenses in the training sample.
%This is why using observed data for training works.

\begin{minipage}{0.9\linewidth}
\makebox[\linewidth]{
\hspace{0.4 in}
\includegraphics[width=.6\linewidth]{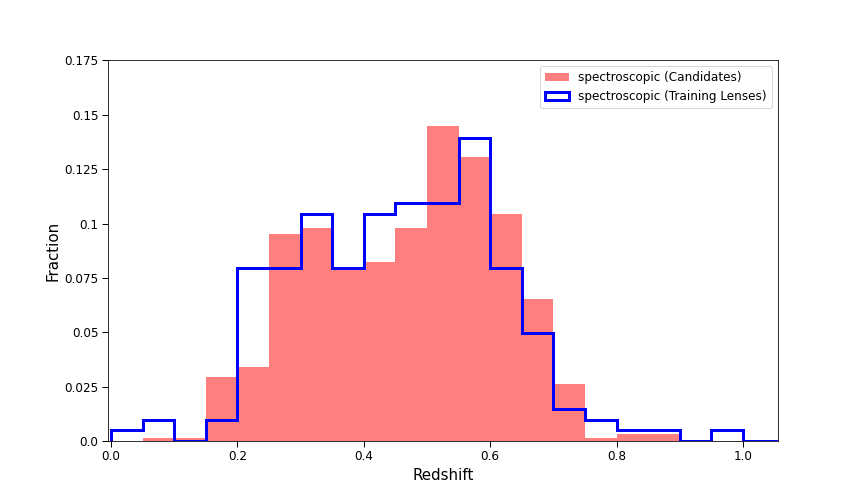}\hspace{-0.2 in}
\includegraphics[width=.6\linewidth]{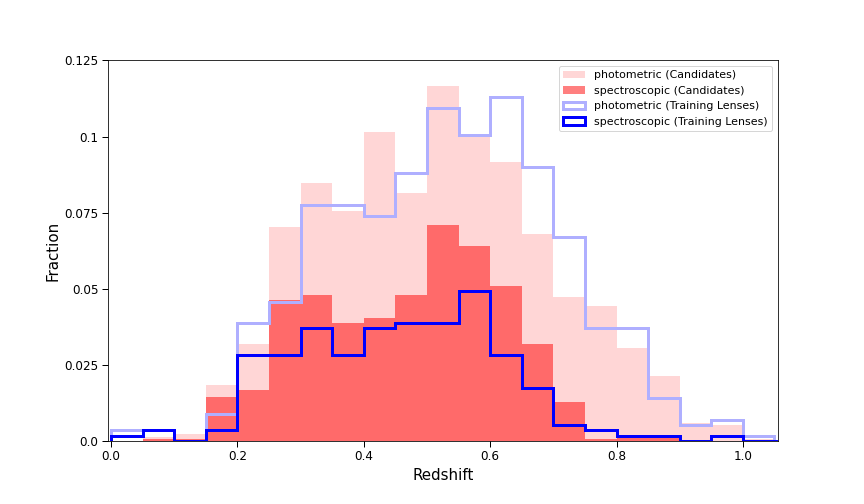}}
\captionof{figure}{
\rf{Distributions of the 642 of the \lenstot candidates
and 201 of the \lenstrain training lenses
%(Table~\ref{tab:all-cands}, first row),
that have spectroscopic redshifts from SDSS DR16 
%\rff{are shown in red columns and blue outline, respectively}.
are shown in the left panel. 
The right panel adds 
%\rff{The light red columns and light blue outline show the respective distributions after adding} 
photometric redshifts from \citet{zhou2020a} for those objects without spectroscopic redshifts.}
%that have spectroscopic redshifts from SDSS DR16 are shown in the left panel. 
%The right panel adds photometric redshifts from \citet{zhou2020a} for those objects without spectroscopic redshifts.
}
\label{fig:redshifts-cands-train}
\end{minipage}

\vspace{0.2 in}

\rf{With photometric redshifts added for objects without spectroscopic redshifts (right panel of Figure~\ref{fig:redshifts-cands-train}), 
the distributions of the candidates again are similar to the lenses in the training sample.
One difference is the peak of the distribution of the training lenses is $\sim 0.1$ higher than our candidates.}

\rf{To understand this difference, 
we show the color comparison between our candidates and the training lenses (Figure~\ref{fig:rz-color-cands-train}).
The training lenses are slightly redder.
% by about 0.5~mag.
A strong majority of the training lenses are from the DES region, 
all of which are of the DC type (see Figure~\ref{fig:lenses-nonlenses-training}).
Given that many of our candidates that are DC are from outside of the DES region (DECaLS but non-DES or MzLS), and the remaining are REX from the entire Legacy Surveys footprint,
it is not a surprise that there is a small color difference.
The vast majority of the DES lenses in the training sample above $z = 0.5$ only have photometric redshifts.
This small offset in redshift distributions is consistent with the color difference between the training lenses and our candidates.
}

\begin{minipage}{0.9\linewidth}
\makebox[\linewidth]{
\includegraphics[width=.6\linewidth]{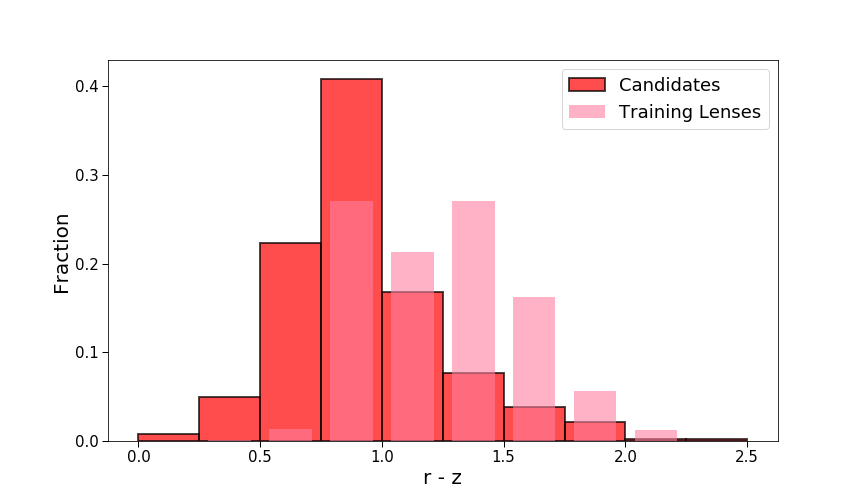}
}	
\captionof{figure}{
\rf{The $r - z$ color distributions of our candidates and the training lenses. 
}}
\label{fig:rz-color-cands-train}
\end{minipage}

\vspace{0.2 in}

\rf{ 
%%%%These plots can help the audiences to understand the advantages and disadvantages of the approaches used in this paper in multiple aspects.
The comparison of redshift distributions for DC and REX are shown in Figure~\ref{fig:redshifts-DC-REX}.
%As mentioned before, generally speaking, 
%DC galaxies have larger apparent sizes than REX and are likely more massive, 
%and correspondingly,
%the REX candidates typically have smaller lens-image separations.
There is clear contrast between these two types. 
For both spectroscopic and photometric redshift distributions, around $\zl \sim 0.5$,
while the DC lenses peak,
we find a ``valley" for the REX.
}
%the magnitude distributions of DC and REX (Figure~\ref{fig:mags-DC-REX}) seem to show that fractionally, 
%there are more candidate lenses with fainter magnitudes, 
%and fewer of them with brighter magnitudes.
%This is not a surprise as if these objects are brighter they would have likely been typed as DC or EXP.

\rf{For spectroscopic redshift distributions,
%, fractionally,there are more REX candidates 
as stated earlier, in the low redshift range ($ \zl \lesssim 0.25$),
a large percentage of DC lenses are known systems and have been included in the training sample, 
leaving fewer new DC lenses to be discovered.
%In contrast, the REX lenses are typically smaller systems, 
%and only a small percentage of them are known lensing systems. 
%Of the \lensknown known systems (but not in our training sample; 
%see Table~\ref{tab:all-cands}) among our candidates, 
%only 24 are REX, of which merely eight are in this redshift range.
%This is a previously ``under-discovered" category.
%% should I also say: also identified by
%Therefore, we expect there are REX lenses in this redshift range, 
%Clearly, our ResNet models are capable of finding them.
%there are no REX lenses in the training sample, 
%and our ResNet models are still capable of finding these typically smaller systems.
In the mid-range ($0.25 \lesssim \zl \lesssim 0.6$), the DC fractions are higher.
For background sources at $z_s \sim 2$, 
the integrated optical depth for strong lensing significantly increases in this redshift range \citep[e.g.,][]{robertson2020a}.
Indeed, most of the training lenses with spectroscopic redshifts are in this range (Figure~\ref{fig:redshifts-cands-train}, left panel).
Therefore it is reasonable to expect that our ResNet models perform well in finding DC lenses
(possibly with a high level of completeness, although as mentioned before, to assess completeness is difficulty to do at this time).
That we have found a large number of DC lenses is consistent with this expectation.
As for the ``valley" for REX, one possible explanation is that given the large number of DC training lenses in this redshift range, 
our neural net models are highly ``tuned" to find DC lenses, 
and consequently somewhat biased against finding REX lens candidates.
We will test this hypothesis by including REX lenses in a training sample for our next search.
%However, since we have not trained a neural network using lenses that are REX or galaxies in the MzLS region,
%(or possibly using a targeted training sample for searches in different categories) will improve the efficiency (or performance?.
If true, this would suggest that there are more REX lenses to be found in this range.
%As we mentioned before, completeness is difficulty to assess at this stage.
%However, this possibility seems to be supported by the fact that our candidates and the training sample are reasonably well-matched in this range 
%(with one exception for photo-z in the bin of 0.4 - 0.45; see 
%(Figure~\ref{fig:redshifts-cands-train}, left panel).
%For $z > 0.6$, there are again more REX fractionally.
%This is likely because for these farther away, fainter galaxies, \tractor can only confidently classify them as spatially extended, or REX. 
%them to a more specific galaxy type (DEV, COMP, or EXP).
With photometric redshifts added for objects without spectroscopic redshifts (right panel of Figure~\ref{fig:redshifts-DC-REX}), 
the same ``valley" for the REX candidates in approximately the same redshift range can be seen.}
%For photometric redshifts, we see a somewhat similar trend.
%The differences are: 1) the dominance of REX at low redshift occurred most notably in the bin of $0.3 < z < 0.35$; 
%2) DC galaxies are slightly less dominant in the mid-range but this dominance extends further into the bin of $0.65 <z < 0.7$;
%For the highest redshift bins ($z > 0.9$) we only find DC candidates.
%The DC lenses in these highest redshift bins tend to be massive elliptical galaxies and most have large lens-image separations (e.g., DESI-072.0873-19.4173 with a photometric redshift of $1.021 \pm 0.061$ has an average lens-image separation of $\sim 3\twopr$; see Figure~\ref{fig:example-cands}).
%REX lensing systems usually have smaller lens-image separations, 
%which make it difficult to detect in an imaging survey at high redshifts.
%This comparison shows that by searching for lenses in REX we not only gain an additional 307 lens candidates,
%but their distributions are complementary to those of the DC candidates.
%especially contributed to 

\begin{minipage}{0.9\linewidth}
\makebox[\linewidth]{
\hspace{0.4 in}
\includegraphics[width=.6\linewidth]{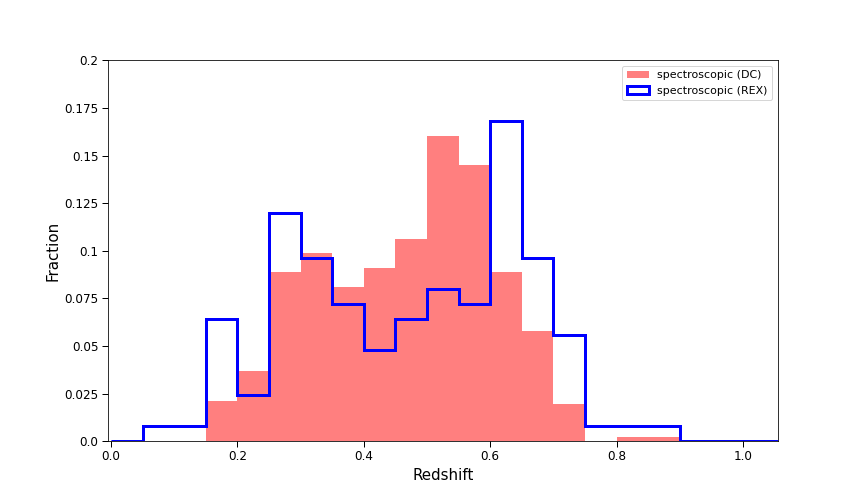}\hspace{-0.2 in}
\includegraphics[width=.6\linewidth]{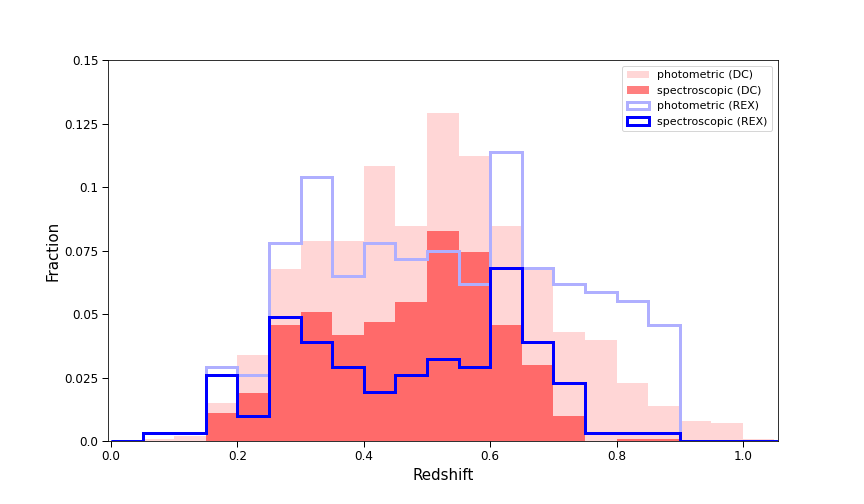}
}	
\captionof{figure}{
\rf{ Distributions of the 517 of the 
\rff{\lensDC} 
DEV and COMP (DC) 
and 125 of the 
\rff{\lensREX} 
REX candidates
%(Table~\ref{tab:all-cands}, first row),
that have spectroscopic redshifts from SDSS DR16 
are shown in the left panel.
%\rff{are shown in red columns and blue outline, respectively}.
%are shown in the left panel. 
%The right panel adds 
%\rff{The light red columns and light blue outline show the respective distributions after adding} 
The right panel adds photometric redshifts from \citet{zhou2020a} for those objects without spectroscopic redshifts.}
%Photometric redshifts from \citet{zhou2020a} for the DC and %REX candidates shown in the right panel.
}
  \label{fig:redshifts-DC-REX}
\end{minipage}

\subsection{Probability Distributions}\label{sec:proba-distr}

\rf{We compare the neural network probabilities for DC and REX candidates in Figure~\ref{fig:proba-DC-REX}.
The difference for the three lowest probability bins is rather striking.
It almost certainly means that at the probability threshold of 0.1, 
%which is what we have done in this paper,
the level of completeness for REX candidates is much lower than for DC.
In our discussion in \S~\ref{sec:redshift-distr},
we suspected that this might be the case.
We can improve the completeness for the REX candidates by lowering the threshold.
But that likely implies a much larger set of images to inspect.
As we have suggested in \S~\ref{sec:redshift-distr}, 
a better remedy may be to include REX lenses in a future training sample, 
and this would also allow us to do a fairer comparison of the probability distributions of DC and REX candidates in our next search.}

\begin{minipage}{0.9\linewidth}
\makebox[\linewidth]{
\includegraphics[width=.6\linewidth]{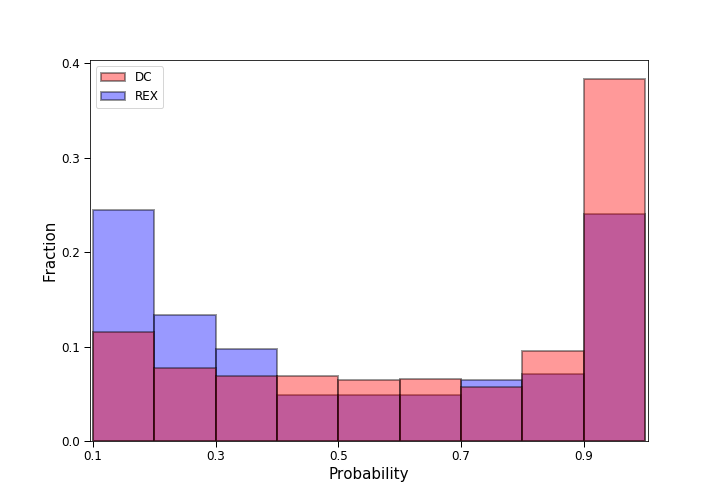}
}	
\captionof{figure}{
\rf{The probability distributions of our DEV+COMP (DC) and REX candidates. 
}}
\label{fig:proba-DC-REX}
\end{minipage}

\vspace{0.2 in}

\subsection{Implications for Future Searches}\label{sec:future-searches}

\rf{As we have noted above,}
for future searches, there is still room for improvement, 
both in terms of algorithm and the construction of the training sample.
On the algorithm side, we have started experimenting with a variety of approaches (see \S~\ref{sec:shielded-model}).  
In this paper we have shown the results from one of them, the ``shielded" model,
with \shieldnewtot \emph{new} candidates found in the same footprint.
This is a promising sign that further exploration is warranted.
In terms of the training sample, 
with the candidates reported in this paper and other recent discoveries \citep[e.g.,][]{canameras2020a}
we can add more lenses and lens candidates to our training sample.
It is possible that further increasing the number of non-lenses in our training sample would help as well, as the current number of 21,000 is still relatively small.

\rf{We also note that in our training sample, 
we did not include any lensed QSO systems.
It is therefore not a surprise that the overwhelming majority of the candidates are galaxy-galaxy lensing systems.
However, we have found a small number of lensed QSO candidate systems, e.g., DESI-273.3831+34.2652 (see Figure~\ref{fig:example-cands})
and DESI-055.7976-28.4777 (\pound68 in our \href{https://sites.google.com/usfca.edu/neuralens/publications/lens-candidates-huang-2020b}{online table}).
In a future search, we will explore the possibility of including known lensed QSOs in the training sample,
%as is presented below.}
%\rf{We would like to see in our future searches whether adding lensed QSOs in our training sample will help find more new systems.
or possibly do a separate lensed QSO search.
}

\rf{
The Rubin Observatory Legacy Survey of Space and Time (LSST), %which will be by far the largest ground-based survey, 
will commence in the near future,
with $\mathcal{O}(10^5)$ strong lenses expected to be discovered \citep[e.g.,][]{collett2015a}.
We have shown that a large number of strong lenses can be discovered by using \lenstrain observed lenses in the training sample.
%and lens candidates found by other groups, 
%We can dramatically increase the number of lenses in our next training sample by including these new strong lens candidates, 
%and this will likely lead to improved performance of the neural network models. Furthermore, 
We have also shown that it is possible to discover lenses that are fainter than the ones in the training sample.
%and increase the number of strong lenses with high redshifts.
Therefore from our experience, using observed images 
%(as opposed to simulated images) 
is a viable
%, and even competitive, %a competitive strategy.
path for LSST. 
One strategy would be to find new lenses iteratively as the observation depth increases over time.
%Especially given that training time is getting ever shorter.
%In the early years of LSST, perhaps it is better to prioritize the neural network purity/efficiency by using a high probability threshold, in order to efficiently increase the number of lenses in the training sample.
As we pointed out at the end of \S~\ref{sec:L18-DC-DECaLS}, 
from the two searches we have conducted in H20 and this paper, respectively,
a larger number of lenses in the training sample is likely a major factor in the improvement of the ResNet's performance.
%The discovery process doubles as the process of creating a labeled training sample.
Thus this kind of ``active learning" strategy will possibly result in the improvement of the neural net model from each successive training, 
as the observation depth increases and more lensing systems become discoverable, 
precisely when higher neural net purity becomes more desirable.
%and likely necessary.
%If, through this process, the purity improves to better than 1 in several neural net recommendations being a lens candidate, 
%discovering $\mathcal{O}(10^5)$ is a realistic goal.
%Otherwise, a citizen science program may be beneficial.
%%  leverage citizen science program
}

\section{Conclusions}\label{sec:conclusion}
%%------------Conclusions----------------

We have carried out a search for strong gravitational lensing systems in the DESI Legacy Surveys data by using a deep residual neural network, 
developed by \citet{lanusse2018a}, 
trained on observed lenses and non-lenses.
We applied our trained neural network to a total of $\sim$~20~million 
cutout images in DR8 with at least three passes in each of the $grz$ bands and a $z$-band magnitude cut of 20.0 for the galaxy at the center of each image.
%%%% We use only real observations for training.  The smallest training sample and yet found lenses with one of the highest human inspection efficiency.
We have found \lensAnew Grade~A, \lensBnew Grade~B, and \lensCnew Grade~C new candidates.
These include \shieldnewtot candidates found by applying a modified neural network to the same data set.
We believe we have held a high standard in grading these candidate systems.
 \lensaboveCplus of our candidates have a human inspection score $\geq 2.5$,
all of which are at least likely lensing systems.

%%\ed{We note the 335 strong lensing candidates (with 159 Grade A's and B's) that we found for Legacy Surveys DR7 \citep{huang2020a} are not included in this paper.}

We note that the candidates reported in this paper do not include the 335 strong lensing candidates (with 159 Grade A's and B's) we already found for Legacy Surveys DR7 \citep{huang2020a}.

%% These candidates do not overlap with the 335 lens candidates (with 159 Grade A's and B's) we found for Legacy Surveys DR7 \citep{huang2020a}.
Compared with efforts by other groups to search for strong lensing systems in other surveys, 
we use a much smaller training sample of 632 lenses and $\sim$21,000 non-lenses from observed data,
for a survey that covers one third of the sky with nonuniform depth and seeing.
We nevertheless have achieved competitive 
\ed{neural network recommendation purity}
%%human inspection efficiencies, 
and in this paper we report the discovery of \lenstotnew new strong lens candidates.

\rf{For our future searches, 
it is important to note that on one hand, our neural network models \textit{are} capable of finding lenses outside the scope of the training sample in multiple aspects, 
and on the other hand, it is also vital to build a ``statistically representative" training sample as much as possible to achieve a good balance between purity and completeness, so as to maximize the reward for human inspection. 
%with slightly worse but still competitive purity.
%Ideally, it would be best to improve the algorithm in this direction.  
%And we are in the process to trying to develop new algorithm.
%Although we don't have alternative training sample
With the lenses we have discovered in this paper, we can now build a training sample that is more statistically representative of the Legacy Surveys for our next search.}

\rf{Thus for future ground-based large surveys, such as the LSST, 
%based on our experience in finding strong lenses in the DESI Legacy Surveys, 
we suggest an iterative search strategy using observed images with increasingly number of lenses in the training sample.
This ``active learning" strategy will allow for discovering lenses not represented (or not well-represented) in the training sample and the continual improvement of the training sample to increase the capacity and enhance the performance of the neural network model.
}

%\rf{For future searches, the $z$-band cut of 20~mag can be relaxed to fainter galaxies.
%The mismatch between the training and candidates will almost certainly be there.  
%Given that we want to find new systems that are fainter and at higher redshift.
%We will choose DC if efficiency is prized.  
%This would be especially true if the efficiency of the neural net is not significantly improved and we expect to find O($10^5$) lenses.  
%Even so, we should leverage citizen science programs.  
%We will add REX to increase completeness.  Given the lower neural net efficiency for this type, certainly citizen science would be needed.
%}

%%%% during revision, can say: the lenses in training were chose from DR7 (b/c DR8 wasn't ready when we started training).  Now with DR7 and with the new lenses discovered here, and with a larger non-lens set, we can construct an even more powerful training sample for future searches.

% we believe it is possible to periodically retrain the model during the inference phase, 
% as more cases of false positives and/or interesting non-lensing astrophysical phenomena are identified by human inspectors. 
% This has the potential of continually improve the purity of the ResNet recommendations, perhaps especially for an even deeper and wider survey such as LSST.
%%new phenomena are bound to be uncovered, to periodically retrain a neural net is an advisable strategy.

\section{Acknowledgement}\label{sec:acknowledgement}
%%%% We thank 
%%%% We are grateful to Joel Brownstein and Lexi Moustakas for granting us access to the Master Lens Database (\url{http://admin.masterlens.org/index.php}).
This research used resources of the National Energy Research Scientific Computing Center (NERSC), a U.S. Department of Energy Office of Science User Facility operated under Contract No. DE-AC02-05CH11231 and the Computational HEP program in The Department of Energy's Science Office of High Energy Physics provided resources through the ``Cosmology Data Repository" project (Grant \#KA2401022).
%%This work was supported in part by the Director, Office of Science, Office of High Energy Physics of the U.S. Department of Energy under Contract No. DE-AC02-05CH11231.  
X.H. acknowledges the University of San Francisco Faculty Development Fund. 
A.D.'s research is supported by \ed{National Science Foundation's National Optical-Infrared Astronomy Research Laboratory}, 
which is operated by the Association of Universities for Research in Astronomy (AURA) under cooperative agreement with the National Science Foundation.

This paper is based on observations at Cerro Tololo Inter-American Observatory, National Optical
Astronomy Observatory (NOAO Prop. ID: 2014B-0404; co-PIs: D. J. Schlegel and A. Dey), which is operated by the Association of
Universities for Research in Astronomy (AURA) under a cooperative agreement with the
National Science Foundation.

This project used data obtained with the Dark Energy Camera (DECam),
which was constructed by the Dark Energy Survey (DES) collaboration.
Funding for the DES Projects has been provided by 
the U.S. Department of Energy, 
the U.S. National Science Foundation, 
the Ministry of Science and Education of Spain, 
the Science and Technology Facilities Council of the United Kingdom, 
the Higher Education Funding Council for England, 
the National Center for Supercomputing Applications at the University of Illinois at Urbana-Champaign, 
the Kavli Institute of Cosmological Physics at the University of Chicago, 
the Center for Cosmology and Astro-Particle Physics at the Ohio State University, 
the Mitchell Institute for Fundamental Physics and Astronomy at Texas A\&M University, 
Financiadora de Estudos e Projetos, Funda{\c c}{\~a}o Carlos Chagas Filho de Amparo {\`a} Pesquisa do Estado do Rio de Janeiro, 
Conselho Nacional de Desenvolvimento Cient{\'i}fico e Tecnol{\'o}gico and the Minist{\'e}rio da Ci{\^e}ncia, Tecnologia e Inovac{\~a}o, 
the Deutsche Forschungsgemeinschaft, 
and the Collaborating Institutions in the Dark Energy Survey.
The Collaborating Institutions are 
Argonne National Laboratory, 
the University of California at Santa Cruz, 
the University of Cambridge, 
Centro de Investigaciones En{\'e}rgeticas, Medioambientales y Tecnol{\'o}gicas-Madrid, 
the University of Chicago, 
University College London, 
the DES-Brazil Consortium, 
the University of Edinburgh, 
the Eidgen{\"o}ssische Technische Hoch\-schule (ETH) Z{\"u}rich, 
Fermi National Accelerator Laboratory, 
the University of Illinois at Urbana-Champaign, 
the Institut de Ci{\`e}ncies de l'Espai (IEEC/CSIC), 
the Institut de F{\'i}sica d'Altes Energies, 
Lawrence Berkeley National Laboratory, 
the Ludwig-Maximilians Universit{\"a}t M{\"u}nchen and the associated Excellence Cluster Universe, 
the University of Michigan, 
{the} National Optical Astronomy Observatory, 
the University of Nottingham, 
the Ohio State University, 
the OzDES Membership Consortium
the University of Pennsylvania, 
the University of Portsmouth, 
SLAC National Accelerator Laboratory, 
Stanford University, 
the University of Sussex, 
and Texas A\&M University.

%%https://www.nersc.gov/users/accounts/user-accounts/acknowledge-nersc/
%%http://legacysurvey.org/

\bibliographystyle{aasjournal}
\bibliography{dustarchive}

\begin{thebibliography}{}
\expandafter\ifx\csname natexlab\endcsname\relax\def\natexlab#1{#1}\fi
\providecommand{\url}[1]{\href{#1}{#1}}
\providecommand{\dodoi}[1]{doi:~\href{http://doi.org/#1}{\nolinkurl{#1}}}
\providecommand{\doeprint}[1]{\href{http://ascl.net/#1}{\nolinkurl{http://ascl.net/#1}}}
\providecommand{\doarXiv}[1]{\href{https://arxiv.org/abs/#1}{\nolinkurl{https://arxiv.org/abs/#1}}}

\bibitem[{{Abbott} {et~al.}(2017){Abbott}, {Abbott}, {Abbott}, {Acernese},
  {Ackley}, {Adams}, {Adams}, {Addesso}, {Adhikari}, {Adya}, \&
  {Affeldt}}]{abbott2017a}
{Abbott}, B.~P., {Abbott}, R., {Abbott}, T.~D., {et~al.} 2017, \nat, 551, 85,
  \dodoi{10.1038/nature24471}

\bibitem[{{Abbott} {et~al.}(2018){Abbott}, {Abdalla}, {Annis}, {Bechtol},
  {Blazek}, {Benson}, {Bernstein}, {Bernstein}, {Bertin}, {Dark Energy Survey
  Collaboration}, \& {South Pole Telescope Collaboration}}]{abbott2018b}
{Abbott}, T.~M.~C., {Abdalla}, F.~B., {Annis}, J., {et~al.} 2018, \mnras, 480,
  3879, \dodoi{10.1093/mnras/sty1939}

\bibitem[{{Auger} {et~al.}(2010){Auger}, {Treu}, {Bolton}, {Gavazzi},
  {Koopmans}, {Marshall}, {Moustakas}, \& {Burles}}]{auger2010a}
{Auger}, M.~W., {Treu}, T., {Bolton}, A.~S., {et~al.} 2010, \apj, 724, 511,
  \dodoi{10.1088/0004-637X/724/1/511}

\bibitem[{{Bolton} {et~al.}(2008){Bolton}, {Burles}, {Koopmans}, {Treu},
  {Gavazzi}, {Moustakas}, {Wayth}, \& {Schlegel}}]{bolton2008a}
{Bolton}, A.~S., {Burles}, S., {Koopmans}, L. V.~E., {et~al.} 2008, \apj, 682,
  964, \dodoi{10.1086/589327}

\bibitem[{{Bolton} {et~al.}(2006){Bolton}, {Burles}, {Koopmans}, {Treu}, \&
  {Moustakas}}]{bolton2006a}
{Bolton}, A.~S., {Burles}, S., {Koopmans}, L.~V.~E., {Treu}, T., \&
  {Moustakas}, L.~A. 2006, \apj, 638, 703, \dodoi{10.1086/498884}

\bibitem[{{Brownstein} {et~al.}(2012){Brownstein}, {Bolton}, {Schlegel},
  {Eisenstein}, {Kochanek}, {Connolly}, {Maraston}, {Pandey}, {Seitz}, {Wake},
  {Wood-Vasey}, {Brinkmann}, {Schneider}, \& {Weaver}}]{brownstein2012a}
{Brownstein}, J.~R., {Bolton}, A.~S., {Schlegel}, D.~J., {et~al.} 2012, \apj,
  744, 41, \dodoi{10.1088/0004-637X/744/1/41}

\bibitem[{{Canameras} {et~al.}(2020){Canameras}, {Schuldt}, {Suyu},
  {Taubenberger}, {Meinhardt}, {Leal-Taixe}, {Lemon}, {Rojas}, \&
  {Savary}}]{canameras2020a}
{Canameras}, R., {Schuldt}, S., {Suyu}, S.~H., {et~al.} 2020, arXiv e-prints,
  arXiv:2004.13048.
\newblock \doarXiv{2004.13048}

\bibitem[{{Carrasco} {et~al.}(2017){Carrasco}, {Barrientos}, {Anguita},
  {Garc{\'{\i}}a-Vergara}, {Bayliss}, {Gladders}, {Gilbank}, {Yee}, \&
  {West}}]{carrasco2017a}
{Carrasco}, M., {Barrientos}, L.~F., {Anguita}, T., {et~al.} 2017, \apj, 834,
  210, \dodoi{10.3847/1538-4357/834/2/210}

\bibitem[{{Choi} {et~al.}(2020){Choi}, {Hasselfield}, {Ho}, {Koopman}, {Lungu},
  {Abitbol}, {Addison}, {Ade}, {Aiola}, {Alonso}, {Amiri}, {Amodeo}, {Angile},
  {Austermann}, {Baildon}, {Battaglia}, {Beall}, {Bean}, {Becker}, {Bond},
  {Bruno}, {Calabrese}, {Calafut}, {Campusano}, {Carrero}, {Chesmore}, {Cho},
  {Clark}, {Cothard}, {Crichton}, {Crowley}, {Darwish}, {Datta}, {Denison},
  {Devlin}, {Duell}, {Duff}, {Duivenvoorden}, {Dunkley}, {D{\"u}nner},
  {Essinger-Hileman}, {Fankhanel}, {Ferraro}, {Fox}, {Fuzia}, {Gallardo},
  {Gluscevic}, {Golec}, {Grace}, {Gralla}, {Guan}, {Hall}, {Halpern}, {Han},
  {Hargrave}, {Henderson}, {Hensley}, {Hill}, {Hilton}, {Hilton}, {Hincks},
  {Hlo{\v{z}}ek}, {Hubmayr}, {Huffenberger}, {Hughes}, {Infante}, {Irwin},
  {Jackson}, {Klein}, {Knowles}, {Kosowsky}, {Lakey}, {Li}, {Li}, {Li},
  {Lokken}, {Louis}, {MacInnis}, {Madhavacheril}, {Maldonado}, {Mallaby-Kay},
  {Marsden}, {Maurin}, {McMahon}, {Menanteau}, {Moodley}, {Morton}, {Naess},
  {Namikawa}, {Nati}, {Newburgh}, {Nibarger}, {Nicola}, {Niemack}, {Nolta},
  {Orlowski-Sherer}, {Page}, {Pappas}, {Partridge}, {Phakathi}, {Prince},
  {Puddu}, {Qu}, {Rivera}, {Robertson}, {Rojas}, {Salatino}, {Schaan},
  {Schillaci}, {Schmitt}, {Sehgal}, {Sherwin}, {Sierra}, {Sievers}, {Sifon},
  {Sikhosana}, {Simon}, {Spergel}, {Staggs}, {Stevens}, {Storer}, {Sunder},
  {Switzer}, {Thorne}, {Thornton}, {Trac}, {Treu}, {Tucker}, {Vale}, {Van
  Engelen}, {Van Lanen}, {Vavagiakis}, {Wagoner}, {Wang}, {Ward}, {Wollack},
  {Xu}, {Zago}, \& {Zhu}}]{choi2020a}
{Choi}, S.~K., {Hasselfield}, M., {Ho}, S.-P.~P., {et~al.} 2020, arXiv
  e-prints, arXiv:2007.07289.
\newblock \doarXiv{2007.07289}

\bibitem[{{Collett} {et~al.}(2019){Collett}, {Montanari}, \&
  {R{\"a}s{\"a}nen}}]{collett2019a}
{Collett}, T., {Montanari}, F., \& {R{\"a}s{\"a}nen}, S. 2019, \prl, 123,
  231101, \dodoi{10.1103/PhysRevLett.123.231101}

\bibitem[{{Collett}(2015)}]{collett2015a}
{Collett}, T.~E. 2015, \apj, 811, 20, \dodoi{10.1088/0004-637X/811/1/20}

\bibitem[{{Collett} {et~al.}(2018){Collett}, {Oldham}, {Smith}, {Auger},
  {Westfall}, {Bacon}, {Nichol}, {Masters}, {Koyama}, \& {van den
  Bosch}}]{collett2018a}
{Collett}, T.~E., {Oldham}, L.~J., {Smith}, R.~J., {et~al.} 2018, Science, 360,
  1342, \dodoi{10.1126/science.aao2469}

\bibitem[{{Cornachione} {et~al.}(2018){Cornachione}, {Bolton}, {Shu}, {Zheng},
  {Montero-Dorta}, {Brownstein}, {Oguri}, {Kochanek}, {Mao},
  {P{\`e}rez-Fournon}, {Marques-Chaves}, \& {M{\`e}nard}}]{cornachione2018a}
{Cornachione}, M.~A., {Bolton}, A.~S., {Shu}, Y., {et~al.} 2018, \apj, 853,
  148, \dodoi{10.3847/1538-4357/aaa412}

\bibitem[{{Dey} {et~al.}(2016){Dey}, {Rabinowitz}, {Karcher}, {Bebek},
  {Baltay}, {Sprayberry}, {Valdes}, {Stupak}, {Donaldson}, {Emmet}, {Hurteau},
  {Abareshi}, {Marshall}, {Lang}, {Fitzpatrick}, {Daly}, {Joyce}, {Schlegel},
  {Schweiker}, {Allen}, {Blum}, \& {Levi}}]{dey2016a}
{Dey}, A., {Rabinowitz}, D., {Karcher}, A., {et~al.} 2016, in \procspie, Vol.
  9908, Ground-based and Airborne Instrumentation for Astronomy VI, 99082C

\bibitem[{{Dey} {et~al.}(2019){Dey}, {Schlegel}, {Lang}, {Blum}, {Burleigh},
  {Fan}, {Findlay}, {Finkbeiner}, {Herrera}, {Juneau}, {Landriau}, {Levi},
  {McGreer}, {Meisner}, {Myers}, {Moustakas}, {Nugent}, {Patej}, {Schlafly},
  {Walker}, {Valdes}, {Weaver}, {Y{\`e}che}, {Zou}, {Zhou}, {Abareshi},
  {Abbott}, {Abolfathi}, {Aguilera}, {Alam}, {Allen}, {Alvarez}, {Annis},
  {Ansarinejad}, {Aubert}, {Beechert}, {Bell}, {BenZvi}, {Beutler}, {Bielby},
  {Bolton}, {Brice{\~n}o}, {Buckley-Geer}, {Butler}, {Calamida}, {Carlberg},
  {Carter}, {Casas}, {Castander}, {Choi}, {Comparat}, {Cukanovaite}, {Delubac},
  {DeVries}, {Dey}, {Dhungana}, {Dickinson}, {Ding}, {Donaldson}, {Duan},
  {Duckworth}, {Eftekharzadeh}, {Eisenstein}, {Etourneau}, {Fagrelius},
  {Farihi}, {Fitzpatrick}, {Font-Ribera}, {Fulmer}, {G{\"a}nsicke},
  {Gaztanaga}, {George}, {Gerdes}, {Gontcho}, {Gorgoni}, {Green}, {Guy},
  {Harmer}, {Hernand ez}, {Honscheid}, {(Wendy Huang}, {James}, {Jannuzi},
  {Jiang}, {Joyce}, {Karcher}, {Karkar}, {Kehoe}, {Jean-Paul}, {Kneib},
  {Kueter-Young}, {Lan}, {Lauer}, {Le Guillou}, {Le Van Suu}, {Lee}, {Lesser},
  {Perreault Levasseur}, {Li}, {Mann}, {Marshall}, {Mart{\'\i}nez-V{\'a}zquez},
  {Martini}, {du Mas des Bourboux}, {McManus}, {Meier}, {M{\'e}nard},
  {Metcalfe}, {Mu{\~n}oz-Guti{\'e}rrez}, {Najita}, {Napier}, {Narayan},
  {Newman}, {Nie}, {Nord}, {Norman}, {Olsen}, {Paat}, {Palanque-Delabrouille},
  {Peng}, {Poppett}, {Poremba}, {Prakash}, {Rabinowitz}, {Raichoor}, {Rezaie},
  {Robertson}, {Roe}, {Ross}, {Ross}, {Rudnick}, {Safonova}, {Saha},
  {S{\'a}nchez}, {Savary}, {Schweiker}, {Scott}, {Seo}, {Shan}, {Silva},
  {Slepian}, {Soto}, {Sprayberry}, {Staten}, {Stillman}, {Stupak}, {Summers},
  {Sien Tie}, {Tirado}, {Vargas-Maga{\~n}a}, {Vivas}, {Wechsler}, {Williams},
  {Yang}, {Yang}, {Yapici}, {Zaritsky}, {Zenteno}, {Zhang}, {Zhang}, {Zhou}, \&
  {Zhou}}]{dey2019a}
{Dey}, A., {Schlegel}, D.~J., {Lang}, D., {et~al.} 2019, \aj, 157, 168,
  \dodoi{10.3847/1538-3881/ab089d}

\bibitem[{{Diaz Rivero} \& {Dvorkin}(2020)}]{diazrivero2020a}
{Diaz Rivero}, A., \& {Dvorkin}, C. 2020, \prd, 101, 023515,
  \dodoi{10.1103/PhysRevD.101.023515}

\bibitem[{{Diehl} {et~al.}(2017){Diehl}, {Buckley-Geer}, {Lindgren}, {Nord},
  {Gaitsch}, {Gaitsch}, {Lin}, {Allam}, {Collett}, {Furlanetto}, {Gill},
  {More}, {Nightingale}, {Odden}, {Pellico}, {Tucker}, {da Costa}, {Fausti
  Neto}, {Kuropatkin}, {Soares-Santos}, {Welch}, {Zhang}, {Frieman}, {Abdalla},
  {Annis}, {Benoit-L{\'e}vy}, {Bertin}, {Brooks}, {Burke}, {Carnero Rosell},
  {Carrasco Kind}, {Carretero}, {Cunha}, {D'Andrea}, {Desai}, {Dietrich},
  {Drlica-Wagner}, {Evrard}, {Finley}, {Flaugher}, {Garc{\'{\i}}a-Bellido},
  {Gerdes}, {Goldstein}, {Gruen}, {Gruendl}, {Gschwend}, {Gutierrez}, {James},
  {Kuehn}, {Kuhlmann}, {Lahav}, {Li}, {Lima}, {Maia}, {Marshall}, {Menanteau},
  {Miquel}, {Nichol}, {Nugent}, {Ogando}, {Plazas}, {Reil}, {Romer}, {Sako},
  {Sanchez}, {Santiago}, {Scarpine}, {Schindler}, {Schubnell},
  {Sevilla-Noarbe}, {Sheldon}, {Smith}, {Sobreira}, {Suchyta}, {Swanson},
  {Tarle}, {Thomas}, {Walker}, \& {DES Collaboration}}]{diehl2017a}
{Diehl}, H.~T., {Buckley-Geer}, E.~J., {Lindgren}, K.~A., {et~al.} 2017, \apjs,
  232, 15, \dodoi{10.3847/1538-4365/aa8667}

\bibitem[{{Flaugher} {et~al.}(2015){Flaugher}, {Diehl}, {Honscheid}, {Abbott},
  {Alvarez}, {Angstadt}, {Annis}, {Antonik}, {Ballester}, {Beaufore},
  {Bernstein}, {Bernstein}, {Bigelow}, {Bonati}, {Boprie}, {Brooks},
  {Buckley-Geer}, {Campa}, {Cardiel-Sas}, {Castander}, {Castilla}, {Cease},
  {Cela-Ruiz}, {Chappa}, {Chi}, {Cooper}, {da Costa}, {Dede}, {Derylo},
  {DePoy}, {de Vicente}, {Doel}, {Drlica-Wagner}, {Eiting}, {Elliott}, {Emes},
  {Estrada}, {Fausti Neto}, {Finley}, {Flores}, {Frieman}, {Gerdes},
  {Gladders}, {Gregory}, {Gutierrez}, {Hao}, {Holland}, {Holm}, {Huffman},
  {Jackson}, {James}, {Jonas}, {Karcher}, {Karliner}, {Kent}, {Kessler},
  {Kozlovsky}, {Kron}, {Kubik}, {Kuehn}, {Kuhlmann}, {Kuk}, {Lahav}, {Lathrop},
  {Lee}, {Levi}, {Lewis}, {Li}, {Mandrichenko}, {Marshall}, {Martinez},
  {Merritt}, {Miquel}, {Mu{\~n}oz}, {Neilsen}, {Nichol}, {Nord}, {Ogando},
  {Olsen}, {Palaio}, {Patton}, {Peoples}, {Plazas}, {Rauch}, {Reil}, {Rheault},
  {Roe}, {Rogers}, {Roodman}, {Sanchez}, {Scarpine}, {Schindler}, {Schmidt},
  {Schmitt}, {Schubnell}, {Schultz}, {Schurter}, {Scott}, {Serrano}, {Shaw},
  {Smith}, {Soares-Santos}, {Stefanik}, {Stuermer}, {Suchyta}, {Sypniewski},
  {Tarle}, {Thaler}, {Tighe}, {Tran}, {Tucker}, {Walker}, {Wang}, {Watson},
  {Weaverdyck}, {Wester}, {Woods}, {Yanny}, \& {DES
  Collaboration}}]{flaugher2015a}
{Flaugher}, B., {Diehl}, H.~T., {Honscheid}, K., {et~al.} 2015, \aj, 150, 150,
  \dodoi{10.1088/0004-6256/150/5/150}

\bibitem[{{Freedman} {et~al.}(2019){Freedman}, {Madore}, {Hatt}, {Hoyt},
  {Jang}, {Beaton}, {Burns}, {Lee}, {Monson}, {Neeley}, {Phillips}, {Rich}, \&
  {Seibert}}]{freedman2019a}
{Freedman}, W.~L., {Madore}, B.~F., {Hatt}, D., {et~al.} 2019, \apj, 882, 34,
  \dodoi{10.3847/1538-4357/ab2f73}

\bibitem[{{Freedman} {et~al.}(2020){Freedman}, {Madore}, {Hoyt}, {Jang},
  {Beaton}, {Lee}, {Monson}, {Neeley}, \& {Rich}}]{freedman2020a}
{Freedman}, W.~L., {Madore}, B.~F., {Hoyt}, T., {et~al.} 2020, \apj, 891, 57,
  \dodoi{10.3847/1538-4357/ab7339}

\bibitem[{{Goldstein} \& {Nugent}(2017)}]{goldstein2017a}
{Goldstein}, D.~A., \& {Nugent}, P.~E. 2017, \apjl, 834, L5,
  \dodoi{10.3847/2041-8213/834/1/L5}

\bibitem[{{Goldstein} {et~al.}(2018{\natexlab{a}}){Goldstein}, {Nugent}, \&
  {Goobar}}]{goldstein2018a}
{Goldstein}, D.~A., {Nugent}, P.~E., \& {Goobar}, A. 2018{\natexlab{a}}, arXiv
  e-prints.
\newblock \doarXiv{1809.10147}

\bibitem[{{Goldstein} {et~al.}(2018{\natexlab{b}}){Goldstein}, {Nugent},
  {Kasen}, \& {Collett}}]{goldstein2018b}
{Goldstein}, D.~A., {Nugent}, P.~E., {Kasen}, D.~N., \& {Collett}, T.~E.
  2018{\natexlab{b}}, \apj, 855, 22, \dodoi{10.3847/1538-4357/aaa975}

\bibitem[{{Goobar} {et~al.}(2017){Goobar}, {Amanullah}, {Kulkarni}, {Nugent},
  {Johansson}, {Steidel}, {Law}, {M{\"o}rtsell}, {Quimby}, {Blagorodnova},
  {Brandeker}, {Cao}, {Cooray}, {Ferretti}, {Fremling}, {Hangard}, {Kasliwal},
  {Kupfer}, {Lunnan}, {Masci}, {Miller}, {Nayyeri}, {Neill}, {Ofek},
  {Papadogiannakis}, {Petrushevska}, {Ravi}, {Sollerman}, {Sullivan}, {Taddia},
  {Walters}, {Wilson}, {Yan}, \& {Yaron}}]{goobar2017a}
{Goobar}, A., {Amanullah}, R., {Kulkarni}, S.~R., {et~al.} 2017, Science, 356,
  291, \dodoi{10.1126/science.aal2729}

\bibitem[{Huang {et~al.}(2020)Huang, Storfer, Ravi, Pilon, Domingo, Schlegel,
  Bailey, Dey, Gupta, Herrera, Juneau, Landriau, Lang, Meisner, Moustakas,
  Myers, Schlafly, Valdes, Weaver, Yang, \& Y{\`{e}}che}]{huang2020a}
Huang, X., Storfer, C., Ravi, V., {et~al.} 2020, The Astrophysical Journal,
  894, 78, \dodoi{10.3847/1538-4357/ab7ffb}

\bibitem[{{Inada} {et~al.}(2003){Inada}, {Becker}, {Burles}, {Castand er},
  {Eisenstein}, {Hall}, {Johnston}, {Pindor}, {Richards}, {Schechter},
  {Sekiguchi}, {White}, {Brinkmann}, {Frieman}, {Kleinman}, {Krzesi{\'n}ski},
  {Long}, {Neilsen}, {Newman}, {Nitta}, {Schneider}, {Snedden}, \&
  {York}}]{inada2003a}
{Inada}, N., {Becker}, R.~H., {Burles}, S., {et~al.} 2003, \aj, 126, 666,
  \dodoi{10.1086/375906}

\bibitem[{{Jacobs} {et~al.}(2017){Jacobs}, {Glazebrook}, {Collett}, {More}, \&
  {McCarthy}}]{jacobs2017a}
{Jacobs}, C., {Glazebrook}, K., {Collett}, T., {More}, A., \& {McCarthy}, C.
  2017, \mnras, 471, 167, \dodoi{10.1093/mnras/stx1492}

\bibitem[{{Jacobs} {et~al.}(2019{\natexlab{a}}){Jacobs}, {Collett},
  {Glazebrook}, {McCarthy}, {Qin}, {Abbott}, {Abdalla}, {Annis}, {Avila},
  {Bechtol}, {Bertin}, {Brooks}, {Buckley-Geer}, {Burke}, {Carnero Rosell},
  {Carrasco Kind}, {Carretero}, {da Costa}, {Davis}, {De Vicente}, {Desai},
  {Diehl}, {Doel}, {Eifler}, {Flaugher}, {Frieman}, {Garc{\'{\i}}a-Bellido},
  {Gaztanaga}, {Gerdes}, {Goldstein}, {Gruen}, {Gruendl}, {Gschwend},
  {Gutierrez}, {Hartley}, {Hollowood}, {Honscheid}, {Hoyle}, {James}, {Kuehn},
  {Kuropatkin}, {Lahav}, {Li}, {Lima}, {Lin}, {Maia}, {Martini}, {Miller},
  {Miquel}, {Nord}, {Plazas}, {Sanchez}, {Scarpine}, {Schubnell}, {Serrano},
  {Sevilla-Noarbe}, {Smith}, {Soares-Santos}, {Sobreira}, {Suchyta}, {Swanson},
  {Tarle}, {Vikram}, {Walker}, {Zhang}, \& {Zuntz}}]{jacobs2019a}
{Jacobs}, C., {Collett}, T., {Glazebrook}, K., {et~al.} 2019{\natexlab{a}},
  \mnras, 484, 5330, \dodoi{10.1093/mnras/stz272}

\bibitem[{{Jacobs} {et~al.}(2019{\natexlab{b}}){Jacobs}, {Collett},
  {Glazebrook}, {Buckley-Geer}, {Diehl}, {Lin}, {McCarthy}, {Qin}, {Odden},
  {Caso Escudero}, {Dial}, {Yung}, {Gaitsch}, {Pellico}, {Lindgren}, {Abbott},
  {Annis}, {Avila}, {Brooks}, {Burke}, {Carnero Rosell}, {Carrasco Kind},
  {Carretero}, {da Costa}, {De Vicente}, {Fosalba}, {Frieman},
  {Garc{\'\i}a-Bellido}, {Gaztanaga}, {Goldstein}, {Gruen}, {Gruendl},
  {Gschwend}, {Hollowood}, {Honscheid}, {Hoyle}, {James}, {Krause},
  {Kuropatkin}, {Lahav}, {Lima}, {Maia}, {Marshall}, {Miquel}, {Plazas},
  {Roodman}, {Sanchez}, {Scarpine}, {Serrano}, {Sevilla-Noarbe}, {Smith},
  {Sobreira}, {Suchyta}, {Swanson}, {Tarle}, {Vikram}, {Walker}, {Zhang}, \&
  {DES Collaboration}}]{jacobs2019b}
---. 2019{\natexlab{b}}, \apjs, 243, 17, \dodoi{10.3847/1538-4365/ab26b6}

\bibitem[{{Jauzac} {et~al.}(2018){Jauzac}, {Harvey}, \& {Massey}}]{jauzac2018a}
{Jauzac}, M., {Harvey}, D., \& {Massey}, R. 2018, \mnras, 477, 4046,
  \dodoi{10.1093/mnras/sty909}

\bibitem[{{Kelly} {et~al.}(2015){Kelly}, {Filippenko}, {Burke}, {Hicken},
  {Ganeshalingam}, \& {Zheng}}]{kelly2015a}
{Kelly}, P.~L., {Filippenko}, A.~V., {Burke}, D.~L., {et~al.} 2015, Science,
  347, 1459, \dodoi{10.1126/science.1261475}

\bibitem[{{Khetan} {et~al.}(2020){Khetan}, {Izzo}, {Branchesi}, {Wojtak},
  {Cantiello}, {Murugeshan}, {Cappellaro}, {Della Valle}, {Gall}, {Hjorth},
  {Benetti}, {Brocato}, {Burke}, {Hiramitsu}, {Howell}, {Tomasella}, \&
  {Valenti}}]{khetan2020a}
{Khetan}, N., {Izzo}, L., {Branchesi}, M., {et~al.} 2020, arXiv e-prints,
  arXiv:2008.07754.
\newblock \doarXiv{2008.07754}

\bibitem[{{Kneib} \& {Natarajan}(2011)}]{kneib2011a}
{Kneib}, J.-P., \& {Natarajan}, P. 2011, \aapr, 19, 47,
  \dodoi{10.1007/s00159-011-0047-3}

\bibitem[{{Kochanek}(1991)}]{kochanek1991a}
{Kochanek}, C.~S. 1991, \apj, 373, 354, \dodoi{10.1086/170057}

\bibitem[{{Koopmans} \& {Treu}(2002)}]{koopmans2002a}
{Koopmans}, L.~V.~E., \& {Treu}, T. 2002, \apjl, 568, L5,
  \dodoi{10.1086/340143}

\bibitem[{{Koopmans} {et~al.}(2006){Koopmans}, {Treu}, {Bolton}, {Burles}, \&
  {Moustakas}}]{koopmans2006a}
{Koopmans}, L.~V.~E., {Treu}, T., {Bolton}, A.~S., {Burles}, S., \&
  {Moustakas}, L.~A. 2006, \apj, 649, 599, \dodoi{10.1086/505696}

\bibitem[{{Lang} {et~al.}(2016){Lang}, {Hogg}, \& {Mykytyn}}]{lang2016a}
{Lang}, D., {Hogg}, D.~W., \& {Mykytyn}, D. 2016, {The Tractor: Probabilistic
  astronomical source detection and measurement}, Astrophysics Source Code
  Library.
\newblock \doeprint{1604.008}

\bibitem[{{Lanusse} {et~al.}(2018){Lanusse}, {Ma}, {Li}, {Collett}, {Li},
  {Ravanbakhsh}, {Mandelbaum}, \& {P{\'o}czos}}]{lanusse2018a}
{Lanusse}, F., {Ma}, Q., {Li}, N., {et~al.} 2018, \mnras, 473, 3895,
  \dodoi{10.1093/mnras/stx1665}

\bibitem[{{Li} {et~al.}(2020){Li}, {Napolitano}, {Tortora}, {Spiniello},
  {Koopmans}, {Huang}, {Vernardos}, {Chatterjee}, {Giblin}, {Getman}, {Covone},
  \& {Kuijken}}]{li2020a}
{Li}, R., {Napolitano}, N.~R., {Tortora}, C., {et~al.} 2020, arXiv e-prints,
  arXiv:2004.02715.
\newblock \doarXiv{2004.02715}

\bibitem[{{Li} {et~al.}(2018){Li}, {Gao}, {Ding}, {Wang}, \& {Zhang}}]{li2018a}
{Li}, Z.-X., {Gao}, H., {Ding}, X.-H., {Wang}, G.-J., \& {Zhang}, B. 2018,
  Nature Communications, 9, 3833, \dodoi{10.1038/s41467-018-06303-0}

\bibitem[{{Linder}(2011)}]{linder2011a}
{Linder}, E.~V. 2011, \prd, 84, 123529, \dodoi{10.1103/PhysRevD.84.123529}

\bibitem[{{Marshall} {et~al.}(2007){Marshall}, {Treu}, {Melbourne}, {Gavazzi},
  {Bundy}, {Ammons}, {Bolton}, {Burles}, {Larkin}, {Le Mignant}, {Koo},
  {Koopmans}, {Max}, {Moustakas}, {Steinbring}, \& {Wright}}]{marshall2007a}
{Marshall}, P.~J., {Treu}, T., {Melbourne}, J., {et~al.} 2007, \apj, 671, 1196,
  \dodoi{10.1086/523091}

\bibitem[{{Meneghetti} {et~al.}(2020){Meneghetti}, {Davoli}, {Bergamini},
  {Rosati}, {Natarajan}, {Giocoli}, {Caminha}, {Metcalf}, {Rasia}, {Borgani},
  {Calura}, {Grillo}, {Mercurio}, \& {Vanzella}}]{meneghetti2020a}
{Meneghetti}, M., {Davoli}, G., {Bergamini}, P., {et~al.} 2020, Science, 369,
  1347, \dodoi{10.1126/science.aax5164}

\bibitem[{{Metcalf} {et~al.}(2018){Metcalf}, {Meneghetti}, {Avestruz},
  {Bellagamba}, {Bom}, {Bertin}, {Cabanac}, {Davies}, {Decenci{\`e}re},
  {Flamary}, {Gavazzi}, {Geiger}, {Hartley}, {Huertas-Company}, {Jackson},
  {Jullo}, {Kneib}, {Koopmans}, {Lanusse}, {Li}, {Ma}, {Makler}, {Li},
  {Lightman}, {Enrico Petrillo}, {Serjeant}, {Sch{\"a}fer}, {Sonnenfeld},
  {Tagore}, {Tortora}, {Tuccillo}, {Valent{\'\i}n}, {Velasco-Forero}, {Verdoes
  Kleijn}, \& {Vernardos}}]{metcalf2018a}
{Metcalf}, R.~B., {Meneghetti}, M., {Avestruz}, C., {et~al.} 2018, arXiv
  e-prints, arXiv:1802.03609.
\newblock \doarXiv{1802.03609}

\bibitem[{{Monna} {et~al.}(2017){Monna}, {Seitz}, {Balestra}, {Rosati},
  {Grillo}, {Halkola}, {Suyu}, {Coe}, {Caminha}, {Frye}, {Koekemoer},
  {Mercurio}, {Nonino}, {Postman}, \& {Zitrin}}]{monna2017a}
{Monna}, A., {Seitz}, S., {Balestra}, I., {et~al.} 2017, \mnras, 466, 4094,
  \dodoi{10.1093/mnras/stx015}

\bibitem[{{Moustakas} {et~al.}(2012){Moustakas}, {Brownstein}, {Fadely},
  {Fassnacht}, {Gavazzi}, {Goodsall}, {Griffith}, {Keeton}, {Kneib},
  {Koekemoer}, {Koopmans}, {Marshall}, {Merten}, {Metcalf}, {Oguri},
  {Papovich}, {Rein}, {Ryan}, {Stewart}, \& {Treu}}]{moustakas2012a}
{Moustakas}, L.~A., {Brownstein}, J., {Fadely}, R., {et~al.} 2012, in American
  Astronomical Society Meeting Abstracts, Vol. 219, American Astronomical
  Society Meeting Abstracts \#219, 146.01

\bibitem[{{Narayan} \& {Bartelmann}(1996)}]{narayan1996a}
{Narayan}, R., \& {Bartelmann}, M. 1996, arXiv e-prints, astro.
\newblock \doarXiv{astro-ph/9606001}

\bibitem[{{Nightingale} {et~al.}(2019){Nightingale}, {Massey}, {Harvey},
  {Cooper}, {Etherington}, {Tam}, \& {Hayes}}]{nightingale2019a}
{Nightingale}, J.~W., {Massey}, R.~J., {Harvey}, D.~R., {et~al.} 2019, \mnras,
  489, 2049, \dodoi{10.1093/mnras/stz2220}

\bibitem[{{Oguri} \& {Marshall}(2010)}]{oguri2010a}
{Oguri}, M., \& {Marshall}, P.~J. 2010, \mnras, 405, 2579,
  \dodoi{10.1111/j.1365-2966.2010.16639.x}

\bibitem[{{Patr{\'\i}cio} {et~al.}(2019){Patr{\'\i}cio}, {Richard}, {Carton},
  {P{\'e}roux}, {Contini}, {Brinchmann}, {Schaye}, {Weilbacher}, {Nanayakkara},
  {Maseda}, {Mahler}, \& {Wisotzki}}]{patricio2019a}
{Patr{\'\i}cio}, V., {Richard}, J., {Carton}, D., {et~al.} 2019, \mnras, 489,
  224, \dodoi{10.1093/mnras/stz2114}

\bibitem[{{Petrillo} {et~al.}(2017){Petrillo}, {Tortora}, {Chatterjee},
  {Vernardos}, {Koopmans}, {Verdoes Kleijn}, {Napolitano}, {Covone},
  {Schneider}, {Grado}, \& {McFarland}}]{petrillo2017a}
{Petrillo}, C.~E., {Tortora}, C., {Chatterjee}, S., {et~al.} 2017, \mnras, 472,
  1129, \dodoi{10.1093/mnras/stx2052}

\bibitem[{{Petrillo} {et~al.}(2019){Petrillo}, {Tortora}, {Vernardos},
  {Koopmans}, {Verdoes Kleijn}, {Bilicki}, {Napolitano}, {Chatterjee},
  {Covone}, {Dvornik}, {Erben}, {Getman}, {Giblin}, {Heymans}, {de Jong},
  {Kuijken}, {Schneider}, {Shan}, {Spiniello}, \& {Wright}}]{petrillo2019a}
{Petrillo}, C.~E., {Tortora}, C., {Vernardos}, G., {et~al.} 2019, \mnras, 484,
  3879, \dodoi{10.1093/mnras/stz189}

\bibitem[{{Philcox} {et~al.}(2020){Philcox}, {Sherwin}, {Farren}, \&
  {Baxter}}]{philcox2020a}
{Philcox}, O. H.~E., {Sherwin}, B.~D., {Farren}, G.~S., \& {Baxter}, E.~J.
  2020, arXiv e-prints, arXiv:2008.08084.
\newblock \doarXiv{2008.08084}

\bibitem[{{Pierel} \& {Rodney}(2019)}]{pierel2019a}
{Pierel}, J.~D.~R., \& {Rodney}, S. 2019, \apj, 876, 107,
  \dodoi{10.3847/1538-4357/ab164a}

\bibitem[{{Planck Collaboration} {et~al.}(2020){Planck Collaboration},
  {Aghanim}, {Akrami}, {Ashdown}, {Aumont}, {Baccigalupi}, \&
  {Ballardini}}]{planck2020a}
{Planck Collaboration}, {Aghanim}, N., {Akrami}, Y., {et~al.} 2020, \aap, 641,
  A6, \dodoi{10.1051/0004-6361/201833910}

\bibitem[{{Pourrahmani} {et~al.}(2018){Pourrahmani}, {Nayyeri}, \&
  {Cooray}}]{pourrahmani2018a}
{Pourrahmani}, M., {Nayyeri}, H., \& {Cooray}, A. 2018, \apj, 856, 68,
  \dodoi{10.3847/1538-4357/aaae6a}

\bibitem[{{Quimby} {et~al.}(2014){Quimby}, {Oguri}, {More}, {More}, {Moriya},
  {Werner}, {Tanaka}, {Folatelli}, {Bersten}, {Maeda}, \&
  {Nomoto}}]{quimby2014a}
{Quimby}, R.~M., {Oguri}, M., {More}, A., {et~al.} 2014, Science, 344, 396,
  \dodoi{10.1126/science.1250903}

\bibitem[{{R{\"a}s{\"a}nen} {et~al.}(2015){R{\"a}s{\"a}nen}, {Bolejko}, \&
  {Finoguenov}}]{rasanen2015a}
{R{\"a}s{\"a}nen}, S., {Bolejko}, K., \& {Finoguenov}, A. 2015, \prl, 115,
  101301, \dodoi{10.1103/PhysRevLett.115.101301}

\bibitem[{{Refsdal}(1964)}]{refsdal1964a}
{Refsdal}, S. 1964, \mnras, 128, 307, \dodoi{10.1093/mnras/128.4.307}

\bibitem[{{Riess} {et~al.}(2019){Riess}, {Casertano}, {Yuan}, {Macri}, \&
  {Scolnic}}]{riess2019a}
{Riess}, A.~G., {Casertano}, S., {Yuan}, W., {Macri}, L.~M., \& {Scolnic}, D.
  2019, \apj, 876, 85, \dodoi{10.3847/1538-4357/ab1422}

\bibitem[{{Ritondale} {et~al.}(2019){Ritondale}, {Vegetti}, {Despali}, {Auger},
  {Koopmans}, \& {McKean}}]{ritondale2019a}
{Ritondale}, E., {Vegetti}, S., {Despali}, G., {et~al.} 2019, \mnras, 485,
  2179, \dodoi{10.1093/mnras/stz464}

\bibitem[{{Robertson} {et~al.}(2020){Robertson}, {Smith}, {Massey}, {Eke},
  {Jauzac}, {Bianconi}, \& {Ryczanowski}}]{robertson2020a}
{Robertson}, A., {Smith}, G.~P., {Massey}, R., {et~al.} 2020, \mnras, 495,
  3727, \dodoi{10.1093/mnras/staa1429}

\bibitem[{{Rodney} {et~al.}(2016){Rodney}, {Strolger}, {Kelly}, {Brada{\v{c}}},
  {Brammer}, {Filippenko}, {Foley}, {Graur}, {Hjorth}, {Jha}, {McCully},
  {Molino}, {Riess}, {Schmidt}, {Selsing}, {Sharon}, {Treu}, {Weiner}, \&
  {Zitrin}}]{rodney2016a}
{Rodney}, S.~A., {Strolger}, L.~G., {Kelly}, P.~L., {et~al.} 2016, \apj, 820,
  50, \dodoi{10.3847/0004-637X/820/1/50}

\bibitem[{{Shajib} {et~al.}(2020){Shajib}, {Treu}, {Birrer}, \&
  {Sonnenfeld}}]{shajib2020b}
{Shajib}, A.~J., {Treu}, T., {Birrer}, S., \& {Sonnenfeld}, A. 2020, arXiv
  e-prints, arXiv:2008.11724.
\newblock \doarXiv{2008.11724}

\bibitem[{{Shajib} {et~al.}(2019){Shajib}, {Birrer}, {Treu}, {Auger},
  {Agnello}, {Anguita}, {Buckley-Geer}, {Chan}, {Collett}, {Courbin},
  {Fassnacht}, {Frieman}, {Kayo}, {Lemon}, {Lin}, {Marshall}, {McMahon},
  {More}, {Morgan}, {Motta}, {Oguri}, {Ostrovski}, {Rusu}, {Schechter},
  {Shanks}, {Suyu}, {Meylan}, {Abbott}, {Allam}, {Annis}, {Avila}, {Bertin},
  {Brooks}, {Carnero Rosell}, {Carrasco Kind}, {Carretero}, {Cunha}, {da
  Costa}, {De Vicente}, {Desai}, {Doel}, {Flaugher}, {Fosalba},
  {Garc{\'\i}a-Bellido}, {Gerdes}, {Gruen}, {Gruendl}, {Gutierrez}, {Hartley},
  {Hollowood}, {Hoyle}, {James}, {Kuehn}, {Kuropatkin}, {Lahav}, {Lima},
  {Maia}, {March}, {Marshall}, {Melchior}, {Menanteau}, {Miquel}, {Plazas},
  {Sanchez}, {Scarpine}, {Sevilla-Noarbe}, {Smith}, {Soares-Santos},
  {Sobreira}, {Suchyta}, {Swanson}, {Tarle}, \& {Walker}}]{shajib2019a}
{Shajib}, A.~J., {Birrer}, S., {Treu}, T., {et~al.} 2019, \mnras, 483, 5649,
  \dodoi{10.1093/mnras/sty3397}

\bibitem[{{Shu} {et~al.}(2017){Shu}, {Brownstein}, {Bolton}, {Koopmans},
  {Treu}, {Montero-Dorta}, {Auger}, {Czoske}, {Gavazzi}, {Marshall}, \&
  {Moustakas}}]{shu2017a}
{Shu}, Y., {Brownstein}, J.~R., {Bolton}, A.~S., {et~al.} 2017, \apj, 851, 48,
  \dodoi{10.3847/1538-4357/aa9794}

\bibitem[{{Sonnenfeld} {et~al.}(2019){Sonnenfeld}, {Jaelani}, {Chan}, {More},
  {Suyu}, {Wong}, {Oguri}, \& {Lee}}]{sonnenfeld2019b}
{Sonnenfeld}, A., {Jaelani}, A.~T., {Chan}, J., {et~al.} 2019, \aap, 630, A71,
  \dodoi{10.1051/0004-6361/201935743}

\bibitem[{{Sonnenfeld} {et~al.}(2015){Sonnenfeld}, {Treu}, {Marshall}, {Suyu},
  {Gavazzi}, {Auger}, \& {Nipoti}}]{sonnenfeld2015a}
{Sonnenfeld}, A., {Treu}, T., {Marshall}, P.~J., {et~al.} 2015, \apj, 800, 94,
  \dodoi{10.1088/0004-637X/800/2/94}

\bibitem[{{Sonnenfeld} {et~al.}(2018){Sonnenfeld}, {Chan}, {Shu}, {More},
  {Oguri}, {Suyu}, {Wong}, {Lee}, {Coupon}, {Yonehara}, {Bolton}, {Jaelani},
  {Tanaka}, {Miyazaki}, \& {Komiyama}}]{sonnenfeld2018a}
{Sonnenfeld}, A., {Chan}, J.~H.~H., {Shu}, Y., {et~al.} 2018, \pasj, 70, S29,
  \dodoi{10.1093/pasj/psx062}

\bibitem[{{Suyu} {et~al.}(2020){Suyu}, {Huber}, {Ca{\~n}ameras}, {Schuldt},
  {Taubenberger}, {Y{\i}ld{\i}r{\i}m}, {Bonvin}, {Chan}, {Courbin}, {Kromer},
  {N{\"o}bauer}, {Sim}, \& {Sluse}}]{suyu2020a}
{Suyu}, S.~H., {Huber}, S., {Ca{\~n}ameras}, R., {et~al.} 2020, arXiv e-prints,
  arXiv:2002.08378.
\newblock \doarXiv{2002.08378}

\bibitem[{{Szegedy} {et~al.}(2014){Szegedy}, {Liu}, {Jia}, {Sermanet}, {Reed},
  {Anguelov}, {Erhan}, {Vanhoucke}, \& {Rabinovich}}]{szegedy2014a}
{Szegedy}, C., {Liu}, W., {Jia}, Y., {et~al.} 2014, arXiv e-prints,
  arXiv:1409.4842.
\newblock \doarXiv{1409.4842}

\bibitem[{{Taubenberger} {et~al.}(2019){Taubenberger}, {Suyu}, {Komatsu},
  {Jee}, {Birrer}, {Bonvin}, {Courbin}, {Rusu}, {Shajib}, \&
  {Wong}}]{taubenberger2019a}
{Taubenberger}, S., {Suyu}, S.~H., {Komatsu}, E., {et~al.} 2019, \aap, 628, L7,
  \dodoi{10.1051/0004-6361/201935980}

\bibitem[{{Tessore} {et~al.}(2016){Tessore}, {Bellagamba}, \&
  {Metcalf}}]{tessore2016a}
{Tessore}, N., {Bellagamba}, F., \& {Metcalf}, R.~B. 2016, \mnras, 463, 3115,
  \dodoi{10.1093/mnras/stw2212}

\bibitem[{{Treu}(2010)}]{treu2010a}
{Treu}, T. 2010, \araa, 48, 87, \dodoi{10.1146/annurev-astro-081309-130924}

\bibitem[{{Treu} {et~al.}(2011){Treu}, {Dutton}, {Auger}, {Marshall}, {Bolton},
  {Brewer}, {Koo}, \& {Koopmans}}]{treu2011a}
{Treu}, T., {Dutton}, A.~A., {Auger}, M.~W., {et~al.} 2011, \mnras, 417, 1601,
  \dodoi{10.1111/j.1365-2966.2011.19378.x}

\bibitem[{{Treu} \& {Marshall}(2016)}]{treu2016a}
{Treu}, T., \& {Marshall}, P.~J. 2016, Astronomy and Astrophysics Review, 24,
  11, \dodoi{10.1007/s00159-016-0096-8}

\bibitem[{{Tzeng} {et~al.}(2017){Tzeng}, {Hoffman}, {Saenko}, \&
  {Darrell}}]{tzeng2017a}
{Tzeng}, E., {Hoffman}, J., {Saenko}, K., \& {Darrell}, T. 2017, arXiv
  e-prints, arXiv:1702.05464.
\newblock \doarXiv{1702.05464}

\bibitem[{{Vanzella} {et~al.}(2020){Vanzella}, {Meneghetti}, {Caminha},
  {Castellano}, {Calura}, {Rosati}, {Grillo}, {Dijkstra}, {Gronke}, {Sani},
  {Mercurio}, {Tozzi}, {Nonino}, {Cristiani}, {Mignoli}, {Pentericci}, {Gilli},
  {Treu}, {Caputi}, {Cupani}, {Fontana}, {Grazian}, \&
  {Balestra}}]{vanzella2020a}
{Vanzella}, E., {Meneghetti}, M., {Caminha}, G.~B., {et~al.} 2020, \mnras, 494,
  L81, \dodoi{10.1093/mnrasl/slaa041}

\bibitem[{{Vegetti} {et~al.}(2018){Vegetti}, {Despali}, {Lovell}, \&
  {Enzi}}]{vegetti2018a}
{Vegetti}, S., {Despali}, G., {Lovell}, M.~R., \& {Enzi}, W. 2018, \mnras, 481,
  3661, \dodoi{10.1093/mnras/sty2393}

\bibitem[{{Vegetti} \& {Koopmans}(2009)}]{vegetti2009a}
{Vegetti}, S., \& {Koopmans}, L.~V.~E. 2009, \mnras, 400, 1583,
  \dodoi{10.1111/j.1365-2966.2009.15559.x}

\bibitem[{{Vegetti} {et~al.}(2014){Vegetti}, {Koopmans}, {Auger}, {Treu}, \&
  {Bolton}}]{vegetti2014a}
{Vegetti}, S., {Koopmans}, L.~V.~E., {Auger}, M.~W., {Treu}, T., \& {Bolton},
  A.~S. 2014, \mnras, 442, 2017, \dodoi{10.1093/mnras/stu943}

\bibitem[{{Verde} {et~al.}(2019){Verde}, {Treu}, \& {Riess}}]{verde2019a}
{Verde}, L., {Treu}, T., \& {Riess}, A.~G. 2019, Nature Astronomy, 3, 891,
  \dodoi{10.1038/s41550-019-0902-0}

\bibitem[{{Williams} {et~al.}(2004){Williams}, {Olszewski}, {Lesser}, \&
  {Burge}}]{williams2004a}
{Williams}, G.~G., {Olszewski}, E., {Lesser}, M.~P., \& {Burge}, J.~H. 2004, in
  \procspie, Vol. 5492, Ground-based Instrumentation for Astronomy, ed.
  A.~F.~M. {Moorwood} \& M.~{Iye}, 787--798

\bibitem[{{Wojtak} {et~al.}(2019){Wojtak}, {Hjorth}, \& {Gall}}]{wojtak2019a}
{Wojtak}, R., {Hjorth}, J., \& {Gall}, C. 2019, arXiv e-prints.
\newblock \doarXiv{1903.07687}

\bibitem[{{Wong} {et~al.}(2018){Wong}, {Sonnenfeld}, {Chan}, {Rusu}, {Tanaka},
  {Jaelani}, {Lee}, {More}, {Oguri}, {Suyu}, \& {Komiyama}}]{wong2018a}
{Wong}, K.~C., {Sonnenfeld}, A., {Chan}, J. H.~H., {et~al.} 2018, \apj, 867,
  107, \dodoi{10.3847/1538-4357/aae381}

\bibitem[{{Wong} {et~al.}(2019){Wong}, {Suyu}, {Chen}, {Rusu}, {Millon},
  {Sluse}, {Bonvin}, {Fassnacht}, {Taubenberger}, {Auger}, {Birrer}, {Chan},
  {Courbin}, {Hilbert}, {Tihhonova}, {Treu}, {Agnello}, {Ding}, {Jee},
  {Komatsu}, {Shajib}, {Sonnenfeld}, {Bland ford}, {Koopmans}, {Marshall}, \&
  {Meylan}}]{wong2019a}
{Wong}, K.~C., {Suyu}, S.~H., {Chen}, G. C.~F., {et~al.} 2019, arXiv e-prints,
  arXiv:1907.04869.
\newblock \doarXiv{1907.04869}

\bibitem[{{Zhou} {et~al.}(2020){Zhou}, {Newman}, {Mao}, {Meisner}, {Moustakas},
  {Myers}, {Prakash}, {Zentner}, {Brooks}, {Duan}, {Landriau}, {Levi}, {Prada},
  \& {Tarle}}]{zhou2020a}
{Zhou}, R., {Newman}, J.~A., {Mao}, Y.-Y., {et~al.} 2020, arXiv e-prints,
  arXiv:2001.06018.
\newblock \doarXiv{2001.06018}

\end{thebibliography}

%%%% -----------  Tables and Figures for Candidates ---------

\begin{minipage}{\linewidth}% to keep image and caption on one page
\makebox[\linewidth]{%        to center the image
\includegraphics[keepaspectratio=true,scale=0.26]{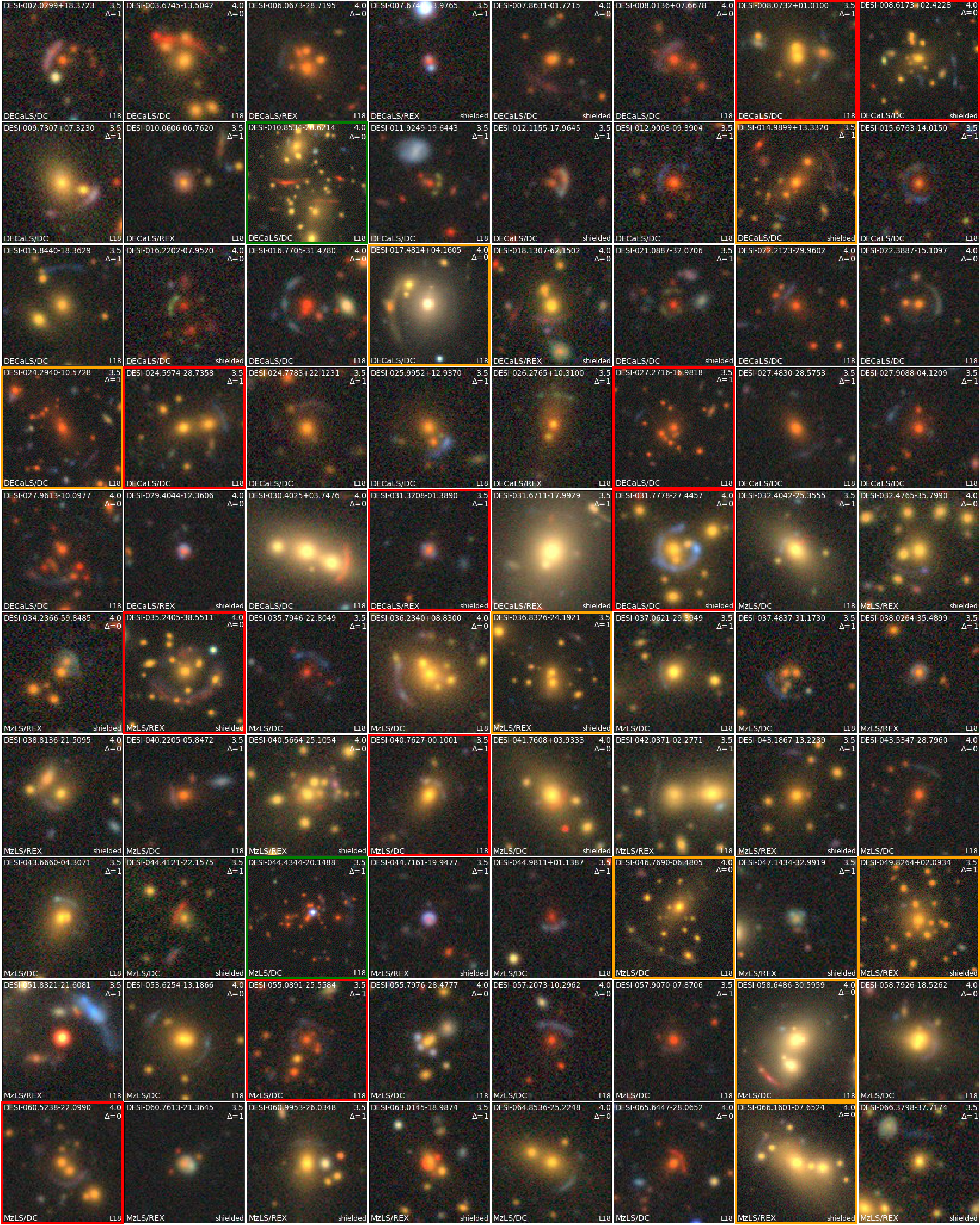}}
\captionof{figure}{ 
%\vspace{-70mm}
\rf{Eighty of t}he \ed{\lensA} Grade A candidates arranged in ascending RA.  
Top right corner indicates the average human inspection score with $\Delta$ being the absolute difference;
bottom left corner, the region and \tractor type (REX or DC = DEV or COMP);
and bottom right, the neural network model.
For each image, N is up, and E to the left.
%For this and all figures that follow:
Images without rims have a width of 101 pixels ($26.5\twopr$);
with orange rims, 151 pixel ($39.6\twopr$); 
and green rims, 201 pixel ($52.7\twopr$).
%; blue rims, 251 pixel ($65.8\twopr$); 
%and purple rims, 351 pixel ($92.0\twopr$).
Images with red rims are known lenses \rf{or candidates} but not included in our training sample, 
\ed{with citations given in Table~\ref{tab:grade-a-1}.
\rf{All \lenstot candidates are shown on the project website: \url{https://sites.google.com/usfca.edu/neuralens}}.}
%Tables~\ref{tab:grade-a} - \ref{tab:grade-c}}.
%unless otherwise noted, these are from \citet{jacobs2019b}.
}\label{fig:grade-a-1}  
 \end{minipage}

\clearpage
\startlongtable
\begin{deluxetable*}{lccccccc}
\tabletypesize{\footnotesize}
\tablecaption{Grade A Candidates\label{tab:grade-a-1}}
\tablehead{
    \colhead{Name} &
    \colhead{Type} &
    \colhead{mag\_g} &
    \colhead{mag\_r} &
    \colhead{mag\_z} &
    \colhead{Probability} &
    \colhead{$z_{spec}$}&
    \colhead{$z_{phot}$}
}
\startdata
          DESI-002.0299+18.3723 &   DC &  20.63 &  19.16 &  17.84 &         0.97 &  0.4724 &                 \\
          DESI-003.6745-13.5042 &   DC &  20.00 &  18.17 &  17.18 &         0.34 &         &  $0.406\pm0.021$ \\
          DESI-006.0673-28.7195 &  REX &  20.80 &  18.86 &  17.55 &         0.97 &         &  $0.533\pm0.025$ \\
          DESI-007.6741-33.9765 &  REX &  21.64 &  20.85 &  19.61 &         0.20 &         &  $0.858\pm0.041$ \\
          DESI-007.8631-01.7215 &   DC &  21.61 &  19.79 &  18.60 &         0.53 &  0.5201 &                 \\
          DESI-008.0136+07.6678 &   DC &  21.14 &  19.57 &  18.10 &         0.99 &  0.5566 &                 \\
    DESI-008.0732+01.0100$^{e}$ &   DC &  20.25 &  18.4 &  17.45 &         0.31 &         &  $0.426\pm0.091$ \\
  DESI-008.6173+02.4228$^{b,e}$ &   DC &  22.62 &  20.73 &  19.81 &         0.54 &  0.4548 &                 \\
          DESI-009.7307+07.3230 &   DC &  18.65 &  17.21 &  16.36 &         0.22 &  0.2547 &                 \\
          DESI-010.0606-06.7620 &  REX &  20.98 &  19.69 &  18.64 &         1.00 &  0.6152 &                 \\
          DESI-010.8534-20.6214 &   DC &  18.93 &  17.4 &  16.56 &         0.11 &  0.3381 &                 \\
          DESI-011.9249-19.6443 &   DC &  21.78 &  20.32 &  18.92 &         0.84 &         &  $0.688\pm0.056$ \\
          DESI-012.1155-17.9645 &   DC &  22.57 &  21.17 &  19.76 &         0.90 &         &  $0.735\pm0.069$ \\
          DESI-012.9008-09.3904 &   DC &  22.30 &  20.33 &  18.64 &         1.00 &  0.4485 &                 \\
          DESI-014.9899+13.3320 &   DC &  20.07 &  18.09 &  16.86 &         0.91 &  0.5163 &                 \\
          DESI-015.6763-14.0150 &   DC &  22.07 &  20.27 &  18.79 &         1.00 &         &  $0.658\pm0.036$ \\
          DESI-015.8440-18.3629 &   DC &  20.42 &  18.71 &  17.83 &         1.00 &         &  $0.364\pm0.019$ \\
          DESI-016.2202-07.9520 &   DC &  23.36 &  21.16 &  19.29 &         0.48 &         &  $0.764\pm0.032$ \\
          DESI-016.7705-31.4780 &   DC &  21.77 &  19.95 &  18.12 &         1.00 &         &  $0.772\pm0.018$ \\
          DESI-017.4814+04.1605 &   DC &  19.66 &  18.53 &  17.85 &         0.19 &         &  $0.281\pm0.140$ \\
          DESI-018.1307-62.1502 &  REX &  20.07 &  18.27 &  17.32 &         0.98 &         &  $0.427\pm0.024$ \\
          DESI-021.0887-32.0706 &   DC &  22.59 &  20.88 &  19.25 &         0.82 &         &  $0.754\pm0.080$ \\
          DESI-022.2123-29.9602 &   DC &  21.73 &  19.98 &  18.50 &         1.00 &         &  $0.649\pm0.023$ \\
          DESI-022.3887-15.1097 &   DC &  21.78 &  20.01 &  18.81 &         1.00 &         &  $0.524\pm0.063$ \\
          DESI-024.2940-10.5728 &   DC &  20.89 &  18.92 &  17.30 &         0.87 &  0.4135 &                 \\
    DESI-024.5974-28.7358$^{d}$ &   DC &  20.42 &  18.60 &  17.62 &         1.00 &         &  $0.414\pm0.022$ \\
          DESI-024.7783+22.1231 &   DC &  21.06 &  19.09 &  17.93 &         0.13 &  0.4684 &                 \\
          DESI-025.9952+12.9370 &   DC &  20.75 &  18.86 &  17.80 &         0.92 &  0.5114 &                 \\
          DESI-026.2765+10.3100 &  REX &  21.56 &  20.16 &  19.51 &         0.28 &         &  $0.462\pm0.138$ \\
    DESI-027.2716-16.9818$^{d}$ &   DC &  20.91 &  19.07 &  17.65 &         0.92 &  0.6916 &                 \\
          DESI-027.4830-28.5753 &   DC &  21.32 &  19.43 &  18.14 &         0.97 &         &  $0.557\pm0.008$ \\
          DESI-027.9088-04.1209 &   DC &  21.89 &  20.11 &  18.58 &         0.81 &  0.6412 &                 \\
          DESI-027.9613-10.0977 &   DC &  22.05 &  20.36 &  18.87 &         1.00 &  0.5234 &                 \\
          DESI-029.4044-12.3606 &  REX &  20.77 &  20.19 &  19.20 &         0.95 &         &  $0.747\pm0.169$ \\
          DESI-030.4025+03.7476 &   DC &  19.16 &  17.93 &  17.16 &         0.64 &  0.1696 &                 \\
    DESI-031.3208-01.3890$^{d}$ &  REX &  21.01 &  20.18 &  19.05 &         1.00 &  0.6992 &                 \\
          DESI-031.6711-17.9929 &  REX &  20.49 &  19.62 &  19.27 &         0.10 &  0.1548 &                 \\
    DESI-031.7778-27.4457$^{d}$ &   DC &  18.78 &  17.26 &  16.37 &         1.00 &         &  $0.294\pm0.018$ \\
          DESI-032.4042-25.3555 &   DC &  18.15 &  16.87 &  16.13 &         0.49 &         &  $0.224\pm0.015$ \\
          DESI-032.4765-35.7990 &  REX &  21.42 &  19.78 &  18.95 &         0.98 &         &  $0.380\pm0.076$ \\
          DESI-034.2366-59.8485 &  REX &  21.01 &  19.34 &  18.29 &         0.66 &         &  $0.498\pm0.017$ \\
    DESI-035.2405-38.5511$^{d}$ &  REX &  18.98 &  17.32 &  16.39 &         0.45 &         &  $0.407\pm0.038$ \\
          DESI-035.7946-22.8049 &   DC &  22.44 &  21.10 &  19.33 &         1.00 &         &  $0.890\pm0.048$ \\
          DESI-036.2340+08.8300 &   DC &  18.15 &  16.50 &  15.61 &         1.00 &         &  $0.326\pm0.035$ \\
          DESI-036.8326-24.1921 &  REX &  19.27 &  17.43 &  16.46 &         0.39 &         &  $0.405\pm0.022$ \\
          DESI-037.0621-29.3949 &   DC &  19.74 &  18.11 &  17.25 &         0.93 &         &  $0.312\pm0.013$ \\
          DESI-037.4837-31.1730 &   DC &  22.06 &  20.28 &  19.17 &         0.99 &         &  $0.466\pm0.075$ \\
          DESI-038.0264-35.4899 &  REX &  21.31 &  20.00 &  18.90 &         1.00 &         &  $0.550\pm0.097$ \\
          DESI-038.8136-21.5095 &  REX &  22.15 &  20.64 &  19.83 &         0.94 &         &  $0.328\pm0.035$ \\
          DESI-040.2205-05.8472 &   DC &  21.68 &  19.81 &  18.52 &         0.97 &  0.5238 &                 \\
          DESI-040.5664-25.1054 &  REX &  21.68 &  20.12 &  19.31 &         0.82 &         &  $0.325\pm0.045$ \\
    DESI-040.7627-00.1001$^{d}$ &   DC &  19.95 &  18.14 &  17.19 &         1.00 &  0.4127 &                 \\
          DESI-041.7608+03.9333 &   DC &  18.53 &  16.99 &  16.14 &         0.30 &  0.2603 &                 \\
          DESI-042.0371-02.2771 &  REX &  20.04 &  18.66 &  17.90 &         0.93 &         &  $0.247\pm0.018$ \\
          DESI-043.1867-13.2239 &  REX &  22.31 &  20.58 &  19.67 &         0.67 &         &  $0.431\pm0.033$ \\
          DESI-043.5347-28.7960 &   DC &  22.19 &  20.30 &  18.82 &         0.93 &         &  $0.632\pm0.021$ \\
          DESI-043.6660-04.3071 &   DC &  19.35 &  17.68 &  16.79 &         1.00 &  0.2880 &                 \\
          DESI-044.4121-22.1575 &   DC &  21.01 &  19.36 &  18.49 &         0.46 &         &  $0.484\pm0.070$ \\
          DESI-044.4344-20.1488 &   DC &  23.07 &  21.09 &  19.42 &         0.49 &         &  $0.711\pm0.025$ \\
          DESI-044.7161-19.9477 &  REX &  20.04 &  19.79 &  19.07 &         0.89 &         &  $0.535\pm0.327$ \\
          DESI-044.9811+01.1387 &   DC &  21.54 &  20.39 &  18.66 &         1.00 &         &  $0.834\pm0.079$ \\
          DESI-046.7690-06.4805 &   DC &  22.11 &  20.70 &  19.80 &         0.32 &         &  $0.480\pm0.105$ \\
          DESI-047.1434-32.9919 &  REX &  20.70 &  19.76 &  19.24 &         0.22 &         &  $0.500\pm0.091$ \\
          DESI-049.8264+02.0934 &  REX &  22.33 &  20.35 &  19.37 &         0.90 &  0.3189 &                 \\
          DESI-051.8321-21.6081 &  REX &  19.43 &  18 &  16.69 &         0.16 &         &  $0.502\pm0.201$ \\
          DESI-053.6254-13.1866 &   DC &  19.51 &  17.72 &  16.79 &         1.00 &         &  $0.362\pm0.010$ \\
    DESI-055.0891-25.5584$^{d}$ &   DC &  21.94 &  20.17 &  18.70 &         1.00 &         &  $0.667\pm0.047$ \\
          DESI-055.7976-28.4777 &   DC &  20.93 &  19.27 &  18.38 &         1.00 &         &  $0.447\pm0.049$ \\
          DESI-057.2073-10.2962 &   DC &  21.97 &  20.33 &  18.65 &         1.00 &         &  $0.745\pm0.075$ \\
          DESI-057.9070-07.8706 &   DC &  21.98 &  20.28 &  18.72 &         1.00 &         &  $0.677\pm0.051$ \\
          DESI-058.6486-30.5959 &   DC &  16.63 &  15.50 &  14.74 &         0.12 &         &  $0.173\pm0.021$ \\
          DESI-058.7926-18.5262 &   DC &  18.30 &  16.76 &  15.93 &         1.00 &         &  $0.282\pm0.011$ \\
    DESI-060.5238-22.0990$^{d}$ &   DC &  21.11 &  19.32 &  18.25 &         1.00 &         &  $0.367\pm0.051$ \\
          DESI-060.7613-21.3645 &  REX &  20.19 &  19.34 &  18.83 &         0.28 &         &  $0.455\pm0.100$ \\
          DESI-060.9953-26.0348 &  REX &  20.63 &  19.74 &  19.12 &         0.80 &         &  $0.289\pm0.083$ \\
          DESI-063.0145-18.9874 &  REX &  22.06 &  20.27 &  19.30 &         0.18 &         &  $0.572\pm0.073$ \\
          DESI-064.8536-25.2248 &   DC &  19.51 &  17.72 &  16.82 &         0.98 &         &  $0.360\pm0.012$ \\
          DESI-065.6447-28.0652 &   DC &  21.63 &  19.81 &  18.31 &         0.53 &         &  $0.641\pm0.036$ \\
          DESI-066.1601-07.6524 &  REX &  21.62 &  20.45 &  19.71 &         0.79 &         &  $0.278\pm0.049$ \\
          DESI-066.3798-37.7174 &  REX &  20.79 &  19.16 &  18.32 &         0.23 &         &  $0.290\pm0.041$ \\
\enddata
\tablecomments{\rf{Eighty of the} 216 Grade~A lens candidates \rf{are listed in this table}.
\rf{All \lenstot candidates are shown on the project website: \url{https://sites.google.com/usfca.edu/neuralens}}.
The spectroscopic redshifts are from SDSS DR16,
all with uncertainties $< \zerrmaxA \times 10^{-4}$.
References for known lenses or candidates are as follows: 
$^{a}$\citet{canameras2020a}, 
$^{b}$\citet{carrasco2017a}, 
$^{c}$\citet{inada2003a}, 
$^{d}$\citet{jacobs2019b}. 
%$^{e}$\citet{jaelani2020a}, 
%$^{f}$\citet{lemon2020a}, 
%$^{g}$\citet{petrillo2019a}, 
%$^{h}$\citet{sharon2020a}, 
%$^{i}$\citet{sonnenfeld2013a}, 
%$^{j}$\citet{sonnenfeld2018a}, 
%$^{k}$\citet{sonnenfeld2020a}, 
%$^{l}$\citet{wong2018a}.
}
\end{deluxetable*}

\end{document}